\documentclass[12pt,twoside,a4paper]{report}
\usepackage[usenames]{color}
\usepackage{fancyheadings}
\usepackage{mystyle}
\usepackage{epsfig}
\usepackage{tabularx}
\usepackage{graphicx}
\usepackage{amssymb,amsfonts,amsmath}
\renewcommand\[{\left[}

\newcommand\eeq{\end{equation}}
\newcommand\beq{\begin{equation}}
\newcommand\eea{\end{eqnarray}}
\newcommand\bea{\begin{eqnarray}}

\newcommand\lsim{\mathrel{\rlap{\lower4pt\hbox{\hskip1pt$\sim$}}
    \raise1pt\hbox{$<$}}}
\newcommand\gsim{\mathrel{\rlap{\lower4pt\hbox{\hskip1pt$\sim$}}
    \raise1pt\hbox{$>$}}}

\def\dslash{\not{\hbox{\kern-2pt $\partial$}}}
\def\Dslash{\not{\hbox{\kern-4pt $D$}}}
\def\Oslash{\not{\hbox{\kern-4pt $O$}}}
\def\Qslash{\not{\hbox{\kern-4pt $Q$}}}
\def\pslash{\not{\hbox{\kern-2.3pt $p$}}}
\def\kslash{\not{\hbox{\kern-2.3pt $k$}}}
\def\qslash{\not{\hbox{\kern-2.3pt $q$}}}
 \newtoks\slashfraction
 \slashfraction={.13}
 \def\slash#1{\setbox0\hbox{$ #1 $}
 \setbox0\hbox to \the\slashfraction\wd0{\hss \box0}/\box0 }

\begin{document}

%%%%%%%%%%%%%%%%%%% Titlepage %%%%%%%%%%%%%%%%%%%%%%%%%%%%%%%%%%%%

\parindent 0cm
%\renewcommand{\baselinestretch}{1.1}
%%%%%%%%%%%%%%           Cover           %%%%%%%%%%%%%%%%%%%%%%%%
\thispagestyle{empty}

\vspace*{-2cm}
\hspace*{-2cm}
\begin{tabular}{lc}
\parbox{3cm}{
\includegraphics[width=3cm]{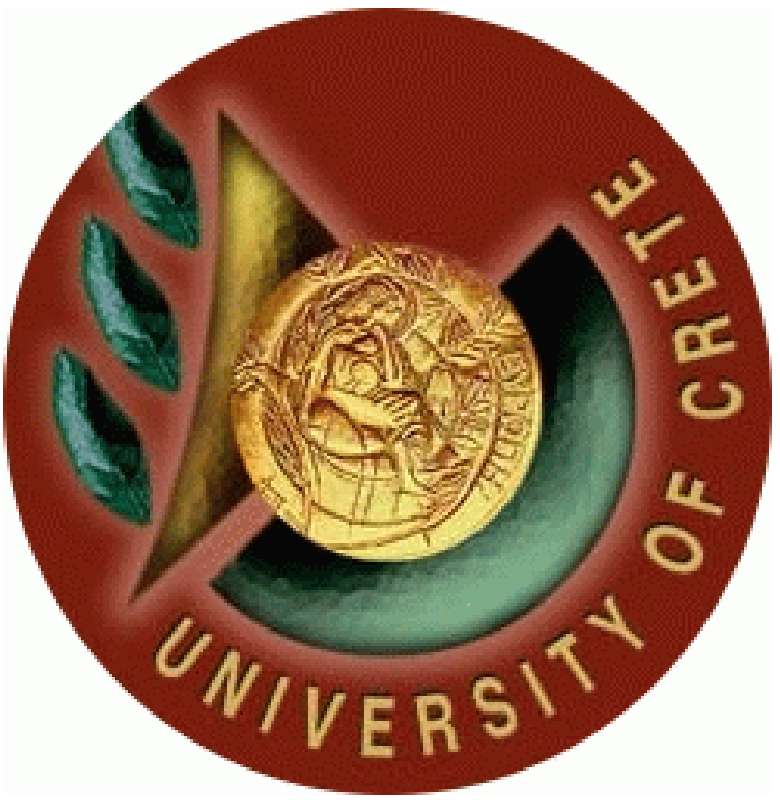}}
&
\parbox{10cm}{
\begin{center}
{\Large \sc University of Crete}\\[2mm]
{\Large \sc Department of Physics} 

\end{center}
}
\end{tabular}

\begin{center}

\vspace{2cm}

{\LARGE  On vortices and solitons in Goldstone }\\[2mm]
{\LARGE  and abelian-Higgs models} \\[2mm]
%{\LARGE } \\[2mm] 

\vspace{5mm}

%\begin{flushright}
%\large A Doctoral Dissertation ~~~~~~~~~
%\end{flushright}

\vspace{15mm}

{\Large C.G.K. Doudoulakis} 

\vspace{1cm}

\vspace{1cm}

{\large Heraklion, Greece, 11 October 2007}

\end{center}

\thispagestyle{empty}
\cleardoublepage

\parindent 0cm
%\renewcommand{\baselinestretch}{1.1}
%%%%%%%%%%%%%%           Cover           %%%%%%%%%%%%%%%%%%%%%%%%

\thispagestyle{empty}

\vspace*{-2cm}
\hspace*{-2cm}
%\begin{tabular}{lc}
%\parbox{3cm}{
%\includegraphics[width=3cm]{UOC_rgb.eps}}
%&
%\parbox{10cm}{
%\begin{center}
%{\Large \sc University of Crete}\\[2mm]
%{\Large \sc Department of Physics} 

%\end{center}
%}
%\end{tabular}

\begin{center}

\vspace{2cm}

{\LARGE  $ E\upsilon\chi a\rho\iota\sigma\tau\acute{\omega}\;\;\; \tau o \;\;\; \Theta\epsilon \acute{o} ... $ }\\[2mm]

%\vspace{5mm}

%\begin{flushright}
%\large A Doctoral Dissertation ~~~~~~~~~
%\end{flushright}

%\vspace{15mm}

%{\Large C.G.K. Doudoulakis} 

%\vspace{1cm}

%\includegraphics[width=7cm]{.eps}

%\vspace{1cm}

%{\large Heraklion, Greece, }

\end{center}

\thispagestyle{empty}
\cleardoublepage

%%%%%%%%%%%%%%%%%%%%%%%%%%%%%%%%%%%%%%%%%%%%%%%%%%%%%%%%%%%%%%%%%
%%%%%%%%%%%%%%            Headers       %%%%%%%%%%%%%%%%%%%%%%%%%
%%%%%%%%%%%%%%%%%%%%%%%%%%%%%%%%%%%%%%%%%%%%%%%%%%%%%%%%%%%%%%%%%
%%%%%%%%%%%%%%           Cover           %%%%%%%%%%%%%%%%%%%%%%%%
\parindent 0.7cm %0.7
\parskip 1.5mm %1.5
\thispagestyle{empty}
\newcommand{\myspace}[0]{78.1pt}

\begin{center}

{\large {Christos G. K. Doudoulakis}} 

\vspace*{\myspace}

%{\Large \sc {Aspects of Gauge Theories: On Vortices and Solitons in Goldstone and Abelian-Higgs Models}}
{\Large \sc {$  \Delta\acute{\iota}\nu\epsilon\varsigma \;\; \kappa\alpha\iota \;\; \sigma o\lambda\iota\tau \acute{o}\nu\iota\alpha \;\; \sigma\tau o \;\;
\mu o\nu\tau\acute{\epsilon}\lambda o \;\; Goldstone$ \\
$\kappa\alpha\iota \;\; \sigma\epsilon \;\;  \alpha\beta\epsilon\lambda\iota\alpha\nu\acute{\alpha}  \;\; \mu o\nu\tau\acute{\epsilon}\lambda\alpha \;\; Higgs$   }}

\vspace*{\myspace} 

\begin{flushright}
\begin{minipage}{310pt}
%\begin{center}
 {Thesis \\
submitted to the Department of Physics, University of Crete \\
 for the Degree of Doctor of Philosophy in Physics}\\
%\end{center}
\end{minipage}
\end{flushright}

\vspace*{\myspace}

\includegraphics[width=3.5cm]{UOC_rgb.eps}

\vspace*{\myspace}

Heraklion, Greece, 11 October 2007
\end{center}

\thispagestyle{empty}
\cleardoublepage

%%%%%%%%%%%%%%%%%%%% Committee %%%%%%%%%%%%%%%%%%%%%%%%%%%%%%%%%%

\thispagestyle{empty}

\begin{center}

\vspace*{2cm}

{\Large 
On vortices and solitons in Goldstone \\ and abelian-Higgs models \\
\vspace{0.8cm}
$\Delta\acute{\iota}\nu\epsilon\varsigma \;\; \kappa\alpha\iota \;\; \sigma o\lambda\iota\tau \acute{o}\nu\iota\alpha \;\; \sigma\tau o \;\;
\mu o\nu\tau\acute{\epsilon}\lambda o \;\; Goldstone$ \\
$\kappa\alpha\iota \;\; \sigma\epsilon \;\;  \alpha\beta\epsilon\lambda\iota\alpha\nu\acute{\alpha}  \;\; \mu o\nu\tau\acute{\epsilon}\lambda\alpha \;\; Higgs$   
}\\

\vspace*{2.5cm}

\begin{tabular}{ll}
Thesis author        &Christos G. K. Doudoulakis                    \\[3em]
%\multirow{2}*
{Thesis supervisor } & Prof. T. N. Tomaras \\ [3em]
{Thesis committee}  & T. N. Tomaras \\
                   & E. Kiritsis \\
                   & N. Tsamis \\
                   & X. Zotos  \\
                   & G. Tsironis \\
                   & P. Ditsas\\
                   & A. Petkou \\
                   
\end{tabular}
%\vspace{12em}
\vspace{2cm}
%\addcontentsline{toc}{section}{Thesis Committee}

Department of Physics, University of Crete\\
Heraklion, Greece \\[15pt]
11 October 2007
\end{center}

% %%%%%%%%%%%%%%%%%%%%%%%%%%% Dedication %%%%%%%%%%%%%%%%%%%%%%%%%%
% \thispagestyle{empty}
% \vspace*{300pt}
% \newenvironment{dedication} {
% %\thispagestyle{empty} 
% \vspace*{\stretch{1}} 
% %\begin{center} \em} {\end{center} \vspace*{\stretch{3}} \clearpage} 
% \begin{flushright} \em} {\end{flushright} \vspace*{\stretch{3}} \clearpage} 
% \begin{dedication} 
%  \end{dedication}
%  \cleardoublepage
%%%%%%%%%%%%%%%%%%%%%%%%%%%%%%%%%%%%%%%%%%%%%%%%%%%%%%%%%%%%%%%%%%

%\setcounter{equation}{0}
%\pacs{75.30.Ds, 75.30.Gw, 75.30.Kz}

\thispagestyle{empty}
\cleardoublepage

%\pagenumbering{roman}

\tableofcontents

%\listoffigures

%\listoftables

%%%%%%%%%%%%%%%%%%%%%%%%%%%%%%%%%%%%%%%%%%%%%%%%%%%%%%%%%%%%%%%%%%

\chapter*{Acknowledgements}

\addcontentsline{toc}{section}{Acknowledgements}

It is a great pleasure to thank my advisor, Professor Theodore N. Tomaras. 
His great experience and  ability to answer my questions in a simple and attractive way, were very helpful for me.

I would like to thank the Physics Department of the University of Crete for its warm hospitality and financial support all these years.

I would like to thank Dr. M.D.Kapetanakis for useful discussions, advice and support in difficult moments. We've passed together 11 years
since our first day in Physics department. I would also like to thank N.Lagos for useful discussions and support in difficult moments.

I would like to thank Dr. E.D.M.Kavoussanaki for collaboration, useful discussions and advice.

Finally, I would like to thank my parents, for their love and support.

\chapter*{Abstract}

\addcontentsline{toc}{section}{Abstract}

In the present work we discuss non-linear physics problems such as
Nielsen-Olesen strings, superconducting bosonic straight strings and static vortex rings.
We start with a toy model. We search for antiperiodic solitons of the Goldstone model
on a circle. Such models provide the basis as well as useful hints for further research on 
three-dimensional more realistic problems.
We proceed with a full research on a $U(1)$ model which admits stable straight string solutions
in a small, numerically determined area. That model has a Ginzburg-Landau potential with a cubic term added to it and can be found
in condensed matter problems as well. The next part of our research, has to do with a $U(1)\times U(1)$ model 
which is the main subject of our interest. There, we search for stable
axially symmetric solutions which are solitons, which can represent particles, the mass of which is of the order of TeV. 
The confirmation or rejection of the existence of those defects is of great interest
if we consider that LHC will work in the same energy range. In our study, we find out that due to current
quenching, these vortex rings seem to be unstable. We also extend the model of vortex rings by adding
higher derivative terms which are favorable for stability. After the extensive analysis we performed, we conclude
that these objects don't seem to be stable. The reasons which brought us to this conclusion are explained.

\chapter{Introduction}

\pagestyle{fancy}
%\pagenumbering{arabic}

In order to study topological defects, in general we need to define the order parameter, to  identify the pattern of broken symmetries
and to classify the defects we have. A {\em defect} is defined as a space region containing the discontinuity of the order 
parameter. In field theory, the order parameter is represented by scalar fields.
The topological defect is the singularity in the continuum that cannot be removed by continuous deformation
of the local region in the vicinity of the singularity. Strings, monopoles, textures, domain walls or combination of them
are some examples of defects. Strings especially are of interest to us.

Current theories of particle physics place chronologically the formation of topological defects
on the violent stages of the early evolution of the universe.
If somehow their existence is observationally proved, that would be of great importance for particle physics 
and cosmology, as the notion of topological defects exists in most interesting models 
of high energy physics trying to describe the early universe. Those defects would be
a direct consequence of events of the very early universe.

\begin{figure}
\centering
\includegraphics[scale=1.1]{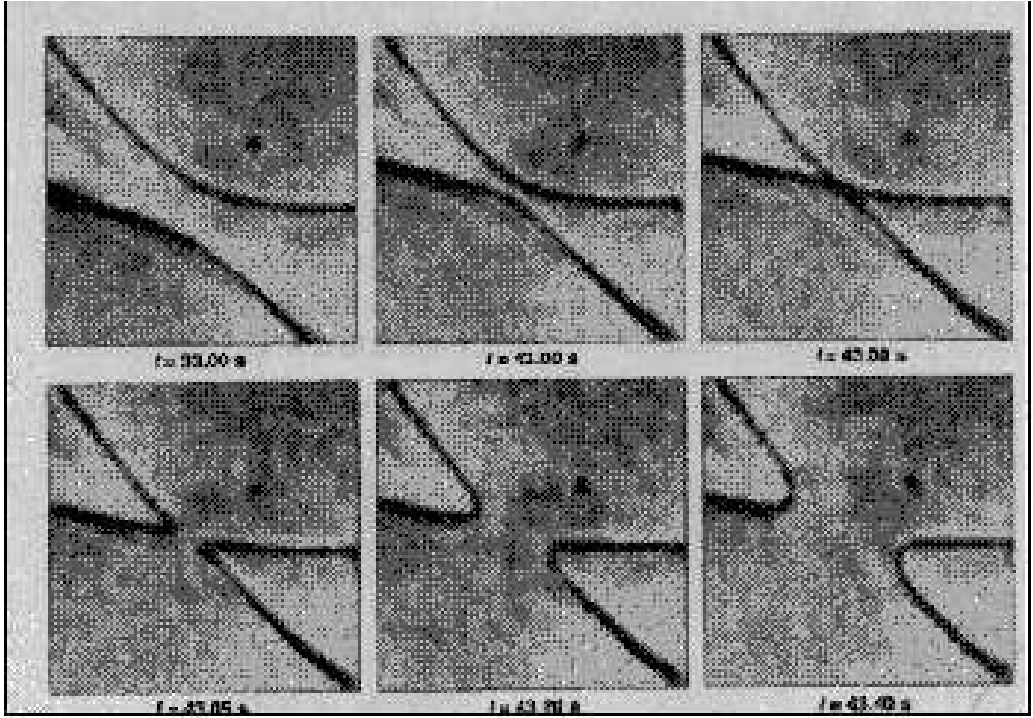}
\caption{\small String intercommutation in a nematic liquid crystal. Further details can be found in \cite{nl}. \label{nematic}}
\end{figure}

\begin{figure}
\centering
\includegraphics[scale=1.6]{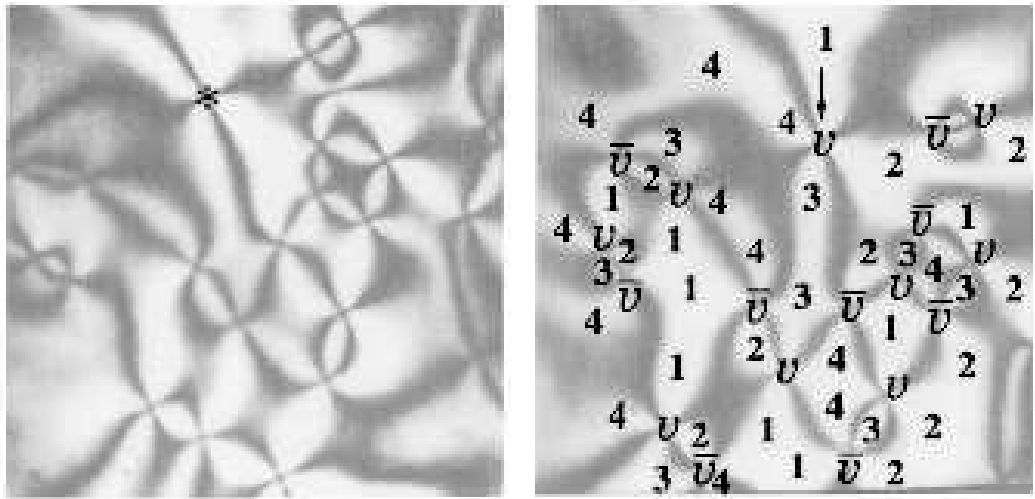}
\caption{\small Vortex-antivortex formation in a nematic liquid crystal. Further details can be found in \cite{nl2}. \label{nematic2}}
\end{figure}

The same notion is also present in condensed matter physics. 
Simple examples are the domains in a ferromagnet or the magnetic vortices in thin ferromagnetic films \cite{paptom} or the
vortices in Ginzburg-Landau superconductors \cite{paptom2}.
Domains are regions in which the magnetic dipoles are aligned, 
separated by domain walls. Liquid crystals exhibit an array of topological defects, such as strings and monopoles. 
Defects can also been found in biochemistry, notably in the process of protein folding.
A more specific example is that when liquid helium is quickly
pressure-quenched around a critical temperature, namely $T_{c}$, Zurek argued that vortices carrying rotation in quantized amounts
form and represent defects analogous to cosmic strings. More recently, the existence of vorton solutions
is involved in sectors such as QCD \cite{condmat2} and high-$T_{c}$ superconductivity \cite{condmat1}.
There are differences from the cosmological case where relativistic dynamics must be used and gravity is important, but their formation
can offer interesting hints for cosmology \cite{s3b,s3c,s3d}. One can also think of Bose-Einstein condensates (BEC) (i.e. see \cite{hau}), 
vortices in superfluid Helium-$3$ and Helium-$4$ \cite{p2,wink} , or nematic liquid crystals \cite{nl,nl2} (see also figs.\ref{nematic}, \ref{nematic2}).

Interest in such defects appears in many branches of theoretical and experimental physics
such as particle physics models, experimental high energy physics, cosmology, observational astrophysics, experimental condensed matter
physics and, recently, superstring theory.
In contrast to other branches of physics, it is established that {\bf no} such topologically stable defects can exist in the Standard Model  (SM)  of
Particle Physics. But, as pointed out in \cite{c1}, there is the possibility of quasi-topological metastable defects in popular extensions of the SM such
as the two-Higgs Standard Model (2HSM) \cite{c1x}. This was proved to be the case for codimension one and two defects \cite{c1x}, but the search for particle-like  solitons
of this kind was not so far succesfull \cite{c3}. Such solitons, if they exist, in the electroweak theory are very interesting. They will have a mass of a
few TeV, and should soon be produced in the LHC at CERN. Search for such objects is the main topic of the present work.

\section{Formation of  defects: The Kibble mechanism }

\noindent
The explanation of the formation of topological defects, uses a very important  notion of particle physics which 
is that of {\em spontaneous symmetry breaking}. This phenomenon appears when the ground state of the system
is characterised by a non-zero expectation value of the Higgs field and does not exhibit the full symmetry
of the Lagrangian. It is believed that the way to unification 
of all fundamental natural forces involves this notion. 
The {\em Higgs field}, named after the British physicist Peter Higgs, is a postulated quantum field, 
mediated by the Higgs boson, which is believed to permeate the entire universe. 
Its presence is said to be required in order to explain the large difference in mass between those particles 
which mediate weak interactions (the $W$ and $Z$ bosons) and that which mediates electromagnetic interactions (the photon).

\begin{figure}
\centering
\includegraphics[scale=0.65]{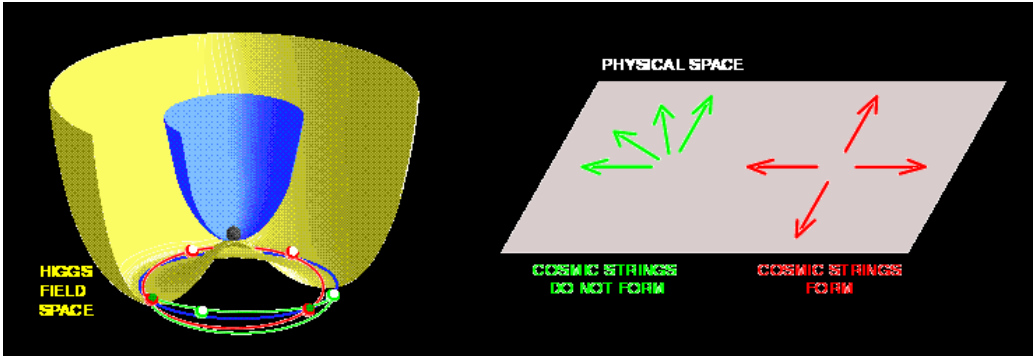}
\caption{\small Higgs field space and physical space. Above $T_{c}$ the potential has a 
minimum at $\phi_{1}=0=\phi_{2}$, but below $T_{c}$ the new form of the potential 
has its minima on a circle. If $m$ is an element of the coset space $G/H$ and $\Psi_{1}$ 
and $\Psi_{2}$ two different states at that circle, then $m \Psi_{1} = \Psi_{2}$. Generally, every element of the coset
space $G/H$ can produce all the other minimum energy states when acting on any of them.
Thus, $G/H = \mathcal{M}$ where $\mathcal{M}$ stands for the vacuum manifold. On the right side of the figure
one can have a picture of how a cosmic string forms when different regions
of space met, having their Higgs field phases as above. Picture from astro-ph/0303504 (A.Gangui).\label{f1}}
\end{figure}
This is a spin $0$ field which
signals the breaking of some symmetry in our theory.
This means that when our system goes from a state of higher symmetry to a final state 
where it obeys a smaller group of symmetry rules, this field represents the 
symmetry breaking {\em order parameter} which acquires a non-zero vacuum expectation value.  
In the stages we described above, 
this field can settle into different ground states or to be more specific,
the phase of the Higgs field can acquire different values.
Topological defects are objects that locally restore the original symmetry.
More details about spontaneous symmetry breaking will be given in the next chapter.

In the context of the standard Big Bang theory, 
the spontaneous breaking of fundamental symmetries is realized through 
phase transitions in the early universe. The time needed for such 
a transition is small compared to the expansion time. 
It is believed that all forces were united in the early stages of our cosmos 
and that these phase transitions play a significant role in order to end up 
to what we today know as four different natural forces. 
It is also believed that these transitions are the source of defect formation 
and if that's true, then there is an important link between particle physics and cosmology.

\begin{figure}
\centering
\includegraphics[scale=0.65]{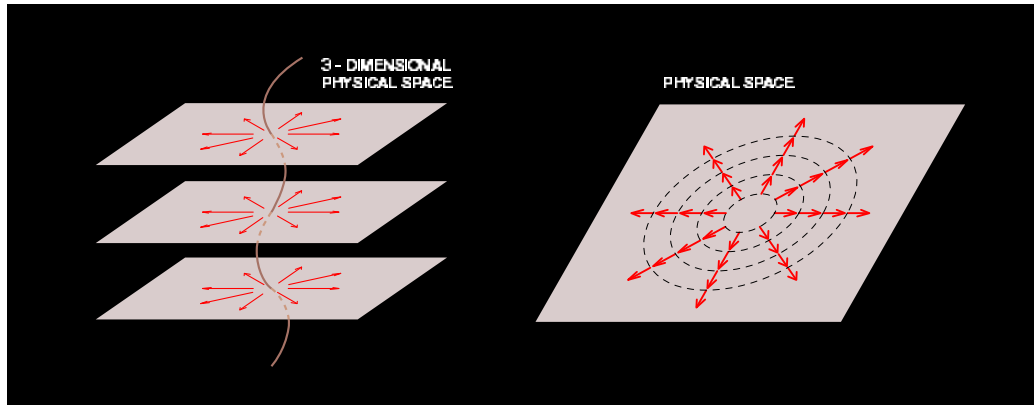}
\caption{\small Cosmic string formation. Picture from astro-ph/0303504 (A.Gangui).\label{f3}}
\end{figure}

Kibble in 1976 \cite{cc1} first noted that, in an expanding universe where are separated regions 
that have no "communication" amongst themselves, due to lack of causal contact, 
we can assume that they have different  values of the phase of the Higgs field. 
As the universe expands and cools down, these regions expand and some of them come in contact with others.
Thus, it is not impossible that by following a closed path through these regions that are in contact,
their phases (orientation of the Higgs field)  vary from $0$ to $2 n \pi$, where $n \neq 0$
is an integer and stands for the {\em winding number} of the field.
Then, the ``meeting point'' between these ``bubbles''  is the core 
of a cosmic string (figs.\ref{f1} and \ref{f3}). In other words, a cosmic string forms when, for purely topological reasons,
the orientation of the arrows cannot be adjusted in order to keep the Higgs field in the minimum 
energy state everywhere.

\section{Classification of defects}

\noindent
The different kinds of defects are due to different symmetries of the
Higgs field itself. For example, the Higgs field may have two possible states to fall into.
Then, when regions that have taken opposite choices meet, their boundary is a {\em domain wall}.
This narrow  region of the defect is called {\em false vacuum} and should not exist but it does 
so, due to the geometry of the Higgs field. Now, if Higgs field has circular symmetry
it can be represented by a ``mexican-hat'' analogy because ground state can lie anywhere
within a circle in field space and its position can be denoted by an arrow. This is the case 
shown in the previous section and the defect involved there was a cosmic string.

\begin{figure}[t]
\centering
\includegraphics[scale=0.55]{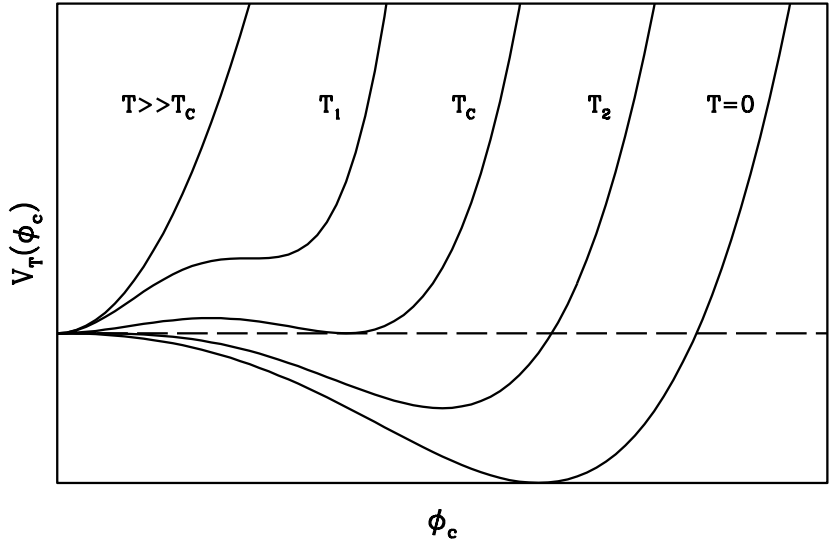}
\caption{\small Temperature dependent potential. Here is an example of such  potential
and how it changes shape as temperature falls. It is clearly visible that above $T_{c}$
the system has one stable minimum at $\phi_{c}=0$. As $T \rightarrow 0$ the potential 
develops a minimum at $\phi_{c}\neq 0$ and when the system
settles there, we say that the initial symmetry of the system is broken. 
The ground state of the model does not respect the original symmetry of the Lagrangian.
Picture from astro-ph/0303504 (A.Gangui).\label{f2}}
\end{figure}

Consider a temperature-dependent potential which at high temperatures 
has a single minimum at $\phi_{c}=0$, while under a critical temperature $T_{c}$ 
it transforms into a  potential which has a minimum for a non-trivial value 
of the field, $\phi_{c}\neq 0$ (fig.\ref{f2}). 
When $T < T_{c}$ the system will try to minimize its energy and will move at the new minimum.
When this is done, and the phase transition 
has finished, our system is ruled 
by a smaller symmetry group, say $H$. Also name the previous, 
bigger symmetry group as $G$. 
Then the coset space $G/H$ produces the vacuum manifold $\mathcal{M}$ and the topology of the latter 
determines the type of defect that will finally arise. Homotopy theory helps us 
by telling us how to map $\mathcal{M}$ into physical space (fig.\ref{f1}). For example $ \pi_{1}(\mathcal{M})\neq {\bf 1} $ indicates that 
the first homotopy group is not trivial and practically informs us for the existence 
of non contractible loops in $\mathcal{M}$, that is to say cosmic strings. Thus, in models where the 
symmetry of the ground state is associated with a non-trivial homotopy group, 
topological defects exist.
 \newline
\begin{tabular}{|l|l|l|}
\hline
\hline
$\pi_{0}(\mathcal{M})\neq {\bf 1}$ & $\mathcal{M} \;$  disconnected & DOMAIN$\;\;$ WALLS \\
$\pi_{1}(\mathcal{M})\neq {\bf 1}$ & non$\;$ contractible$\;$ loops$\;$ in$\;$ $\mathcal{M}$ &  COSMIC$\;\;$ STRINGS \\
$\pi_{2}(\mathcal{M})\neq {\bf 1}$ & non$\;$ contractible$\;$ 2-spheres$\;$ in$\;$ $\mathcal{M}$ &  MONOPOLES \\
$\pi_{3}(\mathcal{M})\neq {\bf 1}$ & non$\;$ contractible$\;$ 3-spheres$\;$ in$\;$ $\mathcal{M}$ &  TEXTURES \\
\hline
\hline
\end{tabular}

\section{Superconducting cosmic strings}

\noindent
{\em Cosmic strings} are linear vortex defects predicted to have formed at a cosmological phase transition 
during which the vacuum manifold was {\em not} simply connected. 
These strings have enormous energy per unit length, namely $\mu$. Roughly speaking,
for a string created at a phase transition characterized by temperature  $T_{c}$, then $\mu \sim T_{c}^2$. 
For GUT strings, $\mu \sim 10^{22} g/cm$.
Their gravitational effects though, are negligible.
The strength of their gravitational interaction is given in terms of the
dimensionless quantity $G\mu\sim (T_{c}/M_{P})^{2}$, with $G$ the Newton's constant
and $M_{P}$ the Planck mass.
There are other significant 
astrophysical effects when, for example, strings have supercurrent.
The latter can make them transform into thin superconducting wires 
with critical current $\sim T_{c}$ thus, interacting strongly with magnetic fields.

It is useful to note here that the notion of {\em superconductivity} is also important
in our research presented in Part II, where one can find details. 
This phenomenon was first discovered by Onnes in 1911 and exhibits many astonishing
and interesting features, two very well known of which are:
\begin{itemize}
\item The superconducting material
\footnote{Not all elements or combinations of elements can be superconducting. In the years passed
since its discovery, superconductivity appears in many other materials and the challenge for the
years to come, is the construction of a material appearing superconducting properties below a $T_{c}$
which will be comparable to every day temperatures, say around 280 $^{o}K$. We are at $\sim 140$ $^{o}K$ for now.}
while being below a critical temperature $T_{c}$, allows current flow in its interior 
with {\em zero} resistance.
\item Magnetic fields generally can not penetrate in the interior of a superconducting material, the
well known {\em Mei$\beta$ner effect}.
\footnote{This effect is of crucial importance for the stability of vortex rings which will be studied
later on.}
\end{itemize}
There are two kinds of superconductors:
\begin{itemize}
\item {\em Type I:} Below a critical field $H_{c}(T)$ there is no penetration of flux inside the 
superconducting material, while above $H_{c}(T)$ the field penetrates perfectly in the material
which has turned into its normal non-superconducting state.
\item {\em Type II:} Below a critical field $H_{a}(T)$ again there is no penetration. The difference 
here is the existence of a second critical field $H_{b}(T)$ above which normal state is restored but
for values of $H_{c}(T)$ where $H_{b}(T)>H_{c}(T)>H_{a}(T)$ is valid, the applied field penetrates
partially in the material which is in a {\em mixed} state where both normal and superconducting regions
coexist.
\end{itemize}

%\clearpage
%\newpage

\section{Brief history of research on defects}

\noindent

\begin{itemize}
\item{The beginning...}
\end{itemize}
Skyrme in 1961 \cite{skyrm}, presented the first three dimensional topological defect solution
arising in a non-linear field theory and proposed that such solutions would have an 
important role in particle physics.
Nambu in 1966, suggested the cosmological significance of defects. Before such proposals could be taken
seriously, one had to establish the existence of stable topological defect solutions in realistic
renormalizable theories. An important step towards that direction, was the discovery of defect solutions
in Higgs and Yang-Mills theories, notably the Nielsen-Olesen vortex in 1973 \cite{c9}, the 't Hooft-Polyakov monopole
in 1974 \cite{hoo} as well as GUT monopoles (i.e \cite{doktom}).

The cosmological implications of symmetry breaking were pointed out by Kirzhnits in 1972 \cite{kir}, who suggested that 
spontaneously broken symmetries in a field theory can be restored at high temperatures just as they are
in condensed matter systems. Zel'dovich, Kozbarev and Okun in 1974 \cite{zel}, argued that domain walls would form at phase transitions
in which a discrete symmetry was broken. Weinberg in 1974 \cite{wein}, noted that cosmic domains could form and Everett \cite{ever}, in
the same year studied possible interactions of domain walls with matter. It was then the pioneering paper of Kibble \cite{cc1}
in 1976, which showed that the existence of defects, depends on the topology of the vacuum manifold classifying them using homotory theory.

The circumstances above, led many researchers to start studying those defects and their role in cosmology.
A new ``golden'' era for topological defects was starting.

\begin{itemize}
\item{The ``booming''...}
\end{itemize}
The first interest in studying them comes from the fact that since a typical GUT predicts a few phase transitions
and because the vacuum structure needed to form strings is generically realized, one can assume, also  by following Vilenkin,
that cosmic strings have a considerable existence probability. 
Cosmic strings appear in many papers especially in the $80$'s and early $90$'s  as having important cosmological implications
in structure formation by providing density fluctuations and thus, becoming 
the seed of galaxies \cite{cc1}-\cite{cosmr9}. Cosmic strings appear either as infinite straight strings
or as rings. However, due to disagreement with the observations of the Cosmic Microwave Background, this theory
was left aside until nowadays, as superstring theory discusses the possibility for the existence of  cosmic superstrings 
in the framework of braneworld cosmologies. Such superstrings could play, under conditions, the role of cosmic strings.
There was also an observational ``adventure'' as we will explain below.
\begin{figure}[t]
\centering
\includegraphics[scale=0.58]{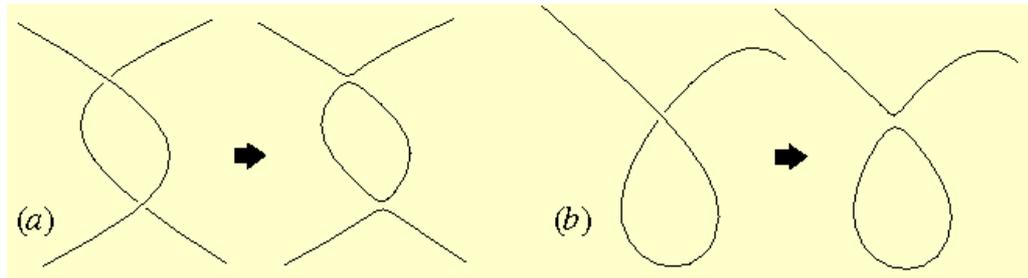}
\caption{\small Apart from the intersection of two different strings (a), self-intersection of a cosmic string can be
another mechanism of loop production (b). Picture from \cite{site}. \label{cr} }
\end{figure}

There are numerous papers on the 
consequences of a possible existence of superconducting vortex rings \cite{l2}-\cite{vort12}.
String loops are predicted to have formed during phase transitions at the early universe. \cite{c3} refers to GUT mass scale $U(1)$ symmetry
breaking occurence which gives rise to strings having important role in many popular cosmological scenarios seen mostly in the $80$'s.
The cosmological role
of loops has been analyzed in many papers such as \cite{c10}-\cite{c19}, where the scenario is that galaxies condensate around oscillating
loops of mass $\approx 10^{9} M_{\odot}$. By the time those strings have radiated away their energy, the matter
density fluctuations are large enough to grow independently. Additional loops 
can be produced by self-intersection  of a string or by intersection of different strings (fig.\ref{cr}). Closed loops are 
doomed to extinction due to oscillation and gradual loss of energy through gravitational radiation \cite{rad1, rad2}, unless perhaps 
they reintersect  longer strings and become reconnected \cite{l2}. On the other hand, it's crucial for closed
loops to have possibility more than $1/2$ to intersect themselves and break up into smaller pieces in order
not to dominate the energy density of a radiation-dominated universe \cite{c10}. Qualitative as well as quantitative
details were revealed by simulations \cite{s1}-\cite{s10} of a cosmic string network consisting of 
infinite straight strings and a small percentage of loops.

\begin{itemize}
\item{Today...}
\end{itemize}
Until today, physicists search for a possible signal of a topological defect from the early universe 
through the ways the theories give them for detecting them \cite{c13}, \cite{d1}-\cite{d5}. Especially for
the case of a ring with supercurrent, when current saturates, such a string will emit particles copiously
and may be seen as an X-ray or $\gamma$-ray source and it is believed to give contributions to X-ray
background and high-energy ($10^{20}$ eV) cosmic rays \cite{det2}. 
In \cite{o1}-\cite{o3} one can find observations which were believed to involve, for the first time, a topological defect through
gravitational lensing while in \cite{o4} the lensing from dark matter was excluded. Later observations
and analysis though, showed that this is not in fact the case \cite{CSL}. In fact, it is a very rare case
of almost identical (up to $99.96$\% !) elliptical galaxies, separated by a misleadingly small, as it proved, distance.
In \cite{c15} Kibble supports the cosmic string cosmological scenario using the above observations and
predictions of fundamental string theory. Concerning the latter, in \cite{css1,css2} one can find details about the 
possibility of the existence of cosmic superstrings, that is to say superstrings of cosmic length, and observation
of them as cosmic strings.
As it concerns the observations, if we ever detect a topological defect,
that would be a discovery of great importance. But even an observed absence of topological defects,
is very useful too (i.e. \cite{loopconst}), since it imposes  constraints on particle physics model building.
For example, the non-abundance of magnetic monopoles inspired the inflationary revolution in cosmology and GUT's models are constrained to provide
the requisite amount of inflation. Reviews on all these matters can be found in \cite{r1}-\cite{r6} and relevant books are \cite{c8,b2}.

\section{Solitons}

\noindent
Topological soliton in general, is a solution of a system of partial differential equations or of a quantum field theory that can be proven to exist 
because the boundary conditions entail the existence of homotopically distinct solutions. 
Typically, this occurs because the boundary on which the boundary conditions are specified, has a non-trivial homotopy group 
which is preserved by differential equations.
The solutions to the differential equations are then topologically distinct, and are classified by their homotopy class.
It is not easy to define precisely what a soliton is. Drazin and Johnson (1989) describe solitons as solutions of nonlinear differential equations which
\begin{itemize}
\item{ represent waves of {\em permanent} form}
\item{ are {\em localised}, so that they decay or approach a constant at infinity}
\item{ can interact strongly with other solitons, but they emerge from the collision {\em unchanged} apart from a phase shift.}
\end{itemize}

There are many equations of mathematical physics which have solutions of the soliton type.
Correspondingly, the phenomena which they describe, be it the motion of waves in shallow water or in an ionized plasma, exhibit solitons. 
The first observation of this kind of wave was made in 1834 by John Scott Russell, who followed on horseback a soliton propagating in the windings of a channel. 
In 1895, D. J. Korteweg and H. de Vries proposed an equation for the motion of waves in shallow waters which possesses soliton solutions, 
and thus established a mathematical basis for the study of the phenomenon. 
Interest in the subject, however, lay dormant for many years, and the major body of investigations began only in the 1950s. 
Researches done by analytical methods and by numerical methods made possible with the advent of computers gradually led to a complete understanding of solitons. 

Eventually, the fact that solitons exhibit particle-like properties, because the energy is at any instant confined to a limited region of space, received attention, 
and solitons were proposed as models for elementary particles. 
However, it is difficult to account for all of the properties of known particles in terms of solitons. 
More recently it has been realized that some of the quantum fields which are used to describe particles and their interactions also have solutions of the soliton type. 
The solitons would then appear as additional particles, and may have escaped experimental detection because their masses are much larger than those of known particles. 
In this context the requirement that solitons emerge unchanged from a collision has been found too restrictive, 
and particle theorists have used the term soliton where traditionally the term solitary wave would be used.

We can classify solitons in three sectors, according to the origin of their stability:
\begin{itemize}
\item{ {\bf Topological solitons:}  They can be found in theories where the vacuum manifold is not simply connected and stability 
comes as a consequence of a topological conservation law.}
\item{ {\bf Non-topological solitons:}  These solitons are stabilized by the confinement of a conserved charge as
a result of  a Noether current arising from a symmetry of the theory and not from a topological conservation law as above \cite{lee}.}
\item{ {\bf Semi-topological solitons:} They are characterized by a winding number and are classically stable, being local minima of the
energy functional, but in contrast to genuine topological defects, they can tunnel quantum mechanically and decay to the trivial vacuum.
They were introduced for the first in \cite{c1, c1x, c3, k1}. They are the only kind of defects that can exist in the SM of Particle Physics.}
\end{itemize}

%==================================================================================================================
%==================================================================================================================
%==================================================================================================================
%==================================================================================================================

\chapter*{PART I}

\section*{THEORETICAL FRAMEWORK }
\addcontentsline{toc}{chapter}{PART I: THEORETICAL FRAMEWORK }

In this part, we present two well known models which provide the basis for our research. These are, the Nielsen-Olesen vortex \cite{c9}
and the straight bosonic superconducting string \cite{c4}. These models also provide the theoretical framework on which we are based. 
The known solutions of these models help us to test our numerical algorithms and to reproduce the solutions.
Before that, we explain the notion of spontaneous symmetry breaking of a global and a local symmetry as well as
some other useful notions such as penetration depth and Meissner effect in superconductors and phase transitions.

\chapter{Cosmic string field theory}

\newpage

\section{Spontaneous symmetry breaking }
The purpose of this section is to introduce the notion of spontaneous symmetry breaking (SB) which 
is very important in our research part. {\em Goldstone theorem} as well as {\em Higgs mechanism} are presented.

\subsection{Spontaneous breaking of a global symmetry}
Consider the following Lagrangian density with a complex scalar $\phi$ field,
\begin{equation}
\label{gold}
\mathcal{L} = \partial_{\mu}\phi \partial^{\mu}\phi^{*}-U(\phi,\phi^{*})
\end{equation}
with 
\begin{equation}
U(\phi,\phi^{*})= m^{2}\phi^{*}\phi + \lambda(\phi^{*}\phi)^{2}
\end{equation}
This Lagrangian is invariant under the following $U(1)$ global transformation
\begin{equation}
\phi \rightarrow e^{ia} \phi
\end{equation}
where $a$ a constant. The ground state of the theory is obtained by minimizing the potential $U$
\begin{equation}
\frac{\partial U}{\partial \phi} = m^{2}\phi^{*} + 2\lambda \phi^{*} (\phi^{*}\phi)
\end{equation}
\begin{itemize}
\item{ $m^{2}>0$: In this case we have a minimum at $\phi^{*}=\phi =0$}
\item{ $m^{2}<0$: In this case, $\phi=0$ is a maximum and the minimum is located at $|\phi|^{2}=-m^{2}/2\lambda \equiv k^{2}$}
\end{itemize}
In the second case, the minima lie along the circle $|\phi|=k$, which form a set of degenerate vacua related to each other by rotation.
We can express the complex field $\phi$ as follows
\begin{equation}
\phi(x) = (r(x)+k) e^{i\theta (x)}
\end{equation}
where $r$ and $\theta$ both have vanishing vacuum expectation values. We substitute in $\mathcal{L}$ and the potential term there becomes
\begin{equation}
U=\lambda r^{4} + 4k\lambda r^{3} + 4\lambda k^{2}r^{2} -\lambda k^{4}
\end{equation}
and we have a mass term for the field $r$, which means that $m_{r}^{2}= 4\lambda k^{2}$, while $\theta$ is a massless
field as there is no mass term for it. Thus, the two massive fields (real parts of $\phi$) became one massive
and one massless field. In general there is the following theorem
\begin{itemize}
\item{ {\bf Goldstone theorem}: The spontaneous breaking of a continuous global symmetry is always accompanied by the 
appearence of one or more massless scalar spin-0 particles known as the Goldstone bosons.}
\end{itemize}

\subsection{Spontaneous breaking of a gauge symmetry}
Now we consider the Lagrangian density (\ref{gold}), which we demand to be invariant under the following $U(1)$ local (gauge) transformation
\begin{equation}
\phi \rightarrow e^{ia(x)}\phi
\end{equation}
This results in the introduction of the electromagnetic field through a covariant derivative and the Lagrangian density becomes
\begin{equation}
\mathcal{L}= -\frac{1}{4}F_{\mu\nu}F^{\mu\nu}+ |D_{\mu}\phi|^{2} -U(\phi,\phi^{*})
\end{equation}
where $|D_{\mu}\phi|^{2}= (\partial_{\mu}+ieA_{\mu})\phi(\partial^{\mu}-ieA^{\mu})\phi^{*}$, $F_{\mu\nu}=\partial_{\mu}A_{\nu} -\partial_{\nu}A_{\mu}$
while the potential is the same as above.
Following the previous steps, in the case we consider $m^{2}<0$, the vacuum is at 
\begin{equation}
|\phi|=k \equiv \Bigg(\frac{-m^{2}}{2\lambda}\Bigg)^{1/2}
\end{equation}
We can set, as above, 
\begin{equation}
\phi (x)= k+\frac{\phi_{1}(x)+i\phi_{2}(x)}{\sqrt{2}}
\end{equation}
and substitute in the Lagrangian which becomes
\begin{equation}
\mathcal{L}=-\frac{1}{4}F_{\mu\nu}F^{\mu\nu}+\frac{1}{2}(\partial_{\mu}\phi_{1})^{2}+\frac{1}{2}(\partial_{\mu}\phi_{2})^{2}
+e^{2}k^{2}A_{\mu}A^{\mu}-2\lambda k^{2}\phi_{1}^{2}+\cdots
\end{equation}
where the dots represent cubic and quartic terms.
There are two massive fields, the $\phi_{1}$ and  $A_{\mu}$ which means that the photon became massive. The $\phi_{2}$ field
is massless and can also be eliminated through an appropriate choice of the vacuum.
The above phenomenon is summarized as follows
\begin{itemize}
\item{ {\bf Higgs mechanism}: The spontaneous breaking of a gauge symmetry makes the massless field to disappear 
and we end up with a massive scalar field representing the Higgs particle, while the gauge field acquires mass.}
\end{itemize}
The Higgs particle is believed to have large mass and this is the reason for not being discovered yet.
LHC, a proton-proton collider will operate in the early $2008$. 
It will work at energies until $7+7$ TeV, something which means
that there are possibilities for finding that particle. To have answers in this subject is, maybe, the greatest challenge
for the years to come.
The following table summarizes the above cases.
 \newline
\begin{tabular}{|l||l|l|}
\hline
\hline
                 & {\bf Goldstone mode}              &    {\bf Higgs mode}                      \\
\hline
{\em Before} SB  & $2$ massive scalars               & $2$ massive scalars$+$$1$ photon         \\
\hline
{\em After} SB   & $1$ massive$+$$1$ massless scalar & $1$ massive scalar$+$$1$ massive photon  \\
\hline
\hline
\end{tabular}
\newline
Finally, if the ground state of the system, obeys the initial symmetry of the Lagrangian,
the system is said to be in the {\em Wigner} mode.

\section{Effective potential}

The discussion of symmetry breaking is somewhat simplistic because we use classical potentials to determine the expectation
value of the Higgs field $\phi$ in the models above. We have to keep in mind that, in reality $\phi$ is a quantum field interacting with itself
and with other quantum fields and the classical potential $V(\phi)$, is modified by {\em radiative corrections}. The corrected potential
for $\phi$, is called effective potential $V_{eff}$ and can be evaluated perturbatively as an expansion in powers of coupling constants
\begin{equation}
V_{eff}(\phi)=V(\phi)+V_{1}(\phi)+V_{2}(\phi)+\cdots
\end{equation}
where $V(\phi)$ is the classical potential and $V_{n}(\phi)$ is the contribution of Feynman diagrams with $n$ closed loops.
In some models, the radiative corrections are negligible, while in others they can alter the character of symmetry breaking.
One example of the latter is the Higgs model
\begin{equation}
\mathcal{L} = -\frac{1}{4}F_{\mu\nu}F^{\mu\nu}+\overline{D_{\mu}\phi} D^{\mu}\phi -V(|\phi|)
\end{equation}
with $V(|\phi|)= \mu_{0}^{2}|\phi|^{2}$, where $\mu_{0}$ a constant.
The one-loop contribution to $V_{eff}(\phi)$ is given by \cite{veff} :
\begin{equation}
V_{1}(\phi)=\frac{3e^{4}}{16\pi^{2}}|\phi|^{4}ln\Bigg(\frac{|\phi|^{2}}{\sigma^{2}}\Bigg)
\end{equation}
where $\sigma$ the renormalization scale. If we introduce a dimensionless quantity
\begin{equation}
v\equiv \frac{16\pi^{2}\mu_{0}^{2}}{3e^{4}\sigma^{2}}
\end{equation}
then, for $v>0.45$ the $V_{eff}(\phi)$ has a single minimum at $\phi=0$, but for $v<0.45$ there is another minimum
at a non-zero value of $\phi$, while for $v<0.37$ the latter minimum goes deeper than the one at $\phi=0$.
In the last case, the absolute minimum is at a non-zero value of $\phi$ and the symmetry is spontaneously broken.

In Chapter $4$, we examine a modified Ginzburg-Landau potential. The modification has to do with the addition of a cubic term.
Such term results from the $1$-loop radiative corrections on the quartic potential \cite{cubic}.

\begin{figure}
%\vspace{0.6 in.}
\centering
\includegraphics[scale=0.4,angle=270]{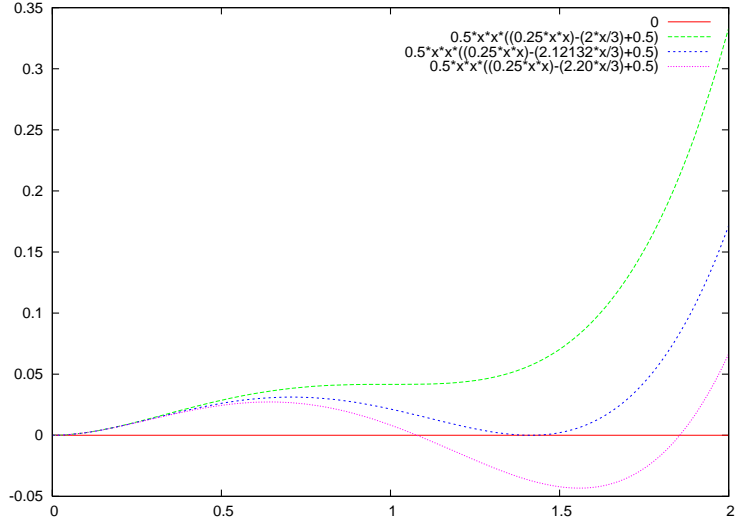}
\caption{\small An example of a first order phase transition. 
The plots are for $\beta =2 <\beta_{crit}$, $\beta = \beta_{crit}$, $\beta=2.2 > \beta_{crit}$. Parameter is $a=1$.\label{pot1order}}
\end{figure}
\section{Phase transitions }

\subsection{First order phase transition} 

In order to observe phase transitions and to find the distinguishing features we suppose appropriate
effective potentials. For a first order phase transition, consider the following potential which
we examine in detail in Chapter $4$:
\begin{equation}
U(|\phi|)=\frac{a}{2}|\phi|^{2}\Bigg( \frac{1}{4}|\phi|^{2}-\frac{\beta}{3}|\phi|+\frac{1}{2}\Bigg)
\end{equation}
This potential has an obvious minimum at $|\phi|=0$. This corresponds to the symmetric phase
of the system.
For $\beta \geq 2$, it develops a secondary minimum at $|\phi|=1$.
The critical value for this potential is $\beta=3/\sqrt{2}\equiv \beta_{crit}$. This happens
because for $\beta\geq\beta_{crit}$, the secondary minimum goes deeper than the one at $|\phi|=0$.
When $\beta =\beta_{crit}$, the secondary minimum makes the potential zero, just like the original one at $|\phi|=0$ (fig.\ref{pot1order}).
The non-zero expectation value of $|\phi|$ is
\begin{equation}
|\phi|=\frac{\beta +\sqrt{\beta^{2}-4}}{2}
\end{equation}

\begin{figure}
%\vspace{0.6 in.}
\centering
\includegraphics[scale=0.4,angle=270]{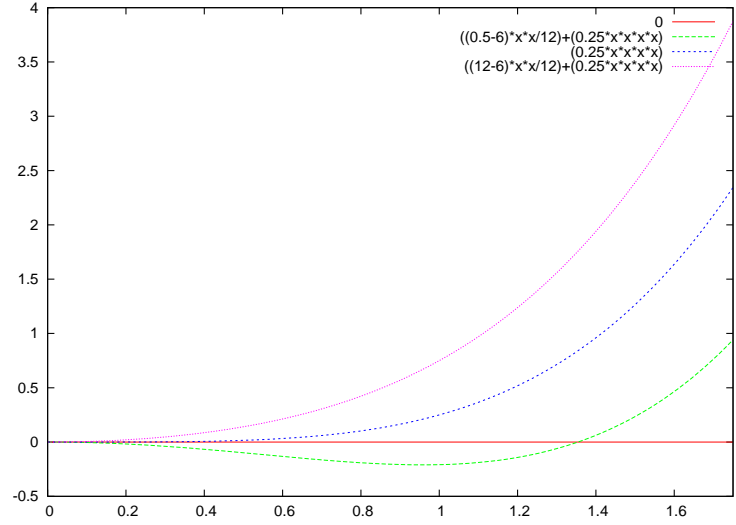}
\caption{\small An example of a second order phase transition. 
The plots are for $T = \sqrt{0.5} <T_{crit}$, $T = T_{crit}$, $T =\sqrt{12} > T_{crit}$. Parameters are $\eta=1,\; \lambda=1$. \label{pot2order}}
\end{figure}

Generally, in a first-order phase transition, a new broken symmetry phase usually nucleates after the temperature falls some degrees below
the critical temperature $T_{c}$. Separated areas of the new phase form independently and expand, resulting in a local selection
of the broken symmetry vacuum. As a result of these independent selections, the final configuration may not be able to get rid of
the locked-out fragments of the original vacuum (before transition).

\subsection{Second order phase transition} 

In order to observe a second order phase transition, consider the following temperature-dependent potential:
\begin{equation}
U(|\phi|,T)=m^{2}(T)|\phi|^{2}+\frac{\lambda}{4}|\phi|^{4}
\end{equation}
with
\begin{equation}
m^{2}(T)=\frac{\lambda}{12}\Big(T^{2}-6\eta^{2}\Big)
\end{equation}
The critical value for this potential is $T_{crit}=\eta\sqrt{6}$. Above this temperature, the mass-squared
term becomes positive, thus the potential is a sum of positive terms with a unique minimum at $|\phi|=0$
which represents the symmetric phase (fig.\ref{pot2order}). When $T<T_{crit}$, then $m^{2}(T)<0$ and the order parameter $|\phi|$
acquires a non-zero expectation value which is
\begin{equation}
|\phi|=\frac{1}{\sqrt{6}}\Big(T_{crit}^{2}-T\Big)^{1/2}
\end{equation}
In second-order phase transitions, the phase transformation occurs almost simultaneously throughout the volume. However, unless the critical temperature
is traversed infinitesimally slowly, the resulting broken symmetry phase will contain many distinct regions with different choices of the local vacuum.
This is because the selection of the new vacuum has to be, somehow, communicated if the same choice is to made elsewhere, and the speed at which this
transformation can be propagated is finite.

The main difference between these phase transitions is that, at the first order phase transition the symmetric phase ($|\phi|=0$) remains metastable
when passing the critical value $\beta_{crit}$ and is often called ``false vacuum''.
A defining feature of the second order phase transition, is that the order parameter $|\phi|$ grows continuously from zero, as  the temperature
is decreased below $T_{crit}$.

\section{Superconductivity and penetration depth}

Ginzburg-Landau theory is used to model superconductivity. It examines the macroscopic properties of a superconductor with the aid of general thermodynamic arguments.
Based on Landau's previously-established theory of second-order phase transitions, Landau and Ginzburg argued that the free energy $F$ 
of a superconductor near the superconducting transition can be expressed in terms of a complex order parameter $\Psi$, 
which describes how deep into the superconducting phase the system is. The free energy has the form
\begin{equation}
F= F_{n}+a|\Psi|^{2}+\frac{\beta}{2}|\Psi|^{4}+\frac{1}{2m}|(-i\hbar\nabla -2e\mathbf{A})\Psi|^{2}+\frac{|\mathbf{B}|^{2}}{2\mu_{0}}
\end{equation}
where $F_{n}$ is the free energy in the normal phase, $a$ and $\beta$ are phenomenological parameters, $\mathbf{A}$ is the electromagnetic vector potential, 
and $\mathbf{B}$ is the magnetic field. By minimizing the free energy with respect to fluctuations in the order parameter and the vector potential, 
one arrives at the Ginzburg-Landau equations.
These equations produce many interesting and valid results. Perhaps the most important of these is its prediction of the existence 
of two characteristic lengths in a superconductor. The first is a coherence length $\xi$, given by
\begin{equation}
\xi=\sqrt{\frac{\hbar^{2}}{2m|a|}}
\end{equation}
which describes the size of thermodynamic fluctuations in the superconducting phase. The second is the penetration depth $\lambda$, given by
\begin{equation}
\lambda=\sqrt{\frac{m}{4\mu_{0}e^{2}\Psi_{0}^{2}}}
\end{equation}
where $\Psi_{0}$ is the equilibrium value of the order parameter in the absence of an electromagnetic field. 
The penetration depth describes the length to which an external magnetic field can penetrate the superconductor.
In general, if one considers a superconducting semi-space at $x > 0$, and weak external magnetic field $B_{0}$ applied along $z$-direction in the empty space $x < 0$, 
then inside the superconductor the magnetic field is given by 
\begin{equation}
\label{magn}
B(x) = B_{0}e^{-x/\lambda}
\end{equation}
In atomic units, one can see that the penetration depth is, in fact,
\begin{equation}
\lambda \propto \frac{1}{e\Psi_{0}}
\end{equation}
We will use this result in Chapter $5$.

The formula (\ref{magn}), can be derived as follows. Suppose that an electric field $\mathbf{E}$ momentarily arises
within a superconductor. The superconducting electrons will be freely accelerated without dissipation and their
mean velocity will be
\begin{equation}
\label{z1}
m\frac{d\mathbf{v}}{dt}=-e\mathbf{E}
\end{equation}
The current density is
\begin{equation}
\mathbf{j}=-en\mathbf{v}
\end{equation}
where $n$ the number of electrons per unit volume, and the above formula, with the help of (\ref{z1}) can be written as
\begin{equation}
\frac{d\mathbf{j}}{dt}=\frac{ne^{2}}{m}\mathbf{E}
\end{equation}
If one substitutes into Faraday's law, this gives
\begin{equation}
\nabla\times\mathbf{E}=-\frac{1}{c}\frac{d\mathbf{B}}{dt}\Rightarrow \nabla\times\mathbf{j}=-\frac{ne^{2}}{mc}\mathbf{B}
\end{equation}
Plug the result into Maxwell's equation and this results to
\begin{equation}
\nabla\times\mathbf{B}=\frac{4\pi}{c}\mathbf{j} \Rightarrow \nabla^{2}\mathbf{B}=\frac{4\pi ne^{2}}{mc^{2}}\mathbf{B}
\end{equation}
Now, rename $mc^{2}/4\pi ne^{2}\equiv \lambda^{2}$ and you get 
\begin{equation}
\nabla^{2}\mathbf{B}=\frac{1}{\lambda^{2}}\mathbf{B}
\end{equation}
where $\lambda$ is the London penetration depth. The last equation can be easily solved to give the result
in (\ref{magn}).
\begin{figure}
%\vspace{0.6 in.}
\centering
\includegraphics[scale=0.58]{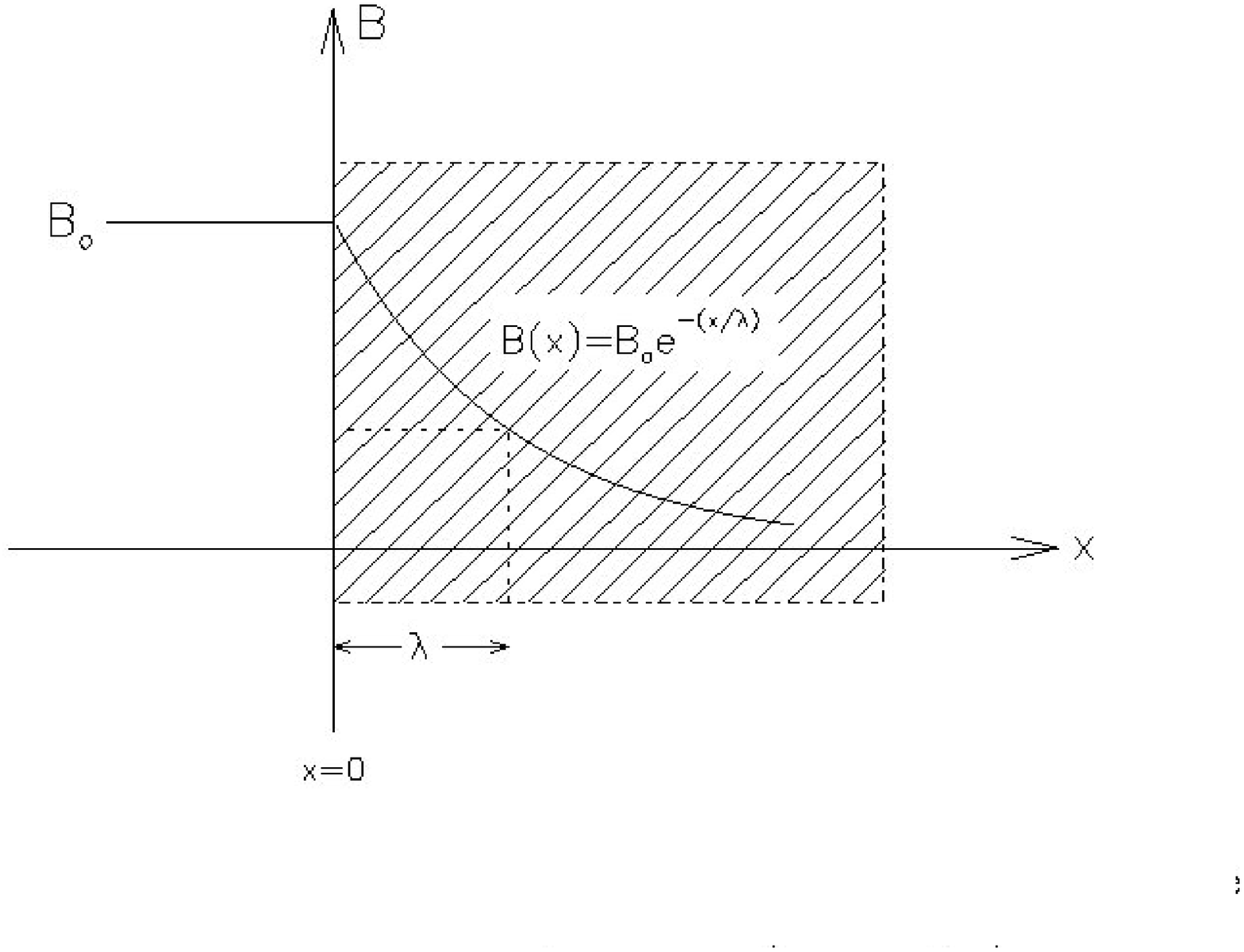}
\caption{\small  The magnetic field drops exponentially inside the superconductor. \label{dep}}
\end{figure}

\newpage

\section{Nielsen-Olesen vortex }

\subsection{Introduction} 

 The simplest theory exhibiting string solutions is that of a complex scalar field $\phi(x)$, in a model described
 by the following Lagrangian density:
 \begin{equation}
 \mathcal{L}=\partial_{\mu}\phi^{*}\partial^{\mu}\phi-V(\phi),\;\;\;V(\phi)=\frac{\lambda}{4} (|\phi|^{2}-\eta^{2})^{2}
 \end{equation}
 which has a global $U(1)$ symmetry under the transformation $\phi \rightarrow \phi e^{ia}$ where
 $a$ a constant. The condensed matter analogue of global strings is the vortices in superfluid Helium-$4$ \cite{gcd}.

 If we move a bid further and consider a gauge symmetry under the transformation $\phi \rightarrow \phi e^{ia(\mathbf{x})}$
 then we are led to the introduction of a vector field $A_{\mu}$ while the replacement of $\partial_{\mu}$
 by the covariant derivative $D_{\mu}=\partial_{\mu}-ieA_{\mu}$ is required. Due to the existence of the vector
 field $A_{\mu}$, a relevant "magnetic" field exists as well. This theory describes the Nielsen-Olesen string \cite{c9}.

\subsection{The model} 
 Consider an Abelian Higgs model in two dimensions with a charged complex scalar field $\phi$ and
 gauge field $A^{\mu}$. The {\bf Lagrangian density} of the model under consideration is:
 \begin{equation}
 \mathcal{L}= \overline{D}_{\mu}\phi^{*}D^{\mu}\phi-\frac{1}{4}F_{\mu\nu}F^{\mu\nu}
 -\frac{\lambda}{4}(|\phi|^{2}-\eta^{2})^{2}
 \end{equation}
 where $D_{\mu}=\partial_{\mu}-ieA_{\mu}$, the covariant derivative and $F_{\mu\nu}=\partial_{\mu}A_{\nu} -\partial_{\nu}A_{\mu}$. 
 The energy of a static vortex configuration in two dimensions follows:
 \begin{equation}
 E= \int d^{2}\rho \Bigg( |D\phi|^{2}
 +\frac{1}{2}(\mbox{\boldmath$E^{2}+B^{2}$})+V(|\phi|)\Bigg)
 \end{equation}
 The vortex energy must be finite which means that field configuration
 must have a form that decreases to $0$ fast enough, as $\rho\rightarrow \infty$. This requirement
 imposes some conditions on the asymptotic field configuration which are:
 \begin{eqnarray*}
 |\phi|\rightarrow \eta   &,&\;   \rho\rightarrow \infty\\
 A_{\varphi}\rightarrow \frac{1}{e\rho}\frac{d\vartheta}{d\varphi}+\cdots   &,&\;   \rho\rightarrow \infty
 \end{eqnarray*}
 where $\vartheta$ is the phase of the Higgs field $\phi$\footnote{Suppose $\phi(\rho)=f(\rho)e^{in\theta}$, $\theta$ the phase of the Higgs field, $n$ the winding number.}. 
 The last condition on the gauge field, can be understood through the requirement that the covariant derivative must be zero as $\rho \rightarrow \infty $
 and $\phi \neq 0$ there.
 When following
 a closed path around the vortex, the Higgs field must return to its original value $\phi(0)=\phi(2\pi)$
 thus  $ 1= \exp 2i \pi n$, which means that the winding number $n$ must be an integer. Also, one expects
 somewhere in the vortex core to have $\phi(x)=0$ for a non-trivial winding number. The fact that
 the winding number is an integer results to quantization of the magnetic flux.
 \begin{equation}
 n=\frac{1}{2\pi}\int_{0}^{2\pi} \frac{d\vartheta}{d\varphi} d\varphi=\frac{e}{2\pi}
 \oint_{S^{1}} \mbox{\boldmath$A$}\cdot  \mbox{\boldmath$dl$}=\frac{e}{2\pi}\int 
 \mbox { \boldmath$B$}\cdot \mbox{\boldmath$dS$}\Rightarrow
 \Phi_{B}=\frac{2\pi n}{e}
 \end{equation}
 where $S^{1}$ is a circle of infinite radius centered on this string. This flux quantization
 is a result of the vanishing of the covariant derivative which determines ${\mathbf A}$ in terms
 of $\partial_{i} \phi$. The phase of $\phi$ must change by $2\pi n$ which forces the flux to be quantized.
 A rough estimation of the size and mass of our vortex can be done. These two characteristic features 
 of the vortex are determined by the regions over which the scalar and gauge fields start to differ from their 
 asymptotic values. These distance scales $\rho_{s}, \rho_{v}$ ($s$ for scalar, $v$ for vector), are related to the
 Higgs and vector particle masses off the string. 
 \begin{eqnarray}
 \rho_{s}&\approx& m_{s}^{-1}=(\sqrt{\lambda}e)^{-1} \\
 \rho_{v}&\approx& m_{v}^{-1}=(\sqrt{2}e\eta)^{-1}
 \end{eqnarray}
 Written in terms of these lengths, and assuming $\rho_{v} > \rho_{s}$ (type II superconductor
 \footnote{Our model can be compared with that of Ginzburg-Landau which has the same Lagrangian with ours
 and $\rho_{s}$ can be thought as the coherence length -having to do with the transition layer from superconducting
 to normal state- and $\rho_{v}$ the {\em London} penetration depth -referring to the exponential decay of the magnetic field
 at the surface of the superconductor- in terms of superconductivity.}), the energy reads:
 \begin{equation}
 E\approx 2\pi \eta^{2} \Big[ ln\Big( \frac{\rho_{v}}{\rho_{s}}\Big)+\frac{1}{e^{2}\eta^{2}\rho_{v}^{2}}
 +\lambda \eta^{2}\rho_{s}^{2}\Big]
 \end{equation}
 where the first term represents the gradient energy of the scalar field, the second is due to the fact that
 the magnetic flux does not wish to be confined and the last term is the cost of the difference of $|\phi|$
 from its vacuum value $\eta$. Finally, this energy is minimized by replacing the characteristic lengths with
 their relative masses as above. Then we are led to the mass of the vortex configuration:
 \begin{equation}
 \mu \approx 2\pi \eta^{2} ln\Big(\frac{m_{s}}{m_{v}}\Big)
 \end{equation}
 a formula that in three dimensions gives the mass per unit length of the string.

 The vortices in the Abelian Higgs model have condensed matter analogues like flux tubes in superconductors
 \cite{cm1} but there are differences \cite{cm2,cm3} because Nielsen-Olesen vortices exist in
 a vacuum background while superconductor vortices are amongst charged bosons called Cooper pairs.

\subsection{Ansatz and equations} 

The {\bf Euler-Lagrange equations} of our model are:
\begin{eqnarray}
(\partial_{\mu}-ieA_{\mu})(\partial^{\mu}-ieA^{\mu})\phi+\frac{\lambda}{2}\phi(\phi\overline{\phi}-\eta^{2})&=&0 \\
\partial_{\mu}F^{\mu\nu}&=&j^{\nu}
\end{eqnarray}
where $j^{\nu}\equiv 2e Im[\overline{\phi}(\partial^{\nu}-ieA^{\nu})\phi]$. We work in the Lorentz gauge 
($\partial_{\mu}A^{\mu}=0$) and we rescaled the fields ($\phi \rightarrow \eta^{-1}\phi$ ,
$A^{\mu}\rightarrow \eta^{-1}A^{\mu}$ and $x \rightarrow \eta x$), while we also set $\eta=1$ for convenience.
A proper {\bf ansatz} for a string lying on the $z$-axis would be:
\begin{eqnarray}
\phi(\mathbf{\rho})&=& f(\rho) e^{in\varphi}\\
A(\mathbf{\rho })&=&\frac{n}{e\rho}a(\rho) \hat{\varphi}
\end{eqnarray}
where $\hat{\rho}$, $\hat{\varphi}$ are the cylindrical unit vectors. We use cylindrical coordinates ($\rho ,\varphi,z)$. According to the above, the appropriate 
asymptotic conditions while $\rho\rightarrow \infty$ and boundary conditions at $\rho \rightarrow 0$
are:
\begin{eqnarray}
f(\rho \rightarrow \infty)\rightarrow 1 &,&\; \; a(\rho \rightarrow \infty) \rightarrow 1 \\
f(\rho =0)=0 &,&\; \; a(\rho =0)=0
\end{eqnarray}
With the above ansatz, we are led to the following system of non-linear differential equations:
\begin{eqnarray}
\frac{d^{2}f}{d\rho^{2}}+\frac{1}{\rho}\frac{df}{d\rho}-\frac{n^{2}f}{\rho^{2}}(a-1)^{2}-\frac{\lambda}{2}f(f^{2}-1)=0 \\
\frac{d^{2}a}{d\rho^{2}}-\frac{1}{\rho}\frac{da}{d\rho}-2e^{2}f^{2}(a-1)=0
\end{eqnarray}
It is helpful to acquire an approximate asymptotic solution as $\rho \rightarrow \infty$. This is:
\begin{eqnarray}
a(\rho)&\approx& 1-\mathcal{O}(\sqrt{\rho}\exp(-\sqrt{2}e\rho)) \\
f(\rho)&\approx& 1-\mathcal{O}(exp(-\sqrt{\lambda}\rho))
\end{eqnarray}
Finally, another way to acquire the solution of the system is through the minimization of the {\bf energy functional} which is
\begin{equation}
E=2\pi\int_{-\infty}^{\infty} \rho d\rho \;\Bigg[ (\partial_{\rho}f)^{2}+\frac{f^{2}}{\rho^{2}}(a-1)^{2}+\frac{(\partial_{\rho}a)^{2}}{4\rho^{2}}+(f^{2}-1)^{2}\Bigg]
\end{equation}

\subsection{Virial theorem} 

Before the numerical results, it is important to have a way to check whether a final configuration
of an algorithm can be accepted or not.

The energy density for a {\em static solution}, which is of interest here, reads:
\begin{equation}
T^{00}=\frac{1}{4}F_{ij}F_{ij}+|D_{i}\phi|^{2}+V(|\phi|)\equiv \varepsilon
\end{equation}
The solution of our physical problem, has to  minimize the energy of the system:
\begin{equation}
\frac{\delta E}{\delta \phi}=\frac{\delta E}{\delta \phi^{*}}=\frac{\delta E}{\delta A_{i}}=0
\end{equation}
Define
\begin{equation}
f_{i}\equiv \frac{\delta E}{\delta \phi}\partial_{i}\phi+\frac{\delta E}{\delta \phi^{*}}\partial_{i}\phi^{*}
+\frac{\delta E}{\delta A_{j}}\partial_{i}A_{j}
\end{equation}
which together with
\begin{equation}
\frac{\delta E}{\delta \Phi}=\frac{\partial T^{00}}{\partial \Phi}
-\partial_{k}\frac{\partial T^{00}}{\partial (\partial_{k} \Phi)}
\end{equation}
can be written in a shorter form $f_{i}=\partial_{j} G_{ij}$ where
\begin{equation}
G_{ij}=\varepsilon \delta_{ij}-\frac{\partial \varepsilon}{\partial (\partial_{j}\phi)}\partial_{i} \phi
-\frac{\partial \varepsilon}{\partial (\partial_{j} \phi^{*})}\partial_{i} \phi^{*} 
-\frac{\partial \varepsilon}{\partial (\partial_{j} A_{k})}\partial_{i}A_{k}
\end{equation}
This means that any static solution of the field equations satisfies
\begin{equation}
\label{eq3} 
\partial_{j} G_{ij}=0
\end{equation}
But, we have 
\begin{eqnarray*}
\frac{\partial \varepsilon}{\partial(\partial_{j}\phi)} &=& (D_{j}\phi)^{*} \\
\frac{\partial \varepsilon}{\partial(\partial_{j}\phi^{*})} &=& D_{j}\phi \\
\frac{\partial \varepsilon}{\partial(\partial_{j}A_{k})} &=& \partial_{j}A_{k} - \partial_{k}A_{j} = F_{jk}
\end{eqnarray*}
which leads to
\begin{eqnarray*}
G_{ij}&=& \varepsilon \delta_{ij}-(D_{j}\phi)^{*}\partial_{i}\phi-D_{j}\phi\partial_{i}\phi^{*}-F_{jk}\partial_{i}A_{k} \\
&=&\varepsilon \delta_{ij}-(D_{j}\phi)^{*}D_{i}\phi-D_{j}\phi(D_{i}\phi)^{*}+ieA_{i}\phi(D_{j}\phi)^{*} \\
&-&ieA_{i}\phi^{*}D_{j}\phi-F_{jk}F_{ik}-F_{jk}\partial_{k}A_{i}\\
&=&\varepsilon \delta_{ij}-(D_{i}\phi)^{*}D_{j}\phi-(D_{j}\phi)^{*}D_{i}\phi -F_{ik}F_{jk}-\partial_{k}(F_{jk}A_{i}) \\
&+&A_{i}\Big( \partial_{k}F_{jk}-ie(\phi^{*}D_{j}\phi-(D_{j}\phi)^{*}\phi)\Big)
\end{eqnarray*}
 where on the second line we just add and subtract the necessary terms in order to derive the first four terms of the third line.
 Now, on the third line we have
 \begin{equation}
 \partial_{j}\partial_{k}(\partial_{j}A_{k}-\partial_{k}A_{j})A_{i}=0\Rightarrow (\partial_{j}^{2}\partial_{k}A_{k}-\partial_{k}^{2}\partial_{j}A_{j})A_{i}=0 \\
 \end{equation}
and from the field equations, the last term of the third line is also zero as
 \begin{equation}
\frac{\delta E}{\delta A_{j}}=0=\frac{\delta T^{00}}{\partial(\partial_{k} A_{j})}=-ie(\phi^{*}D_{j}\phi-\phi (D_{j}\phi)^{*})+\partial_{k}F_{jk}=0
\end{equation}
 thus
 \begin{equation}
 G_{ij}=\varepsilon \delta_{ij} -F_{ik}F_{jk}-(D_{i}\phi)^{*}D_{j}\phi-(D_{j}\phi)^{*}D_{i}\phi
 \end{equation}
 but
 \begin{equation}
 F_{ik}F_{jk}= \epsilon_{nik}B_{n}\epsilon_{mjk}B_{m}={\mathbf B^{2}} \delta_{ij}- B_{i}B_{j}
 \end{equation}
 which means that
 \begin{equation}
 \label{eq2}
 G_{ij}=\Bigg(-\frac{{\mathbf B^{2}}}{2}+|D\phi|^{2}+V\Bigg)\delta_{ij}+B_{i}B_{j}-(D_{i}\phi)^{*}D_{j}\phi-(D_{j}\phi)^{*}D_{i}\phi
 \end{equation}
 From (\ref{eq3}) one obtains
 \begin{equation}
 \label{eq1} 
 \int d^{3}x G_{ik}=0=\int d^{3}x \partial_{j}(x_{k}G_{ij})=\int dS_{j}x_{k}G_{ij}
 \end{equation}
 The last equality follows from Gauss theorem, while the first is due to the fact that
 \begin{equation}
 \partial_{j}(x_{k}G_{ij})=\frac{\partial x_{k}}{\partial x_{i}}G_{ij} + x_{k}(\partial_{j}G_{ij})=\delta_{kj}G_{ij}=G_{ik}
 \end{equation}
 where we also used (\ref{eq3}).
\newpage
 \begin{itemize}
 \item {\bf Virial relation for Nielsen-Olesen string model}
 \end{itemize}
 The formulas above are general and can be applied to any model which is being examined for static solutions.
 For the model at hand, take the trace of equation (\ref{eq1})
 \begin{equation}
 \int d^{3}x TrG =\int dS_{j}x_{i} G_{ij}=0
 \end{equation}
 while equation (\ref{eq2}) leads to
 \begin{equation}
 TrG = -\frac{1}{2}\mbox{\boldmath$B$}^{2}+|D\phi|^{2}+3V
 \end{equation}
 Thus, by taking a cylindrical surface of integration around the infinite string with cylindrical symmetry
 we are led to a virial relation 
 \begin{equation}
 \int d^{2}x \Bigg( \frac{1}{2} B^{2}-V\Bigg) =0
 \end{equation}
 Using $\nabla \times \mbox{\boldmath $A$} = \mbox{\boldmath $B$}$ and remembering that in our case
 \begin{equation}
 \mbox{\boldmath $A$}=\frac{a(\rho)}{e\rho} \hat{\varphi},\;\;\;\;V(|\phi|)=\frac{\lambda}{4}(|\phi|^{2}-m_{0}^{2})^{2}
 \end{equation}
 we acquire the formula for the magnetic field 
\begin{equation}
\mbox{\boldmath $B$}= \frac{1}{e\rho}\frac{da(\rho)}{d\rho}\hat{z}
\end{equation}
and rewrite in a
 more convenient way for our numerical purposes, the {\bf virial} relation
 \begin{equation}
 2\pi \int \rho d\rho \Bigg(\frac{1}{2e^{2}\rho^{2}}\Bigg(\frac{da}{d\rho}\Bigg)^{2}-\frac{\lambda}{4}(f^{2}-m_{0}^{2})^{2}\Bigg)=0
 \end{equation}
 Numerically though, we ``break'' the integral into two parts. Consider for example
\begin{equation}
I_{1}\equiv  2\pi \int \rho d\rho \Bigg(\frac{1}{2e^{2}\rho^{2}}\Bigg(\frac{da}{d\rho}\Bigg)^{2}\Bigg),\;\;\; 
I_{2}\equiv  -2\pi \int \rho d\rho \Bigg(\frac{\lambda}{4}(|\phi|^{2}-m_{0}^{2})^{2}\Bigg)
\end{equation}
Then we must have $I_{1}+I_{2}=0$. We define the index $V=\frac{||I_{1}|-|I_{2}||}{|I_{1}|+|I_{2}|}$ and
we want this index as small as possible.

\subsection{Numerical solution}

\begin{figure}
\vspace{0.6 in.}
\centering
\includegraphics[scale=0.55]{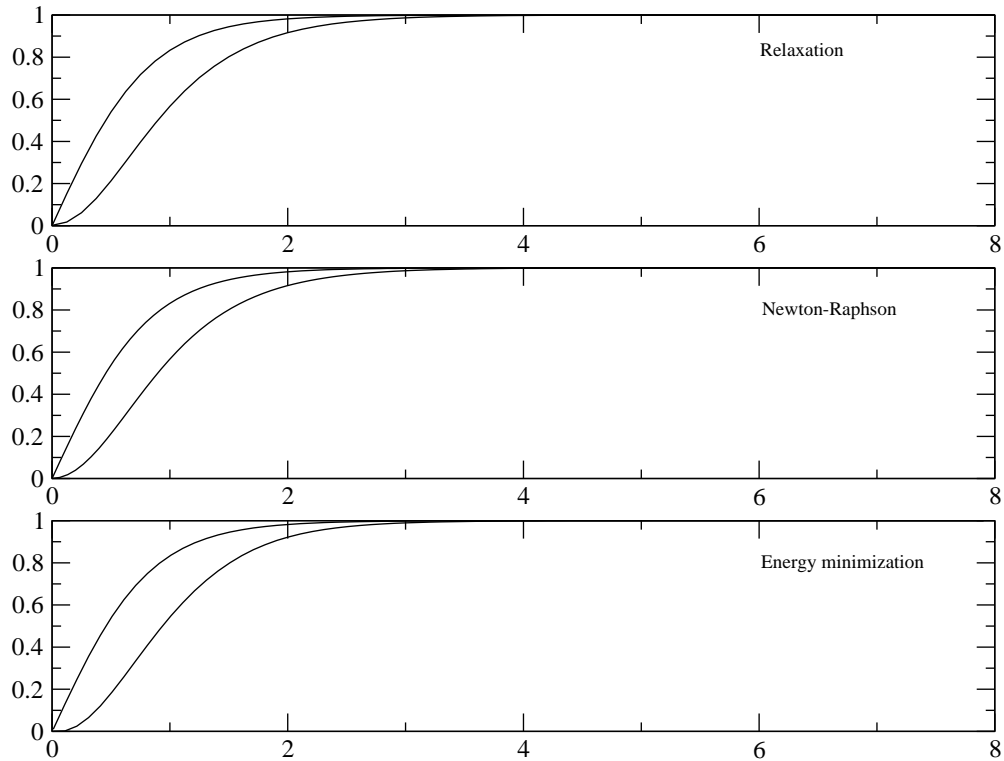}
\caption{\small Nielsen-Olesen string. The graph on top is the output of a relaxation program. The graph 
in the middle is the result of  a Newton-Raphson algorithm and the bottom graph is done through energy minimization. 
All graphs show $f(\rho)$ and $a(\rho)$, for $n=1,\;\lambda=2e^{2}=4$.\label{no}}
\end{figure}

There are many different numerical methods for solving the above system of equations. In figure \ref{no} one can
see the result of three different algorithms. Details concerning the Relaxation algorithm can be
found on page $762$ of \cite{c6} and a brief outline in chapter $7$ (Part III). Details about the Newton-Raphson
algorithm can also be found in chapter $7$. Finally, as it concerns minimization algorithm, many details can 
be found in chapter $7$, while we use minimization subroutines from page $425$ of \cite{c6}.

\newpage

\section{Straight bosonic superconducting string}

\subsection{Introduction}

Superconductivity can be understood as a spontaneously broken electromagnetic gauge invariance. When
the gauge invariance is broken, then any magnetic field applied at the boundary of a superconductor
decays exponentially towards its interior. This happens due to the screening of the superconducting 
current which flows along the boundary. This is known as Mei$\beta$ner effect. 
Having this mechanism in mind, one can support the idea that cosmic strings can become superconductors and behave
as  thin superconducting wires with an enormous critical current. For
this to happen, a charged scalar field must acquire a non-zero expectation value in the neighborhood of the
string core.
Practically, this points out the need of two complex scalar fields $\phi$ and $\sigma$ interacting with
seperate $U_{\phi}(1)$ and $U_{\sigma}(1)$ gauge fields $R_{\mu}$ and $A_{\mu}$ respectively. The 
breaking of $U_{\phi}(1)$ is responsible for the existence of vortices, while if we identify 
$U_{\sigma}(1)$ with electromagnetism, then its breaking inside the string will result to supercurrents 
flowing along the string.
If more fields are added to the Abelian Higgs model describing an Nielsen-Olesen vortex, then the appearence of currents in the core
of the defect is possible.
The notions involved in this case, which is one step further from the previous
one, will be the basis for the case of vortex rings later on.

\subsection{The model} 

 The above description takes us to the following {\bf Lagrangian density} for this model:
 \begin{equation}
 \mathcal{L}=|\tilde{D}_{\mu}\phi|^{2}+|D_{\mu}\sigma|^{2}-\frac{1}{4}B_{\mu\nu}B^{\mu\nu}
 -\frac{1}{4}F_{\mu\nu}F^{\mu\nu}-V(|\phi|, |\sigma|)
 \end{equation}
 with potential of the following form:
 \begin{equation}
 V(|\phi|, |\sigma|)=\frac{\lambda_{\phi}}{8}(|\phi|^{2}-\eta^{2})^{2}+\frac{\lambda_{\sigma}}{4}\Bigg(|\sigma|^{2} - \frac{m^{2}}{\lambda_{\sigma}}\Bigg)^{2}
 +\frac{v}{2}|\phi|^{2}|\sigma|^{2}-\frac{m^{4}}{4\lambda_{\sigma}}
  \end{equation}
 and $D_{\mu}\sigma =(\partial_{\mu}+ie A_{\mu})\sigma,\;\tilde{D}_{\mu}\phi =(\partial_{\mu}+ig R_{\mu})\phi$, $F_{\mu\nu}=\partial_{\mu}A_{\nu}-\partial_{\nu}A_{\mu}$,
$B_{\mu\nu}=\partial_{\mu}R_{\nu}-\partial_{\nu}R_{\mu}$.

 Away from the string, we have the vacuum of the theory where $U_{\phi}(1)$ is broken. There, 
 we have $<\phi>=\eta$ and the potential has the form:
 \begin{equation}
 V(|\phi|=\eta, |\sigma|)= \frac{\lambda_{\sigma}|\sigma|^{2}}{4}\Big(|\sigma|^{2}+2(v \eta^{2}-m^{2})\Big)
 \end{equation}
 which takes us to the condition:
 \begin{equation}
 \label{ccond1}
 v \eta^{2}-m^{2}\ge 0 \Rightarrow v \eta^{2} \ge m^{2}
 \end{equation}
 The latter ensures that away from the string: $<\sigma>=0$.

 A topological defect, here a string, is a region where the original bigger symmetry of our system 
 is restored, as mentioned in the introduction. There, we have $<\phi>=0$ and the  potential becomes:
\begin{equation}
V(|\phi|=0, |\sigma|)= \frac{\lambda_{\phi}}{8} \eta^{4}+ \frac{\lambda_{\sigma}}{4}|\sigma|^{4}
-\frac{m^{2}}{2}|\sigma|^{2}
\end{equation}
Breaking of $U_{\sigma}(1)$ inside the string translates to 
\footnote{$<\sigma>\neq 0$ inside the string and $\frac{dV(\phi=0,\sigma)}{d\sigma}=0 
\rightarrow
\sigma=\frac{m}{\sqrt{\lambda_{\sigma}}}$.}
$<\sigma>=\frac{m}{\sqrt{\lambda_{\sigma}}}$ and
the requirement of a non-zero potential
means another condition:
\begin{eqnarray}
\label{ccond2}
\frac{\lambda_{\phi}}{8}\eta^{4}-\frac{m^{4}}{4\lambda_{\sigma}} > 0 \Rightarrow{}
   \nonumber\\
{}\frac{\lambda_{\phi}}{2}\eta^{4}>\frac{m^{4}}{\lambda_{\sigma}}
\end{eqnarray}

 The {\bf Euler-Lagrange} equations follow:
 \begin{eqnarray}
 D_{\mu}D^{\mu}\phi+\frac{\lambda_{\phi}}{4}\Big(|\phi|^{2}-\Big(\eta^{2}
 -\frac{2 v}{\lambda_{\phi}}|\sigma|^{2}\Big)\Big)\phi=0 \\
 D_{\mu}D^{\mu}\sigma+\frac{\lambda_{\sigma}}{2}\Big(|\sigma|^{2}+\frac{v}{\lambda_{\sigma}}\Big(|\phi|^{2}-
 \frac{m^{2}}{v} \Big)\Big)\sigma=0 \\
 \partial_{\mu}B^{\mu\nu}=\tilde{J}^{\nu}=-ig\Big(\overline{\phi}\tilde{D}^{\nu}\phi-\phi
 \overline{\tilde{D}}^{\nu}\overline{\phi}\Big) \\
 \partial_{\mu}F^{\mu\nu}=J^{\nu}=-ie\Big(\overline{\sigma}D^{\nu}\sigma-\sigma\overline{D}^{\nu}\overline{\sigma}\Big)
 \end{eqnarray}
 Exploiting the equations with the currents above, useful and general relations can be extracted which
 can be helpful in order to write the differential equations for the gauge fields 
 $A_{\mu}$ and $R_{\mu}$. Maxwell equation $\nabla \times \mbox{\boldmath $B$}=\mbox{\boldmath$J$}$ 
 will be also used.
 Suppose a very general ansatz in three dimensions for $\phi$ and $\sigma$ such as 
 \begin{eqnarray} 
\phi=\phi_{0}(x,y,z)exp(i\varphi_{\phi}(x,y,z))\\
\sigma=\sigma_{0}(x,y,z)exp(i\varphi_{\sigma}(x,y,z))
\end{eqnarray}
 Then we get:
 \begin{eqnarray}
 -\Bigg(\frac{\partial^{2}A_{z}}{\partial x^{2}}+\frac{\partial^{2}A_{z}}{\partial y^{2}}\Bigg)
 +\frac{\partial^{2}A_{x}}{\partial z\partial x}+\frac{\partial^{2}A_{y}}{\partial z\partial y}
 =J_{z}=-2e\sigma_{0}^{2}(eA_{z}-\partial_{z}\varphi_{\sigma}) \\
 -\Bigg(\frac{\partial^{2}A_{y}}{\partial x^{2}}+\frac{\partial^{2}A_{y}}{\partial z^{2}}\Bigg)
 +\frac{\partial^{2}A_{x}}{\partial y\partial x}+\frac{\partial^{2}A_{z}}{\partial z\partial y}
 =J_{y}=-2e\sigma_{0}^{2}(eA_{y}-\partial_{y}\varphi_{\sigma}) \\
 -\Bigg(\frac{\partial^{2}A_{x}}{\partial y^{2}}+\frac{\partial^{2}A_{x}}{\partial z^{2}}\Bigg)
 +\frac{\partial^{2}A_{y}}{\partial x\partial y}+\frac{\partial^{2}A_{z}}{\partial x\partial z}
 =J_{x}=-2e\sigma_{0}^{2}(eA_{x}-\partial_{x}\varphi_{\sigma}) \\
 -\Bigg(\frac{\partial^{2}\tilde{A}_{z}}{\partial x^{2}}+\frac{\partial^{2}\tilde{A}_{z}}{\partial y^{2}}\Bigg)
 +\frac{\partial^{2}\tilde{A}_{x}}{\partial z\partial x}+\frac{\partial^{2}\tilde{A}_{y}}{\partial z\partial y}
 =\tilde{J}_{z}=-2g\phi_{0}^{2}(g\tilde{A}_{z}-\partial_{z}\varphi_{\phi}) \\
 -\Bigg(\frac{\partial^{2}\tilde{A}_{y}}{\partial x^{2}}+\frac{\partial^{2}\tilde{A}_{y}}{\partial z^{2}}\Bigg)
 +\frac{\partial^{2}\tilde{A}_{x}}{\partial y\partial x}+\frac{\partial^{2}\tilde{A}_{z}}{\partial z\partial y}
 =\tilde{J}_{y}=-2g\phi_{0}^{2}(g\tilde{A}_{y}-\partial_{y}\varphi_{\phi}) \\
 -\Bigg(\frac{\partial^{2}\tilde{A}_{x}}{\partial y^{2}}+\frac{\partial^{2}\tilde{A}_{x}}{\partial z^{2}}\Bigg)
 +\frac{\partial^{2}\tilde{A}_{y}}{\partial x\partial y}+\frac{\partial^{2}\tilde{A}_{z}}{\partial x\partial z}
 =\tilde{J}_{x}=-2g\phi_{0}^{2}(g\tilde{A}_{x}-\partial_{x}\varphi_{\phi}) 
 \end{eqnarray}
 The first three equations lead to a differential equation for $A_{\mu}$ and the last three 
 for $R_{\mu}$.

\subsection{Ansatz and equations} 

 The {\bf ansatz} we use for the fields in the case of an infinite straight superconducting string lying along
 the $z$-axis, is the following:
 \begin{eqnarray}
 \sigma&=&\sigma(\rho)  \\
 A_{\mu} &=&\frac{I(\rho)}{e}\partial_{\mu}z \\
 \phi& =& f(\rho)e^{in\varphi} \\
 R_{\mu}&=&\frac{P(\rho)}{g}\partial_{\mu}\varphi
 \end{eqnarray}
 where $\varphi$ is the polar angle and $n$ the winding number of the field $\phi$. 
 We use cylindrical coordinates $(\rho,\varphi,z)$.
These gauge fields have "magnetic" fields of the following form:
 \begin{eqnarray}
 \mbox{\boldmath $B$}&=&-\frac{1}{e}\frac{dQ(\rho)}{d\rho} \mbox{\boldmath $\hat{\varphi}$} \\
 \mbox{\boldmath $B_{R}$}&=&\frac{1}{g\rho}\frac{d P(\rho)}{d\rho} \mbox{\boldmath $\hat{z}$}
 \end{eqnarray}
 where $\hat{\rho}, \hat{\varphi}, \hat{z}$ are the cylindrical unit vectors.
 In order to have finite energy, the function $P(\rho)$ must vanish for large $\rho$ while the regularity condition
 at small $\rho$ demands that $P(0)=1$ and $\phi (0)=0$. The scalar field $\sigma$ does not have dependence on $\varphi$
 at the origin or else we would have a singularity there. The vector potential $A_{\mu}$ describes the effects of
 any current present in the string. For large $\rho$ the function $I(\rho)\propto I_{tot}ln(\rho)$, where $I_{tot}$ is
 the total current inside the string.

 The natural unit of length is the radius of the string $\delta= 1/\eta\sqrt{\lambda_{\phi}}$ so we choose to 
 rescale our fields: $\sigma =Y\eta,\;\phi =X\eta,\;I=Q\eta$ and $\rho \rightarrow x=\sqrt{\lambda_{\phi}}\eta \rho $.
 Thus, the system of {\bf field equations} to be solved becomes:
 \begin{eqnarray}
 X^{''}+\frac{1}{x}X^{'}-\frac{(P-1)^{2}}{x}X -\frac{1}{2}X^{3}+\frac{1}{2}X-\bar{v}XY^{2}=0 \\
 Q^{''}+\frac{1}{x}Q^{'}-\bar{e}^{2}Y^{2}Q=0 \\
 Y^{''}+\frac{1}{x}Y^{'}-YQ^{2}-\bar{\lambda}Y^{3}+\bar{m}^{2}Y-\bar{v}X^{2}Y=0 \\
 P^{''}-\frac{1}{x}P^{'}-\bar{g}^{2}X^{2}(P-1)=0
 \end{eqnarray}
 where prime denotes differentiation with respect to $x$ and $\bar{\lambda}=\lambda_{\sigma}/\lambda_{\phi}$,
 $\bar{v}=v/\lambda_{\phi}$, $\bar{g}^{2}=g^{2}/\lambda_{\phi}$, $\bar{e}^{2}=e^{2}/\lambda_{\phi}$,
 $\bar{m}^{2}=m^{2}/\lambda_{\phi}\eta^{2}$. According to these changes the conditions (\ref{ccond1}), (\ref{ccond2})
  become:
 \begin{equation}
 \bar{v} \ge \bar{m}^{2},\;\;\; \bar{\lambda} \ge 2 \bar{m}^{4}
 \end{equation}
 Notice that if we approximate $X(x)$ by $\tanh(x/2)$ for an ordinary cosmic string and search for a solution
 with vanishing electromagnetic current ($Q=0$) then the $Y$-equation decouples from the others and becomes:
 \begin{equation}
 Y^{''}+\frac{1}{x}Y^{'}-\bar{\lambda}Y^{2}-\bar{v}\tanh^{2}(x/2)Y+\bar{m}^{2}Y=0
 \end{equation}

Finally, the {\bf energy functional} which one has to minimize in order to get the final field configuration
of this model, is:
\begin{eqnarray*}
E=2\pi\int_{0}^{\infty}\rho d\rho\;\;\Bigg[ (\partial_{\rho}X)^{2}+(\partial_{\rho}Y)^{2}+\frac{1}{2\eta\bar{e}^{2}}(\partial_{\rho}P)^{2}
+\frac{1}{2\bar{g}^{2}\rho^{2}}(\partial_{\rho}Q)^{2}+{}
   \nonumber\\
{}+\frac{1}{\rho^{2}}(Q-1)^{2}X^{2}+\frac{1}{\eta}Y^{2}P^{2}+\frac{\bar{\lambda}}{8}(X^{2}-1)^{2}+\frac{\bar{\lambda}}{4}Y^{4}
+\frac{\bar{v}}{2}Y^{2}\Bigg(Q^{2}-\frac{\bar{m}^{2}}{v}\Bigg)\Bigg]
\end{eqnarray*}

\subsection{Numerical solution} 

We used Newton-Raphson algorithm (chapter $7$) to solve the system of equations above with an initial guess that
satisfies the boundary conditions and asymptotics of the fields. The virial relation for the 
problem at hand, changes slightly because of the existence of the scalar condensate $\sigma(x)$ and another 
"magnetic" term is added. Thus the virial relation  we check, is:
\begin{equation}
\int d^{2}x \Bigg( \frac{1}{2} B^{2} + \frac{1}{2} B_{R}^{2}-V\Bigg) =0
\end{equation}

\begin{figure}
\vspace{0.6 in.}
\centering
\includegraphics[scale=0.532]{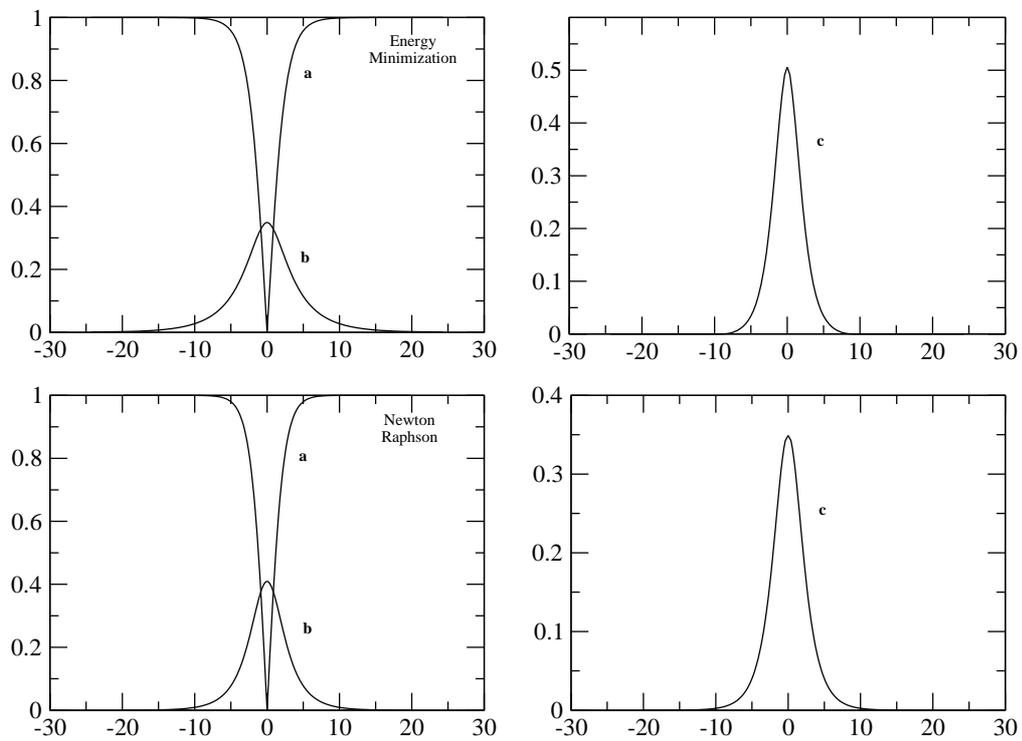}
\caption{\small Straight bosonic superconducting string. Both graphs on top are the output of energy
minimization algorithm.  Parameters are $(\bar{\lambda},\bar{m}^{2},\bar{g}^{2}, \bar{v}, e)$
=$(1, 0.65, 0.25, 0.75, 1)$. 
Bottom graphs are the output of Newton-Raphson algorithm. Parameters are $(\bar{\lambda},\bar{m}^{2},\bar{g}^{2}, \bar{v}, e)$
=$(1, 0.62, 0.25, 0.75, 1)$.
 Letter $a$ denotes the scalar field, $b$ the charge condensate and $c$ the magnetic field associated with the $a$ field. \label{ss}}
\end{figure}

We also found a solution to this model through 
minimization algorithm. The output of both algorithms is exhibited in figure \ref{ss} and the small differences one may observe are due to 
small changes in the  parameters.
A very helpful review and analysis concerning 
especially the $Y$ field, which is the charge condensate, can be found in \cite{cc2}.

%==================================================================================================================================
%==================================================================================================================================
%==================================================================================================================================
%==================================================================================================================================
%==================================================================================================================================
%==================================================================================================================================
%==================================================================================================================================
%==================================================================================================================================

\chapter*{PART II}

\section*{SEARCH FOR STRINGS AND SOLITONS}
\addcontentsline{toc}{chapter}{PART II: SEARCH FOR STRINGS AND SOLITONS }

This part includes the research. Briefly, the first section involves a search on antiperiodic solitons on $S^{1}$ which
is a based on \cite{k1}. The second section does a full numerical analysis of a straight string
with a Ginzburg-Landau potential with a cubic term added to it. Such a potential is used in condensed matter
physics as well \cite{paramos}. The third section presents a numerical search on static bosonic superconducting vortex rings
which is based on \cite{c1},\cite{c3} and can be seen as a continuation of \cite{c3}, where spherically symmetric
solutions were examined. The final section is based on the model of static vortex rings which is now extended by
adding higher derivative terms. We analyze the behavior of the solutions of the model under these modifications.

\chapter{Antiperiodic solitons of the Goldstone model on  $S^{1}$}

\newpage

\section{Introduction}

The purpose of this section is to present the complete list of static
classical solutions of the Goldstone model on a circle $S^{1}$ of
radius $L$ in $1+1$ dimensions and the corresponding bifurcation tree together with a
study of the stability of our solutions. It comes as a supplement
to a previous note \cite{k1} which was searching for such solutions
but with periodicity condition imposed there. Jacobi elliptic and standard trigonometric functions
are used to express the solutions found and stability analysis of
the latter is what follows. Classically stable quasi-topological solitons
are identified. Many of the results
have obvious similarities with those in \cite{k1}. We notice those and
make comparisons with this reference. We also check the case of mixed
boundary conditions and place a note with our conclusions in the end.
The following analysis has both mathematical \cite{k8} and physical
interest as it can be useful for the search of stable solitons in
the two-Higgs standard model (2HSM) or the minimal supersymmetric
standard model.

\section{The classical solutions}

The {\bf Langrangian density} of our model is
\begin{equation}
\label{lagr}
\mathcal{L}=\frac{1}{2}(\partial _{\mu }\phi _{1})^{2}+\frac{1}{2}(\partial _{\mu }\phi _{2})^{2}-V(\phi _{1},\phi _{2})\: ,\quad \mu =0,1
\end{equation}
 with $\phi _{1}$ and $\phi _{2}$ real Higgs fields and potential
of the following form:
\begin{equation}
V(\phi _{1},\phi _{2})=\frac{1}{4}(\phi _{1}^{2}+\phi _{2}^{2}-1)^{2}
\end{equation}
 We impose antiperiodic boundary conditions on the scalar fields with
coordinates $x\in [0,2\pi L].$
\begin{equation}
\label{anticond}
\phi _{i}\left(x+2\pi L\right)=-\phi _{i}\left(x\right)\: ,\quad i=1,2
\end{equation}
The general static solution of model (\ref{lagr}) can be expressed in terms
of Jacobi elliptic functions:

\begin{equation}
\label{eqa1}
\phi _{1}=\sqrt{R}\: \cos \Omega ,\quad \phi _{2}=\sqrt{R}\: \sin \Omega 
\end{equation}

\begin{equation}
\label{eqa2}
R(x)=a_{1}+a_{2}sn^{2}(\sqrt{2}\: \Lambda (x-x_{0})\: ,k)
\end{equation}

\begin{equation}
\label{eqa3}
\Omega (x)=C\! \int _{\xi }^{x}\! \frac{1}{R(y)}dy
\end{equation}
where $\! C\left(a_{1}\left(k,\Lambda \right),a_{2}\left(k,\Lambda \right)\right),\! x_{0},\! \xi $
are constants while $sn$ denotes the Jacobi elliptic function; $sn(z,k),$
$sn^{2}(z,k)$ are periodic functions on the real axis with periods
$4K(k)$ and $2K(k)$, respectively. $K(k)$ is the complete elliptic
integral of the first kind. Inserting (\ref{eqa1})-(\ref{eqa3}) into the equations corresponding
to (\ref{lagr}) leads to several conditions on the parameters and as a consequence
to three different types of non-trivial solutions. We start with the
simplest, the trivial solution.

\section{The trivial solution}

The energy functional for static configurations is given by 
\begin{equation}
E=\int _{0}^{2\pi L}dx\: \Bigg[ \frac{1}{2}\Bigg(\frac{\partial \phi _{1}}{\partial x}\Bigg)^{2}
+\frac{1}{2}\Bigg(\frac{\partial \phi _{2}}{\partial x}\Bigg)^{2}+V(\phi _{1},\phi _{2})\Bigg]
\end{equation}
and the field equations are
\begin{eqnarray}
\label{fieldeq}
-\frac{d^{2}\phi_{1}}{dx^{2}}+\phi_{1}\Big(\phi_{1}^{2}+\phi_{2}^{2}-1\Big)=0 \\
-\frac{d^{2}\phi_{2}}{dx^{2}}+\phi_{2}\Big(\phi_{1}^{2}+\phi_{2}^{2}-1\Big)=0 
\end{eqnarray}

Apart from the vacuum solutions which have $E_{vac}=0$, one can immediately
think of the simplest solution, the trivial one
\begin{equation}
\label{triv}
\phi _{1}=\phi _{2}=0\; ,\; E_{0}=\frac{L\pi }{2}
\end{equation}
which exists for all values of $L$. The corresponding small oscillation
eigenmodes, labeled by $j$, have
\begin{equation}
\label{om}
\tilde{\omega }^{2}(j)=\frac{1}{L^{2}}(\, \left(j+1/2\right)^{2}-L^{2})\: ,\quad j=0,1,2...
\end{equation}
 This solution is stable until $L=1/2$ because $\tilde{\omega }^{2}(0)<0$
for $L>1/2$. Many additional solutions, some of which were discussed
in \cite{c1} bifurcate from the solution $\phi _{1}=\phi _{2}=0$
at critical values of $L$.

\section{Three types of non-trivial solutions}

Now, we present without many details the three non-trivial solutions
and mention their similarities with those in \cite{k1}. The simplest
case one can think, is to set $a_{2}=0$ so that from (\ref{eqa2}) we obtain
$R(x)=a_{1}$ and the solution becomes
\begin{equation}
\label{s10}
\phi _{1}=\sqrt{1-\frac{\left(N+\frac{1}{2}\right)^{2}}{L^{2}}}\; \cos \left(\frac{\left(N+\frac{1}{2}\right)x}{L}\right)
\end{equation}
\begin{equation}
\label{s11}
\phi _{2}=\sqrt{1-\frac{\left(N+\frac{1}{2}\right)^{2}}{L^{2}}}\; \sin \left(\frac{\left(N+\frac{1}{2}\right)x}{L}\right)
\end{equation}
where $N$ is an integer. This solution is called type-I. One can
observe that the above solution reduces to (\ref{triv}) when $L=N+1/2$ i.e.
when one of the $\tilde{\omega }$ of equation (\ref{om}) crosses zero. The
Higgs field winds $(2N+1)/2$ times around the top of the Mexican
hat. It's energy is given by

\begin{equation}
E_{I}\left(L,N\right)=\frac{\pi }{L}\left(N+\frac{1}{2}\right)^{2}-\frac{\pi }{2L^{3}}\left(N+\frac{1}{2}\right)^{4}.
\end{equation}
 If we denote the above result which corresponds to the antisymmetric
case we study here with $E_{I}^{A}$ and the result of \cite{k1} for
the same type of solutions with $E_{I}^{S}$ where $S$ means {}``Symmetric''
and $A$ {}``Antisymmetric'', then we notice that 
\begin{eqnarray*}
E_{I}^{A}\left(N\, ,L\right) & = & E_{I}^{S}\left(N+\frac{1}{2}\: ,L\right)
\end{eqnarray*}
Another choice is to set $a_{1}=0$ on (\ref{eqa2}) so the solution now involves
the Jacobi elliptic functions. $C$ becomes zero as well and the solution
is
\begin{equation}
\label{sol2}
\phi _{1}=2k\Lambda sn\left(\sqrt{2}\Lambda x,\! k\right)\: ,\quad \quad \phi _{2}=0\: ,\quad \quad \Lambda ^{2}=\frac{1}{2\left(1+k^{2}\right)}
\end{equation}
This solution is called type-II. In fact, it corresponds to an oscillation
of the Higgs field in the $\phi _{2}=0$ plane about the origin $\phi _{1}=0=\phi _{2}$.
If we take account of the antiperiodicity condition (\ref{anticond}), then the
argument $k$ of the Jacobi elliptic function is related to the radius
$L$ of $S^{1}$ through the following formula

\begin{equation}
L=\frac{\left(2m+1\right)K(k)}{\pi }\sqrt{1+k^{2}}
\end{equation}
 for some integer $m$. When we reach the limit of $k\rightarrow 0$
(i.e. $L\rightarrow m+\frac{1}{2}$) the solution (\ref{sol2}) approaches
(\ref{triv}). The energy of the solution above is given by the integral 

\begin{equation}
E_{II}=\frac{4\left(2m+1\right)}{\sqrt{2}\Lambda \left(1+k^{2}\right)^{2}}\! \! \int _{0}^{K(k)}dy\! \left[\left(k^{2}sn^{2}\left(y,k\right)-\frac{1+k^{2}}{2}\right)^{2}+\frac{2k^{2}-1-k^{4}}{8}\right]
\end{equation}
 and by means of Elliptic integrals it becomes

\begin{equation}
E_{II}=\frac{4\left(2m+1\right)}{\sqrt{2}\Lambda \left(1+k^{2}\right)^{2}}\; \left[\frac{K(k)}{24}\: \left(3k^{4}+2k^{2}-5\right)+\frac{E(k)}{3}\: \left(k^{2}+1\right)\right]
\end{equation}
Two specific values of $k$ with the corresponding results follow
for $k=0$
\begin{equation}
E_{II}\left(L=m+1/2,m\right)=\frac{\left(2m+1\right)\pi }{4}
\end{equation}
and for $k=1$
\begin{equation}
E_{II}\left(L=\infty ,m\right)=\frac{4\left(2m+1\right)}{3\sqrt{2}}
\end{equation}
The comparison of $E_{II}^{A}$ which represents our solutions, with
$E_{II}^{S}$ which represents the symmetric case of the same type
of solutions studied in \cite{k1}, implies that 
\begin{equation}
E_{II}^{A}(m)  =  E_{II}^{S}\Big(m+\frac{1}{2}\Big)
\end{equation}
If none of $a_{1},a_{2}$ are zero then we are led to type-III solutions
where we have the following conditions for $a_{1},a_{2}$ and $C^{2}$
\begin{figure}
\centering
\includegraphics[ scale=0.5]{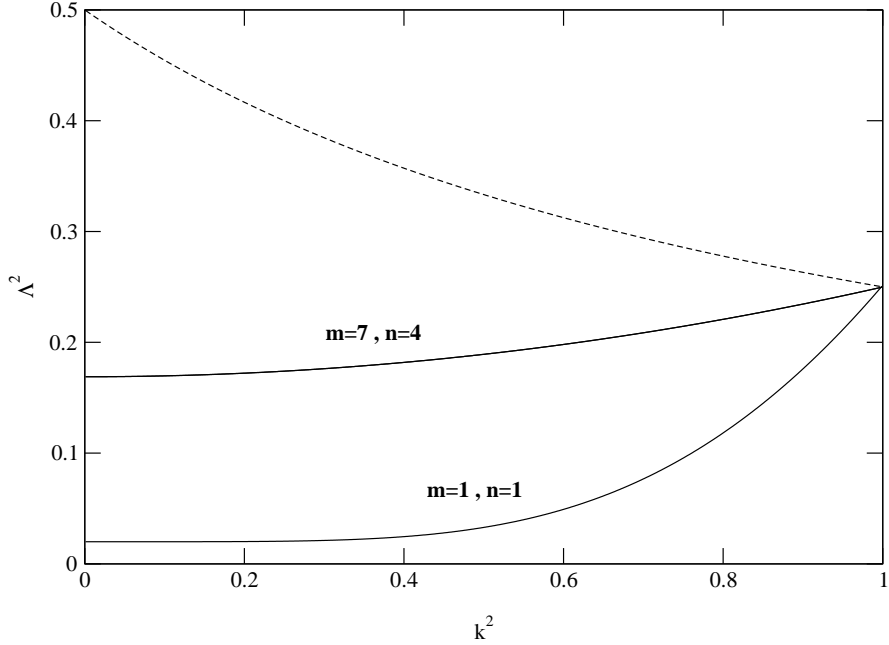}
\caption{\small  {\bf Goldstone model:}  The solutions of equations (\ref{s21}) and (\ref{s22}) are plotted as functions
of $k^{2}$ for two values of $(n,m)$. The dashed curve indicates the limit (\ref{s20}).  \label{pica1}}
\end{figure}

\begin{equation}
\label{s18}
a_{1}=\frac{2}{3}\left(1-2\Lambda ^{2}\left(1+k^{2}\right)\right),\qquad a_{2}=4k^{2}\Lambda ^{2}
\end{equation}

\begin{eqnarray}
\label{s19}
C^{2}&=&  \frac{4}{27}\left(1+\left(4k^{2}-2\right)\Lambda ^{2}\right)\left(1+\left(4-2k^{2}\right)\Lambda ^{2}\right)\left(1-2\Lambda ^{2}\left(1+k^{2}\right)\right)={}
                     \nonumber \\
{}&=& \frac{2}{9}\left(1+\left(4k^{2}-2\right)\Lambda ^{2}\right)\left(1+\left(4-2k^{2}\right)\Lambda ^{2}\right)a_{1}
\end{eqnarray}
Here, one can explicitly observe the fact that when $a_{1}=0$ then
$C^{2}=0$ as well. In addition, $R$ and $C^{2}$ must not be negative.
This means another condition 

\begin{equation}
\label{s20}
\Lambda ^{2}\leq \frac{1}{2\left(1+k^{2}\right)}
\end{equation}
 with the equality leading to type-II solutions.

In order for $\Omega $ to satisfy $\Omega \left(x+2\pi L\right)=\Omega \left(x\right)+\pi $
and $R$ to be periodic (so for the solution to be antiperiodic as
we want) on $[0,2\pi L]$ the following equations must hold:

\begin{equation}
\label{s21}
C\! \int _{0}^{2\pi L}\frac{1}{R(y)}dy=\left(2n+1\right)\pi 
\end{equation}

\begin{equation}
\label{s22}
L=\frac{\left(2m+1\right)K(k)}{\sqrt{2}\pi \Lambda }
\end{equation}
 where $m,n$ are positive integers which respectively determine the
number of oscillations of the modulus of the Higgs field and the number
of the Higgs field windings around the origin $\phi _{1}=0=\phi _{2}$
in a period $2\pi L$.

Now we return to (\ref{s18}) and (\ref{s19}) and analyze further type-III solutions.
Solving these equations for $k=0$, remembering that $K(0)=\pi /2$,
we find the critical values of $L$ where these solutions start to
exist:

\begin{equation}
\label{s23}
\Lambda ^{2}=\frac{m^{2}}{6\left(2n+1\right)^{2}-4m^{2}}\: \: \Rightarrow \: \: L^{2}=\frac{3}{4}\left(2n+1\right)^{2}-\frac{m^{2}}{2}
\end{equation}
The expression for $\Lambda ^{2}$ together with (\ref{s20}) leads to the
condition $2n+1>m$ on the integers $m$ and $n$ ($m$ and $n$ $\neq 0$). 
For any $n$ and $m\neq 2n+1$ the coefficient of $sn^{2}$ in the
integral (\ref{eqa3}) is proportional to $k^{2}$ and if one expands the solution in
powers of $k^{2}$ is led to the following formulae

\begin{equation}
\Lambda ^{2}=\frac{m^{2}}{6\left(2n+1\right)^{2}-4m^{2}}\! \! \left(1+\frac{k^{2}}{2}\right)+\mathcal{O}\left(k^{4}\right)
\end{equation}

\begin{equation}
L^{2}=\frac{3}{4}\left(2n+1\right)^{2}-\frac{m^{2}}{2}+\mathcal{O}\left(k^{4}\right)
\end{equation}

\begin{equation}
C^{2}=\frac{4}{3}\frac{\left(\left(2n+1\right)^{2}-m^{2}\right)^{2}\left(2n+1\right)^{2}}{\left(3\left(2n+1\right)^{2}-2m^{2}\right)^{3}}+\mathcal{O}\left(k^{4}\right)\! .
\end{equation}
This expansion is necessary as there is no closed form for integral
(\ref{s21}). In the limit $k=0$ we observe that the $2n$ solutions of type-III
yield the type-I solution with $N=n+1/2$ at $L^{2}=\left(3/4\right)\left(2n+1\right)^{2}-m^{2}/2$
, where $m=1,2,...,2n$. For fixed values of $k,\, n,\, m$ we find
a single solution $\Lambda ^{2}\left(k,\: \left(2n+1\right)/m\right)$
obeying the following properties
\begin{eqnarray}
\Lambda ^{2}\left(k=0,\: \left(2n+1\right)/m\right)&=&\frac{m^{2}}{6\left(2n+1\right)^{2}-4m^{2}} {}
\nonumber\\
{}\Lambda ^{2}\left(k=1,\: \left(2n+1\right)/m\right)&=&1/4
\end{eqnarray}
This is illustrated in figure \ref{pica1} for two different values of the pair
$(n,m)$ by the solid curves, the dashed curve representing the limit
(\ref{s20}).

The energy of the general type-III solution is given by the integral

\begin{equation}
E_{III}=  \frac{2m+1}{\sqrt{2}\: \Lambda }
 \cdot  \int _{0}^{K(k)}dy\: \left[\left(a_{1}-1+a_{2}sn^{2}(y,k)\right)^{2}+\frac{1}{6}\: \left(1+16\Lambda ^{4}\left(k^{2}-1-k^{4}\right)\right)\right]
\end{equation}
which, by means of $K(k),\: E(k)$ becomes

\begin{eqnarray}
E_{III}&=& \frac{2m+1}{\sqrt{2}\: \Lambda }\: \Bigg[ K(k)\: \left(\frac{8}{9}\, \Lambda ^{2}k^{2}\left(1+\Lambda ^{2}-\Lambda ^{2}k^{2}\right)-\frac{8}{9}\, 
\Lambda ^{2}\left(2+\Lambda ^{2}\right)+\frac{5}{18}\right)+{}
  \nonumber\\
 {}&+&\frac{8}{3}\, \Lambda ^{2}E(k)\Bigg]
\end{eqnarray}
For $k=1$ one obtains

\begin{equation}
E_{III}\left(1,m,n\right)=\frac{4\left(2m+1\right)}{3\sqrt{2}}=E_{II}\left(1,m\right).
\end{equation}
This shows that solution III approaches solution II in the limit
$k\rightarrow 1$, a fact which also appears in \cite{k1}.

\begin{figure}[t]
\centering
\includegraphics[scale=0.5]{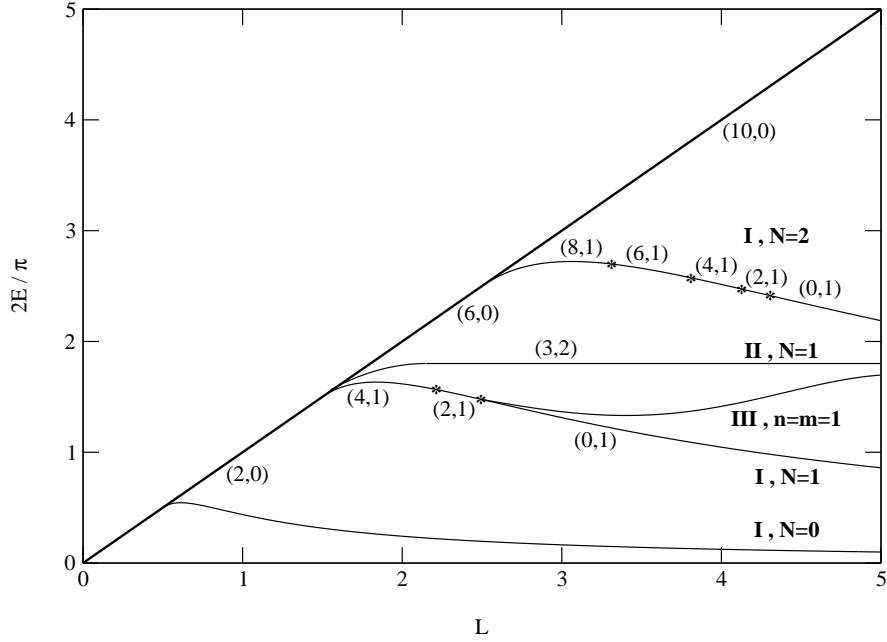}
\caption{\small  {\bf Goldstone model:} The energies of some of the solutions are plotted as functions of the parameter $L$.
The four stars on the upper branch indicate the four bifurcation points on the $N=2$ type-I solution.
The two lower stars indicate the two bifurcation points on the $N=1$ type-I solution. The two numbers in parentheses refer
to the number of negative and zero modes of the corresponding solution.   \label{pica2}}
\end{figure}

The energies of the solutions of some low-lying branches are presented
in figure \ref{pica2} as functions of $L$ . For $L>5/2$ all four types of
solutions coexist and satisfy

\begin{equation}
E_{I}\left(L,N=0\right)<E_{I}\left(L,N=1\right)<E_{III}\left(L,m=1,n=1\right)<E_{II}\left(L,m=1\right)<E_{0}\left(L\right).
\end{equation}
The four stars on the upper part of the figure show the position
of the four bifurcation points $L=3.279,$ $3.775,4.093,4.272$ of
the $n=2,\: m=3,2,1,0$ solutions respectively from the $N=2$ type-I
solution. The two lower stars show the bifurcation values $\left(L=2.179,\: 2.5\right)$
of the $n=1,\: m=2,1$ solutions respectively from the $N=1$ type-I
solution. 

\section{Stability Analysis}

\subsection{Type-I solutions}

In order to analyze the stability of type-I solutions we follow the
steps done in \cite{k1} which uses notions that can be found in \cite{k10}
and \cite{k15}. Thus perturbing the fields $\phi _{a}\: ,a=1,2$ around
the classical solution (\ref{s10})-(\ref{s11}), denoted here by $\phi _{a}^{cl}$,

\begin{equation}
\label{subst}
\phi _{a}\left(x\right)=\phi _{a}^{cl}\left(x\right)+\eta _{a}\left(x\right)\exp \left(-i\omega t\right)
\end{equation}
we are led to the following equation for the normal modes:

\begin{equation}
A\left(\tilde{N},L\right)\left(\begin{array}{c}
 \eta _{1}\\
 \eta _{2}\end{array}
\right)=\omega ^{2}\left(\begin{array}{c}
 \eta _{1}\\
 \eta _{2}\end{array}
\right)
\end{equation}

\begin{equation}
A\left(\tilde{N},L\right)\equiv -\frac{d^{2}}{dx^{2}}+2\left(1-\left(\frac{\tilde{N}}{L}\right)^{2}\right)\left(\begin{array}{cc}
 c^{2} & sc\\
 sc & s^{2}\end{array}
\right)-\left(\frac{\tilde{N}}{L}\right)^{2}
\end{equation}
 where $\omega ^{2}$ is the eigenvalue and 
\begin{equation}
c\equiv \cos \left(\tilde{N}x/L\right),\;\; s\equiv \sin \left(\tilde{N}x/L\right),\;\; N+1/2\equiv \tilde{N}
\end{equation}
We derived the above formula by substituting (\ref{subst}) into the 
field equations (\ref{fieldeq}). Terms $\mathcal{O}(\eta_{a}^{s})$
with $s \geq 2$ are considered small.

The complete list of eigenvalues of the operator $A\left(\tilde{N},L\right)$
for $\tilde{N}\geq 1$ can be obtained by classifying its invariant
subspaces. This can be done by using the Fourier decomposition. Type-I
solution has a twisted translational invariance and one can check
that for an integer $n\geq \tilde{N}$ the following finite-dimensional
vector spaces%
\footnote{These vector spaces should have been the same as in (34) of \cite{k1}
(with $N\rightarrow \tilde{N}$) but they are not, due to a misprint
in \cite{k1}. Here we correct this by writing these spaces explicitly
in equations (\ref{s34}) and (\ref{s35}).%
} are preserved by $A\left(\tilde{N},L\right)$ 

\begin{equation}
\label{s34}
V_{n}=Span\left\{ a_{p}\cos \frac{px}{L}\: ,\: \beta _{p}\sin \frac{px}{L}\: ,\: p-n=0\left(mod2\tilde{N}\right)\: ,\: \left|p\right|\leq n\right\} ,
\end{equation}

\begin{equation}
\label{s35}
\tilde{V_{n}}=Span\left\{ a_{p}\cos \frac{px}{L}\: ,\: -\beta _{p}\sin \frac{px}{L}\: ,\: p-n=0\left(mod2\tilde{N}\right)\: ,\: \left|p\right|\leq n\right\} 
\end{equation}
under the following condition
\begin{equation}
a_{p}=\beta _{p}\qquad \quad \quad \textrm{if}\qquad p-n+2\tilde{N}>0
\end{equation}
where $a_{p}$ , $\beta _{p}$ are arbitrary constants. The operator
$A\left(\tilde{N},L\right)$ can then be diagonalized on each of the
finite-dimensional vector spaces above, leading to a set of algebraic
equations.

To be more specific, define $\lambda _{1}\equiv 2\left(1-\lambda _{2}\right)$,
$\lambda _{2}\equiv \left(\tilde{N}/L\right)^{2}$. Also, consider
the vector

\begin{equation}
\left(\begin{array}{c}
 V_{\tilde{N}+k}\\
 V_{\tilde{N}-k}
\end{array}\right)\; 
\textrm{with}\quad \; \: V_{\tilde{N}+k}\equiv \left(\begin{array}{c}
 \cos \left(\frac{\tilde{N}+k}{L}\right)x\\
 \sin \left(\frac{\tilde{N}+k}{L}\right)x\end{array}
\right)\: ,\: V_{\tilde{N}-k}\equiv \left( \begin{array}{c}
 \cos \left(\frac{\tilde{N}-k}{L}\right)x \\
 \sin \left(\frac{\tilde{N}-k}{L}\right)x\end{array}
\right)
\end{equation}

Acting with the operator $A\left(\tilde{N},L\right)$ on one of the
above vectors (say $V_{\tilde{N}+k}$) we have the following steps:

\begin{eqnarray*}
\left\{ -\frac{d^{2}}{dx^{2}}+\frac{\lambda _{1}}{2}\: \left(\begin{array}{cc}
 2c^{2} & 2sc\\
 2sc & 2s^{2}\end{array}
\right)-\lambda _{2}-\omega ^{2}\right\} V_{\tilde{N}+k}=0\quad \Rightarrow 
\end{eqnarray*}

\begin{figure}[t]
\centering
\includegraphics[scale=0.5]{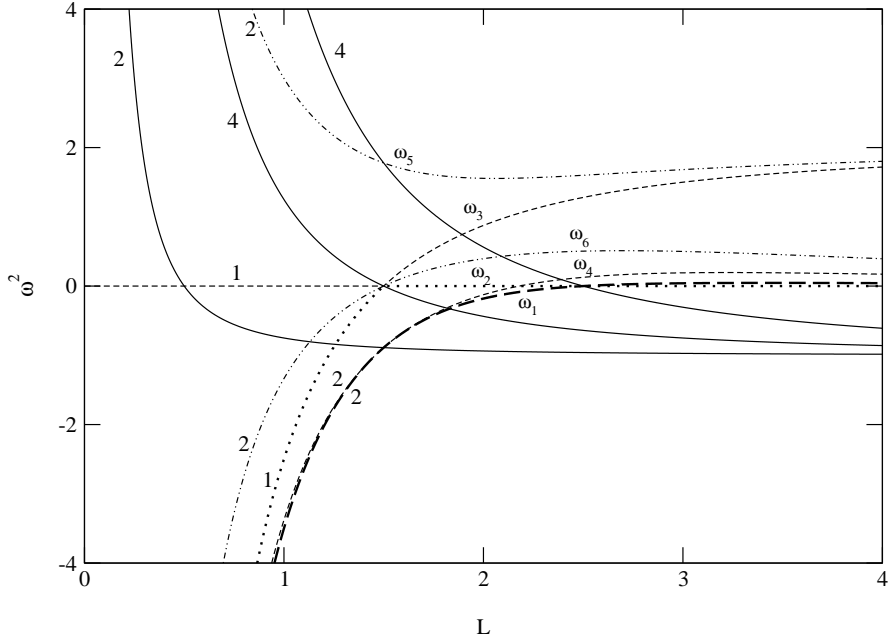}
\caption{\small  {\bf Goldstone model:}  The values (\ref{om}) are plotted as functions of $L$ for $j=0,1,2$ (solid curves),
together with the values $\omega_{1}^{2}(1,1,-1)$, $\omega_{2}^{2}(1,0,-1)$, $\omega_{3}^{2}(1,0,1)$, $\omega_{4}^{2}(1,2,-1)$, $\omega_{5}^{2}(1,1,1)$,
$\omega_{6}^{2}(1,3,-1)$ of (\ref{s37}) (dashed and dotted curves). The numbers indicate the multiplicity of the eigenvalues. \label{pica3}}
\end{figure}

\begin{equation}
\Rightarrow \left\{ -\frac{d^{2}}{dx^{2}}+\frac{\lambda _{1}}{2}\, \mathbf{1}-\lambda _{2}-\omega ^{2}+\frac{\lambda _{1}}{2}\, \mathbf{M}\right\} V_{\tilde{N}+k}=0\, 
\end{equation}
where $\mathbf{1}$ is the $2\times 2$ unit matrix and 
\begin{eqnarray*}
\mathbf{M}\equiv \left(\begin{array}{cc}
 \cos \left(\frac{2\tilde{N}x}{L}\right) & \sin \left(\frac{2\tilde{N}x}{L}\right)\\
 \sin \left(\frac{2\tilde{N}x}{L}\right) & -\cos \left(\frac{2\tilde{N}x}{L}\right)\end{array}
\right)
\end{eqnarray*}
which has the action
\begin{equation}
\mathbf{M}V_{\tilde{N}+k}=V_{\tilde{N}-k}\:  
\end{equation}

The same happens if we choose $V_{\tilde{N}-k}$ instead. In that
case, the equation above is $\mathbf{M}V_{\tilde{N}-k}=V_{\tilde{N}+k}$
as expected. A matrix which has exactly the same action as above can
be found easily and this is $\left(\begin{array}{cc}
 0 & \lambda _{1}/2\\
 \lambda _{1}/2 & 0\end{array}
\right)$. 

Finally, we observe that

\begin{eqnarray*}
\left(\begin{array}{cc}
 \left(\frac{\tilde{N}+k}{L}\right)^{2}+\frac{\lambda _{1}}{2}-\lambda _{2}-\omega ^{2} & \frac{\lambda _{1}}{2}\\
 \frac{\lambda _{1}}{2} & \left(\frac{\tilde{N}-k}{L}\right)^{2}+\frac{\lambda _{1}}{2}-\lambda _{2}-\omega ^{2}\end{array}
\right)V_{\tilde{N}+k}=0
\end{eqnarray*}

which implies the condition that the determinant of the above matrix
must be zero in order to acquire the non-zero eigenvalues.

The eigenvalues of $A\left(\tilde{N},L\right)$ on $V_{n}$ are

\begin{eqnarray}
\label{s37}
\omega ^{2}\left(N,k,\pm 1\right)&=&  \frac{1}{L^{2}}\: \: \left[L^{2}-\left(N+\frac{1}{2}\right)^{2}+k^{2}\right] \pm {}
   \nonumber\\
{}&\pm&   \frac{1}{L^{2}}\, \sqrt{\left(L^{2}-\left(N+\frac{1}{2}\right)^{2}\right)^{2}+4k^{2}\left(N+\frac{1}{2}\right)^{2}}
\end{eqnarray}
for $k=0,1,2,...$ and similarly on $\tilde{V_{n}}$ for $k=1,2,3,...$. 
Notice that if we replace $N+1/2$ by $\tilde{N}$ then the above
result is the same with eq.(36) of \cite{k1}.

For $N$ having a specific value and $L$ slightly greater than $N$,
the solution (\ref{s10})-(\ref{s11}) possess $4N$ negative modes corresponding
to $k=1,2,3,...,2N$ in (\ref{s37}) (minus sign). When $L$ increases, the
number of positive eigenmodes increases as well

\begin{equation}
L^{2}=\frac{3}{4}\left(2N+1\right)^{2}-\frac{m^{2}}{2}\! ,\quad \quad \quad m=1,2,...,2N,
\end{equation}
 i.e. (cf(\ref{s23})) at those values of $L$ where the solutions of type-III
with $n=N$ bifurcate from the solution of type-I. For 

\begin{equation}
L^{2}\geq L_{cr}^{2}\left(N\right)\equiv \frac{3}{4}\left(2N+1\right)^{2}-\frac{1}{2}
\end{equation}
all the modes are positive and (\ref{s10})-(\ref{s11}) are classically stable solitons.
These results are illustrated in figure \ref{pica2} for $N=1$ and $N=2$. The
numbers in parentheses represent the number of negative and zero modes
of the corresponding branch.

\subsection{Type-II and Type-III solutions}
The equations for the fluctuations about the solution (\ref{sol2}) decouple
to take the form of Lame equations:

\begin{equation}
\label{s40}
\left\{ -\frac{d^{2}}{dy^{2}}+6k^{2}sn^{2}\left(y,k\right)\right\} \eta _{1}=\Omega _{1}^{2}\eta _{1}
\end{equation}

\begin{equation}
\label{s41}
\left\{ -\frac{d^{2}}{dy^{2}}+2k^{2}sn^{2}\left(y,k\right)\right\} \eta _{2}=\Omega _{2}^{2}\eta _{2}
\end{equation}
 where

\begin{equation}
\label{s42}
x=\sqrt{1+k^{2}}\: y\: ,\quad \quad \Omega _{a}^{2}\equiv \left(\omega _{a}^{2}+1\right)\left(k^{2}+1\right)\! ,\quad \quad a=1,2
\end{equation}
 and $\omega _{a}^{2}$ is the effective eigenvalue of the relevant
operator. Equations (\ref{s40}) and (\ref{s41}) admit two and one algebraic modes,
respectively, with corresponding eigenvalues

\begin{equation}
\Omega _{1}^{2}:\: 4+k^{2}\: ,\: 1+4k^{2}\qquad \quad and\qquad \quad \Omega _{2}^{2}:\: 1+k^{2}
\end{equation}
The corresponding values of $\omega ^{2}$ follow immediately from
(\ref{s42}); they have signature $(+,+)$ and ($0$), respectively.

It's a property of the Lame equation that the solutions determined
algebraically correspond to the solutions of the lowest eigenvalues.
The remaining part of the spectrum therefore consists of positive
eigenvalues. The spectrum of the equation (\ref{s40}) was studied perturbatively
in \cite{k9}, while the relation between the Lame equation and the Manton-Samols
sphalerons was first pointed out in \cite{k11}-\cite{k13}.

Concerning the stability equation of type-III solution, it seems to
be more difficult to deal with as the presence of the Jacobi Elliptic
functions prevent us from having an analytic expression for $\Omega (x)$
in (\ref{eqa1}). Thus, we are not able to follow the steps done in the case
of type-I solutions (i.e. find the \textbf{M} matrix) where trigonometric
functions are easier to handle. Detailed analysis especially for the
case of Lame equations can be found in \cite{k12} and \cite{k15}.

\section{Conclusions}

A detailed analytical study of the static solutions of the Goldstone
model on a circle has been given in \cite{k1} and we followed the same
path here for our boundary conditions. Many results of \cite{k1} are
connected with ours by a simple change on variables used. We write
them explicitly above whenever is necessary. We also note our effort
to impose mixed boundary conditions as well. Specifically, we enforced
$\phi _{1}$ to be antiperiodic and $\phi _{2}$ periodic but there
was no solution to satisfy this choice so it's needless to extend
beyond this small remark here.

Many details were presented on the stability analysis 
for the solutions found. Classically stable solitons were identified,
together with the range of the parameter $L$ for which they are stable. 

This simpler model we present here and many others can be connected
and can also give us the experience to deal with realistic $(3+1)$-dimensional
particle physics models in our search for possible metastable localized
solitons. Ref. \cite{k1} on which this note was based, has connections
with \cite{k9} and \cite{k10} as it concerns the branches of their solutions,
also with \cite{c1}, while there are also interesting physical applications
\cite{k14} as it is already mentioned in \cite{k1}. Soliton solutions
in $(3+1)$-dimensional models with antiperiodic boundary condition imposed
on one spatial dimension, which is compactified on $S^{1}$, are analyzed
in \cite{k15}-\cite{k17} where supersymmetry breaking is examined.

\clearpage
\newpage

\section{{\normalsize APPENDIX}}

\subsection{{\normalsize Part I: Jacobi elliptic functions}}

The following serves as a small and helpful supplement on the mathematical functions we used above,
as well as some of their basic properties, which were useful.

Consider the following elliptic integral of first kind
\begin{equation}
u=\int_{0}^{\varphi} \frac{d\theta}{\sqrt{1-k\sin^{2}\theta}}
\end{equation}
Then we have the following definitions of the Jacobi elliptic functions
\begin{eqnarray}
sn\; u&=&\sin\varphi \\
cn\; u&=&\cos\varphi \\
dn\; u&=&\sqrt{1-k\sin^{2}\varphi}
\end{eqnarray}
There are other nine Jacobi elliptic functions which are constructed from the above three but we don't use them here.
The parameter $k$ is called elliptic modulus and $0\leq k \leq 1$, so the elliptic functions can be thought of as being given by two variables,
the amplitude $\varphi$ and the parameter $k$.

From the above definition, it is easy to see that
\begin{equation}
cn^{2}+sn^{2}=1
\end{equation}
is valid. Also,
\begin{equation}
dn^{2}+k^{2}sn^{2}=1
\end{equation}

The derivatives of these functions are
\begin{eqnarray}
\frac{d}{dz}\; sn (z,k)&=& cn(z,k) dn(z,k) \\
\frac{d}{dz}\; cn (z,k)&=& -sn(z,k) dn(z,k) \\
\frac{d}{dz}\; dn (z,k)&=& -k^{2}sn(z,k) cn(z,k) 
\end{eqnarray}
where $z$, in fact, depends on $u$.

Practically, $sn (x,k)$ is the solution of the following differential equations
\begin{equation}
\frac{d^{2}y}{dx^{2}}+(1+k^{2})y-2k^{2}y^{3}=0, \;\;\; \Bigg(\frac{dy}{dx}\Bigg)^{2}=(1-y^{2})(1-k^{2}y^{2})
\end{equation}
and $cn (x,k)$ is the solution of the following  equations
\begin{equation}
\frac{d^{2}y}{dx^{2}}+(1-2k^{2})y+2k^{2}y^{3}=0, \;\;\; \Bigg(\frac{dy}{dx}\Bigg)^{2}=(1-y^{2})(1-k^{2}+k^{2}y^{2})
\end{equation}
while $dn (x,k)$ is the solution of the following  equations
\begin{equation}
\frac{d^{2}y}{dx^{2}}-2k^{2}y+2y^{3}=0, \;\;\; \Bigg(\frac{dy}{dx}\Bigg)^{2}=-(1-y^{2})(1-k^{2}-y^{2})
\end{equation}

%===============================================================================================================
%===============================================================================================================
%===============================================================================================================
%===============================================================================================================
%===============================================================================================================
%===============================================================================================================
%===============================================================================================================

\chapter{$U(1)$ model for straight strings}

\newpage

\section{{\normalsize Introduction}}

The evolution of the Universe is believed to involve several symmetry breaking phase transitions, out of which,
topological defects, such as strings, are created \cite{cc1, p1, cc1b}. These transitions can be examined in the framework of
condensed matter systems. Although there are differences from the cosmological case where, 
relativistic dynamics must be used and gravity is important,  the formation of such defects in the laboratory \cite{zurek}, can
provide helpful hints for cosmology. Topologically stable knots and vortex-like structures in general, 
are of wide interest in condensed matter physics.
For example, one can think of Bose-Einstein condensates (BEC) (i.e. see \cite{hau}), 
vortices in superfluid Helium-$3$ and Helium-$4$ \cite{p2} , or nematic liquid crystals \cite{nl, nl2}.

Also, in the framework of high energy physics, future experiments in LHC could answer whether metastable particle-like
solitons exist in minimal supersymmetric Standard Model or two-Higgs Standard Model (2HSM) or not. In \cite{c3, c1},
work on classically stable, metastable quasi-topological domain walls and strings in simple topologically trivial models,
as well as in the 2HSM has been done. These solutions are local minima of the energy functional and can quantum mechanically 
tunnel to the vacuum, not being protected by an absolutely conserved quantum number. One can also find other interesting subjects
involving superconducting vortex rings such as, rotating superconducting rings \cite{vort1}, electroweak strings \cite{c8a, c8b}
or work on such rings in $SU(2)$ non-Abelian Yang-Mills-Higgs model \cite{c8c}. Finally, twisted semilocal vortices
examined in \cite{forg} can be connected to the models we present below while one can also search if stable rings can be made out of these strings.

\begin{figure}
\vspace{1.0 in.}
\centering
\includegraphics[scale=0.52]{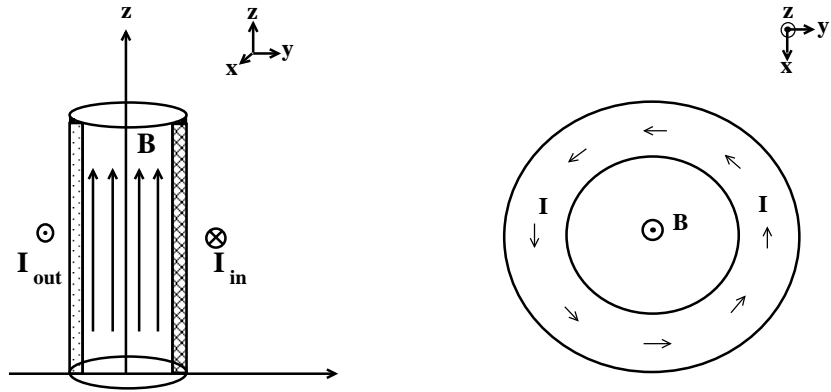}
\caption{\small {\bf $U(1)_{A}$ model:} The relative position of the supercurrent as well as of the magnetic field on
a $xy$-profile of the system on the right. The left picture is strongly reminiscent of an infinite solenoid.\label{sole}}
\end{figure}

In this chapter, we consider a $U(1)_{A}$ model with a modified Ginzburg-Landau (GL) potential. The modification has to do with the addition
of a cubic term. In Thermal Field Theory, such  term comes from the 1-loop radiative corrections to the GL potential \cite{cubic}.
We search whether this model can admit stable strings or not. The features of such strings, if they exist, are the supercurrent
which flows on the surface of the defect within a certain finite width, as well as,  a magnetic flux in the interior of the defect (fig.\ref{sole}).
This magnetic flux is a consequence of the existence of the supercurrent. One has to keep in mind that
the magnetic field can penetrate in a certain depth inside the superconducting regions where the supercurrent flows. If the penetration depth is greater
than the width of the superconducting surface, then the defect becomes unstable and can be destroyed.
The GL potential we use here, is used in condensed matter physics as well (see \cite{paramos} and references therein).

Stable defects of this $U(1)_{A}$ model, can also be used to form torus-like strings and study their stability. This can happen by taking a piece of such  straight string
and periodically connect its ends together. These string loops are examined in \cite{toros} (chapter $5$) but in the frame of a $U(1)\times U(1)$ model,
where the existence of the defect is ensured for topological reasons. That model, is a continuation of previous work \cite{c3, c1}.
In chapter $6$, we will examine an extended version of the model in \cite{toros} (chapter $5$).

\section{{\normalsize The  $U(1)_{A}$ model}}

This model consists of a complex scalar field $\psi$ and a gauge field $A_{\mu}$. The  {\bf Lagrangian density} describing our system is:
\begin{equation}
\label{lagg}
\mathcal{L} = -\frac{1}{4}F_{\mu\nu}^{2}+|D_{\mu}\psi|^{2}-U(|\psi|)
\end{equation}
where the covariant derivative is $D_{\mu}\equiv \partial_{\mu}+ieA_{\mu}$, the strength of the field is $F_{\mu\nu}=\partial_{\mu}A_{\nu}-\partial_{\nu}A_{\mu}$,
while $e$ is the $U(1)_{A}$ charge. The Lagrangian (\ref{lagg}), is invariant under the following $U(1)$ gauge transformation
\begin{equation}
\psi \rightarrow e^{ib(\mathbf{x})} \psi, \;\;\; A_{\mu}\rightarrow A_{\mu}-\frac{1}{e}\partial_{\mu}b(\mathbf{x})
\end{equation}
where $b(\mathbf{x})$ is a position-dependent phase.
We choose the potential
\begin{equation}
U(|\psi|)=\frac{a}{2}|\psi|^{2}\Bigg(\frac{1}{4}|\psi|^{2}-\frac{\beta}{3} |\psi|+\frac{\gamma}{2}\Bigg)
\end{equation}
where $a,\beta, \gamma$ constants. We can set $\gamma =1$. Since $\sqrt{\gamma}$ has dimensions of mass,
we count the energy of the system in units of $\sqrt{\gamma}$.
The vacuum is $|\psi|=0$. This vacuum leaves unbroken the gauge symmetry $U(1)_{A}$.
\begin{figure}
\vspace{0.4 in.}
\centering
\includegraphics[scale=0.45]{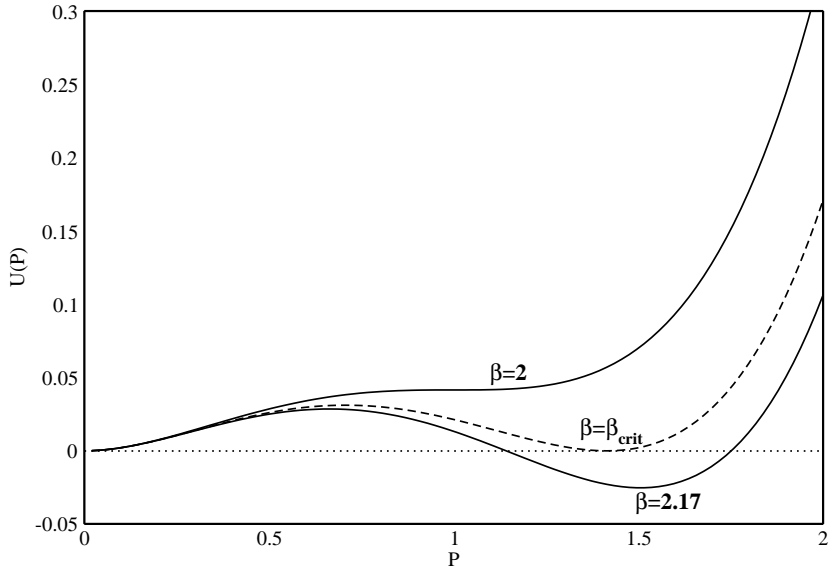}
\caption{\small {\bf $U(1)_{A}$ model:} The potential for $a=1$ and $\beta =2$, $\beta =\beta_{crit} =\frac{3}{\sqrt{2}}$ (dashed line) and  $\beta =2.17$ (bottom line). 
The plot is $U$ vs. $P$.\label{pot}}
\end{figure}
When 
\begin{equation}
\label{eqmin}
|\psi|=  |\psi_{0}| \equiv \frac{\beta +\sqrt{\beta^{2} -4}}{2}
\end{equation}
(for $\beta > 2$), we have $U(1)_{A}\rightarrow \mathbf{1}$
giving non-zero mass to $A$. Thus, one may generate an electric current flowing along regions where $|\psi|\neq  0$.
In fig.\ref{pot} one can see the shape of the potential. The equation (\ref{eqmin}) gives the position of the minimum of interest
for every $\beta >2$.
When $\beta =2$, the secondary (non-trivial) minimum of the potential disappears, at $|\psi|=1$ position.
When $2 < \beta < \frac{3}{\sqrt{2}} \equiv \beta_{crit}$, it becomes zero only at $|\psi|=0$, while another
minimum with non-zero $|\psi|$ forms.
When $\beta = \beta_{crit}$, the potential  has another zero at $|\psi|=\sqrt{2}$ which is also a local minimum. 
Finally, when $\beta > \beta_{crit}$ it has two more zeros at
\begin{equation}
\label{ppmm}
P_{\pm}=\frac{2\beta}{3}\pm \sqrt{\frac{4\beta^{2}}{9}-2}
\end{equation}
between which, it becomes negative (fig.\ref{neg}).
The mass spectrum is 
\begin{equation}
m_{A}=0,\;\;\; m_{\psi}^{2}=\frac{a}{2}
\end{equation}

\section{{\normalsize The $U(1)_{A}$ model: Search for stable vortices}}

We are interested in configurations having cylindrical symmetry, that is, infinite straight strings.
The field $\psi$ can be non-vanishing on a cylindrical surface of specific radius. 
At infinity $(\rho\rightarrow \infty)$, we have the vacuum of the theory $|\psi|=0$.
The {\bf ansatz} for the fields is:
\begin{equation}
\psi(\rho, \varphi ,z)= P(\rho)e^{iM\varphi}, \;\;\; \mathbf{A}(\rho, \varphi ,z) = \frac{A_{\varphi}(\rho)}{\rho} \hat{\varphi}
\end{equation}
where $M$ the winding number of the field $\psi$ and $\hat{\rho}$, $\hat{\varphi}$, $\hat{z}$ are the cylindrical unit vectors.
We use cylindrical coordinates $(t,\rho,\varphi,z)$, with space-time metric
$g_{\mu\nu}=diag(1,-1,-\rho^{2},-1)$. We work in the $A^{0}=0$ gauge. For the gauge field we suppose the above form based on 
the following thought: The $\mathbf{A}$ field is the one produced by the supercurrent flowing on the cylindrical surface. The 
current is in the $\hat{\varphi}$ direction thus, we expect the non-vanishing component to be $A_{\varphi}$ and the amplitude $P$ of $\psi$
to be independent of $\varphi$. As it concerns
the scalar field $\psi$, since it follows the geometry of the cylindrical defect, we expect that its amplitude is independent of $z$ as well.

With the above ansatz, the {\bf energy functional} for minimization takes the form
\begin{equation}
\label{funcsol}
E= 2\pi  \int_{0}^{\infty} \rho d\rho \Bigg[ \frac{1}{2\rho^{2}} (\partial_{\rho} A_{\varphi})^{2} +(\partial_{\rho} P)^{2}+\frac{P^{2}}{\rho^{2}}(eA_{\varphi} +M)^{2}+U(P)\Bigg]
\end{equation}
and the potential is
\begin{equation}
U(P)=\frac{a}{2}P^{2}\Bigg(\frac{1}{4}P^{2}-\frac{\beta}{3} P+\frac{1}{2}\Bigg)
\end{equation}
The gauge field $\mathbf{A}$ has a magnetic field of the form
\begin{equation}
\mathbf{B}= \frac{1}{\rho}\frac{\partial A_{\varphi}}{\partial \rho} \hat{z}
\end{equation}
while, the {\bf field equations} are
\begin{equation}
\label{fieq}
\partial_{\rho}^{2}P+\frac{1}{\rho}\partial_{\rho}P-\frac{P}{\rho^{2}}(eA_{\varphi}+M)^{2}-\frac{aP}{4}\Big(P^{2}-\beta P +1\Big)=0
\end{equation}
\begin{equation}
\label{fieq2}
\partial_{\rho}^{2}A_{\varphi} -\frac{1}{\rho}\partial_{\rho}A_{\varphi} - 2eP^{2}(eA_{\varphi}+M)=0 
\end{equation}

The usual rescaling arguments lead to the {\bf virial} relation:
\begin{equation}
\label{virsol}
2\pi \int_{0}^{\infty} \rho d\rho \Bigg( \frac{B^{2}}{2}-U\Bigg) =
2\pi \int_{0}^{\infty} \rho d\rho \Bigg(\frac{(\partial_{\rho}A_{\varphi})^{2}}{2\rho^{2}}-\frac{aP^{2}}{2}\Bigg(\frac{P^{2}}{4}-\frac{\beta P}{3}+\frac{1}{2}\Bigg)\Bigg)=0
\end{equation}
Define
\begin{eqnarray*}
I_{1}&\equiv& 2\pi \int_{0}^{\infty} \rho d\rho \; \frac{1}{2\rho^{2}} \Bigg(\frac{\partial A_{\varphi}}{\partial \rho}\Bigg)^{2} \\
I_{2}&\equiv& -2\pi \int_{0}^{\infty} \rho d\rho \; \frac{aP^{2}}{2}\Bigg(\frac{P^{2}}{4}-\frac{\beta P}{3}+\frac{1}{2}\Bigg)
\end{eqnarray*}
For a solution of the model, we theoretically must have $I_{1}+I_{2}=0$. In fact,
we define the index $V\equiv \frac{||I_{1}|-|I_{2}||}{|I_{1}|+|I_{2}|}$. 
We want this index as small as possible.
Other virial relations can be derived as follows.
For example, one can consider the double rescaling of $\rho \rightarrow \lambda \rho$ and either $P\rightarrow \mu P$ or $A_{\varphi} \rightarrow \mu A_{\varphi}$
or even both of the fields and then demand $\partial_{\lambda}E|_{\lambda=1=\mu}=0=\partial_{\mu}E|_{\lambda=1=\mu}$.

\section{{\normalsize The $U(1)_{A}$ model: Numerical results }}

We use a standard minimization algorithm to minimize the energy functional  (\ref{funcsol}). The algorithm is written in C.
One can find details about the algorithm used, on page 425 of \cite{c6} but, briefly, the basic idea is this:
Given an appropriate initial guess, there are several corrections to it, having as a criterion the minimization of the energy in every step. When the corrections
at the value of the energy are smaller than $\approx 10^{-8}$ the program stops and we get the final results.
We are interested in final configurations having non-trivial energy. This signals the existence of a stable
vortex with that energy. We check our results through virial relation (\ref{virsol}). Finally, our results must also
satisfy the field equations (\ref{fieq}), (\ref{fieq2}).
\begin{figure}[t]
\vspace{0.4 in.}
\centering
\includegraphics[scale=0.45]{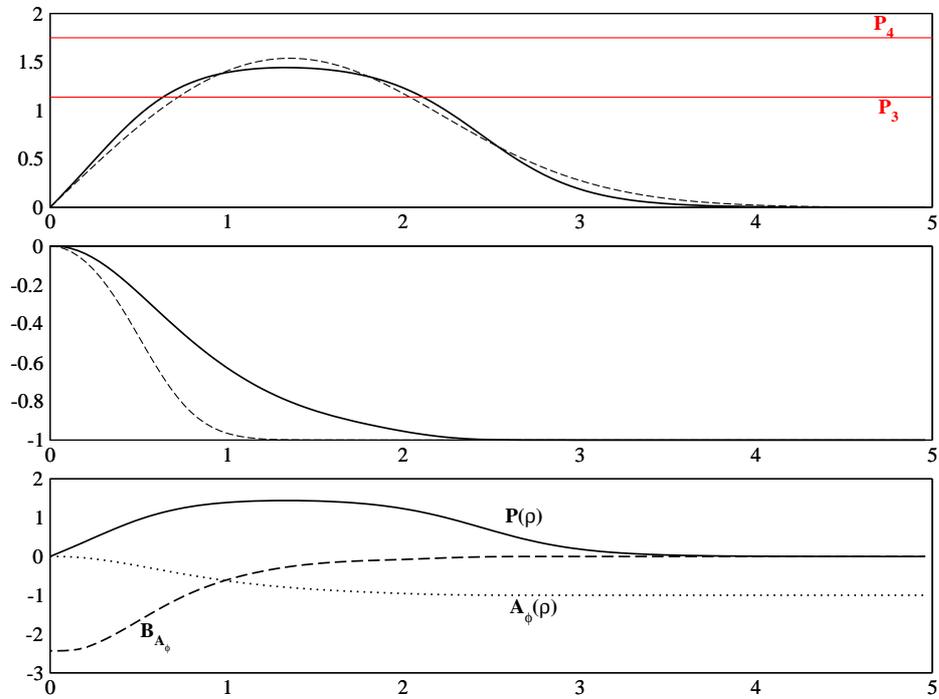}
\caption{\small {\bf $U(1)_{A}$ model:}  In the first two graphs we have the initial guess (dashed lines - - - ) 
as well as the final configuration of fields (solid lines --- ) $P(\rho),\;A_{\varphi}(\rho)$. We chose $M=1$, $e=1$, $\beta= 2.17$, $a=43.7$. 
The energy $E= 35.3$ and virial is $10^{-4}$. The bottom graph gathers all the fields.
For the field $P$, the area between the horizontal lines $P_{3}$ and $P_{4}$ is energetically favorable (see also fig.\ref{neg}). \label{pa217}}
\end{figure}

\begin{figure}[t]
\vspace{0.4 in.}
\centering
\includegraphics[scale=0.45]{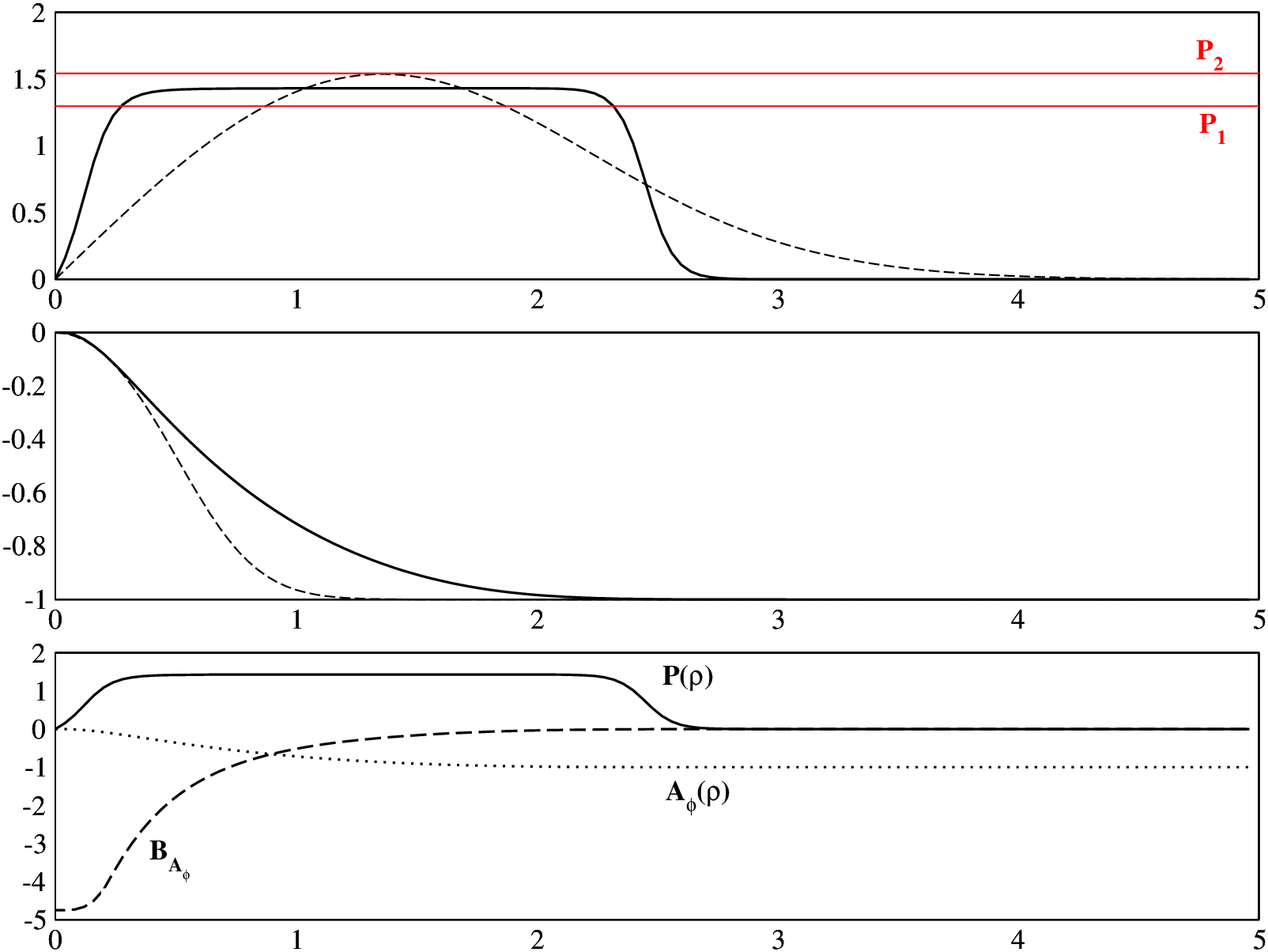}
\caption{\small {\bf $U(1)_{A}$ model:} In the first two graphs we have the initial guess (dashed lines - - - ) 
as well as the final configuration of fields (solid lines --- ) $P(\rho),\;A_{\varphi}(\rho)$. We chose $M=1$, $e=1$, $\beta= 2.13$, $a=1104$. 
The energy $E= 115.2$ and virial is $2\cdot 10^{-3}$. The bottom graph gathers all the fields.
For the field $P$, the area between the horizontal lines $P_{1}$ and $P_{2}$ is energetically favorable (see also fig.\ref{neg}). \label{pa213}}
\end{figure}

The {\bf initial guess} we use for our computation is:
\begin{eqnarray*}
P(\rho) & = & \xi_{1} \rho^{M} (1-\tanh(0.2\rho^{2})) \\
A_{\varphi} & = & -\tanh(\xi_{2} \rho^{2})
\end{eqnarray*}
where $\xi_{1}, \xi_{2}$ are constants, the value of which, depends also on the location of the minimum (say $|\psi|=|\psi_{0}|$) of the potential.
For the final configurations we present in the figures \ref{pa217}, \ref{pa213}, we chose $\xi_{1}=1.75,\; \xi_{2}=2$.

The initial guess also satisfies the appropriate asymptotics
\begin{itemize}
\item{near $\rho=0$: $P\sim \rho^{M},\;\;\; A_{\varphi}\sim \rho^{2}$
}
\item{at infinity: $P\rightarrow 0$, exponentially}
\end{itemize}
while $P$ must be non-zero somewhere between $\rho =0$ and $\rho \rightarrow \infty$.
One must make a careful choice of the initial guess. That is to say, the maximum value of $P$ in the initial guess (dashed lines) must
be inside the favorable area denoted by the horizontal lines in figs.\ref{pa217}-\ref{pa213}.
This area is dictated by the form of the potential and especially by its negative sectors (see fig.\ref{neg}).

\begin{figure}
\vspace{0.4 in.}
\centering
\includegraphics[scale=0.45]{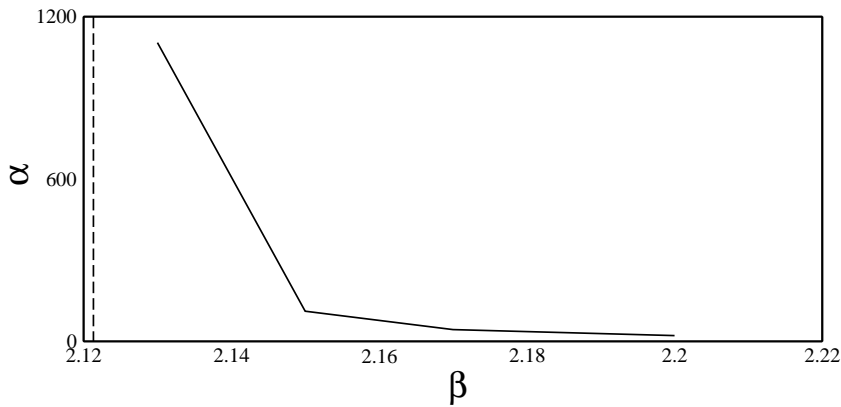}
\caption{\small {\bf $U(1)_{A}$ model:} The plot exhibits the relation of the parameter $a$ with respect to $\beta$. We observe that
as we approach the limit $\beta_{crit}$, we need an increasingly deeper minimum of the potential which is expressed through
the fastly increasing value of $a$. The dashed vertical line signals the position of $\beta_{crit}=\frac{3}{\sqrt{2}}$.\label{ab}}
\end{figure}

Since, for values of $\beta$ where $2\leq \beta \leq \beta_{crit}$ is valid,  we find no non-trivial solution,
we searched and tried to find out what happens when we make the minimum of interest deeper. 
For that reason we searched in the region where $\beta >\beta_{crit}$. 
It is possible
to find solutions to this model until $\beta$ reaches $2.13$ (from above). Under this value, this is difficult if not impossible.
Even at $\beta=2.13$ (fig.\ref{pa213}) we use great values of $a$ in order to find the solution exhibited. 
The relation between $a$ and $\beta$ can be found in fig.\ref{ab}.
The solutions of the model for two different values of $\beta > \beta_{crit}$, are shown in figs.\ref{pa217}-\ref{pa213} and
a comparison between them in fig.\ref{comp} in order to observe their different features. 

\subsection{{\normalsize Analysis for  $2.13 \leq \beta \leq 2.2$ }}

We observe that for a specific $\beta$,
there is a small range for the parameter $a$ of the potential, where the model exhibits the solution presented in the figures.
Out of this small range and for greater $a$, we end up to a final configuration of negative energy. This happens because,
the bigger the parameter $a$ becomes, the stronger the potential is, thus the field $P$ strongly prefers to acquire the value
where the non-trivial minimum of the potential is (see fig.\ref{pot}), in order to decrease further  the energy. But,
for $\beta >\beta_{crit}$ we have $U<0$ at the position of that minimum, which also
enforces the total energy $E$ to be negative in this case. We have to note that, virial relation of such a final configuration
is {\em not} satisfied due to the great values the term $(\partial_{\rho} P)^{2}$ acquires around $\rho =0$ and $\rho \rightarrow \infty$.
This is clear if one directly observes (\ref{virsol}), which can not be satisfied for $U<0$.

On the other hand, for smaller values of $a$, we get the trivial configuration ($P=0, A_{\varphi}=0$), as the benefit
from the potential term is no longer satisfactory in order to have a non-trivial $P$.

\subsection{{\normalsize Reasons for instability when $\beta > 2.2$ }}

Now, for high values of $\beta$ (i.e. $\beta > 2.2$) we faced difficulties in finding a  solution.
We believe that this has the following explanation: as $\beta$ grows, the area of values where
the potential $U$ is negative, increases as well.
\begin{figure}
\vspace{0.4 in.}
\centering
\includegraphics[scale=0.52]{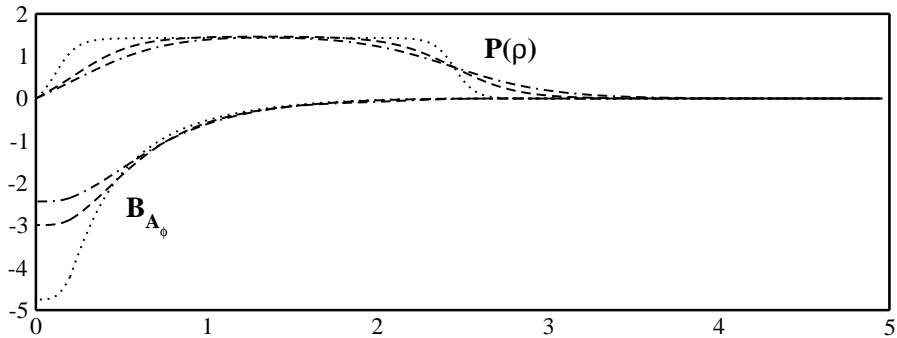}
\caption{\small {\bf $U(1)_{A}$ model:} Here we have a plot  in order to compare the changes of $P(\rho)$ and $B_{A}(\rho)$ as
$a$ goes from 2.17 (dashed and dotted - $\cdot$ - ), to 2.15 (dashed - - - ) and 2.13 (dotted $\cdot \cdot \cdot$).\label{comp}}
\end{figure}

If the parameter $a$ is large, then the potential becomes a strong factor in reducing the energy, as it is
highly negative. Thus, for large $a$ it is energetically favorable
to decrease further the potential value. The latter happens by converting $P$ function in such a way, so as to be, as much of it as possible,
inside the energetically favorable area (it is denoted by the  horizontal lines in the figs.\ref{pa217}-\ref{pa213}). Thus, we
end up to a final configuration with potential $U << 0$ and virial relations can not be satisfied (see for example eq.\ref{virsol}) as
we have a sum of positive terms.

On the other hand, for lower $a$ the potential is no longer a strong factor for reducing the energy.
In this case, it is energetically favorable  to reduce the value of the $(\partial_{\rho} P)^{2}$
and $(\partial_{\rho} A_{\varphi})^{2}$ terms as the system can gain more from these. The consequence is that $P$ leaves the area of stability as
its peak lowers in order to reduce the two terms above and 
there is only one possibility: to end up to zero energy, that is to say, the trivial configuration.

The difference in the stable solutions we have found above, is that the values of $\beta$ are such, that the potential is negative
but not strongly negative while the changes on the terms $(\partial_{\rho} P)^{2}$ and $(\partial_{\rho} A_{\varphi})^{2}$ can deform
the field $P$ in such a way, so that it can still be inside the favorable area. Then, the potential term has
the possibility to change in such a way, so it can satisfy virial as well.
\begin{figure}
\vspace{0.4 in.}
\centering
\includegraphics[scale=0.45]{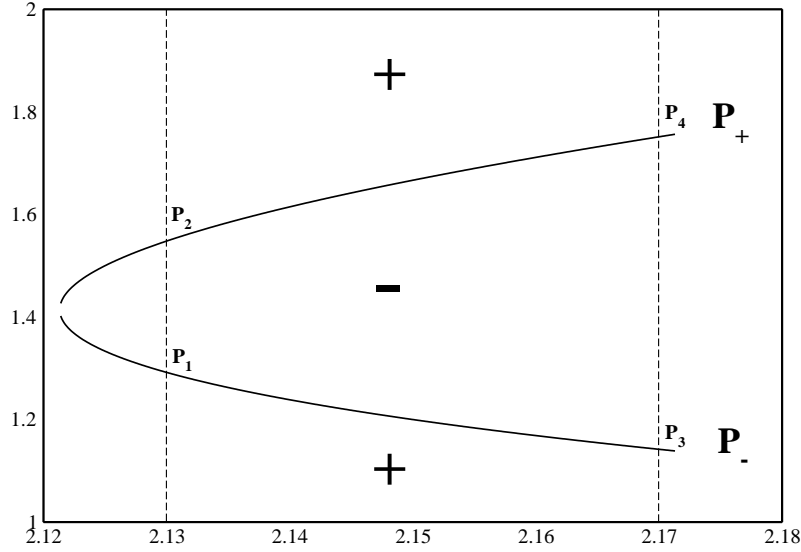}
\caption{\small {\bf $U(1)_{A}$ model:} The $P_{+}$ and $P_{-}$ solution of eq.(\ref{ppmm}), for which the potential becomes zero. Between these lines
the potential gets negative values (see also fig.\ref{pot}). The plot is $P$ vs. $\beta$ for $\beta > \beta_{crit.} =\frac{3}{\sqrt{2}}$.\label{neg}}
\end{figure}

\subsection {{\normalsize Reasons for instability when $\beta \leq \beta_{crit}$ }}

In the following explanation we will use fig.\ref{comp} \& \ref{neg}. In fig.\ref{comp} one can observe
that as we get closer to the critical value $\beta =3/\sqrt{2} \equiv \beta_{crit}$, the $P$ field tends
to acquire everywhere the value $P=P_{0}$ (location of the non-trivial minimum of the potential).
This leads to greater values of $\partial_{\rho}P$. 

The above has a reasonable explanation which can be found in fig.\ref{neg}. As we get closer to $\beta_{crit}$, the space
within the lines, where the potential can get negative values, becomes smaller. Under $\beta_{crit}$ the potential can be either positive
or zero (the latter for $P=0$ only).

Observe the energy functional to be minimized:
\begin{equation}
E= 2\pi  \int_{0}^{\infty} \rho d\rho \Bigg[ \frac{1}{2\rho^{2}} (\partial_{\rho} A_{\varphi})^{2} +(\partial_{\rho} P)^{2}+\frac{P^{2}}{\rho^{2}}(eA_{\varphi} +M)^{2}+U(P)\Bigg]
\end{equation}
The main target of minimization  is to ``fix'' all the above terms in order to have the minimum possible value for the energy. 
Below, we analyze the possible cases.
\begin{itemize}
\item{$\beta \rightarrow \beta^{+}_{crit}$: }

In that case, all terms except for the potential term, can be either positive or zero. The potential term (as we saw in fig.\ref{neg}) can become negative
for a range of values of $P$. But, as $\beta$ decreases, this range becomes narrower (as we see in fig.\ref{neg} as well as in figs.\ref{pa217}-\ref{pa213}
where this range is represented by the space between the two horizontal lines) and $(\partial_{\rho}P)^{2}$ increases.
This happens because the energy functional
tends to decrease its value through the negative values of the potential term. 
We believe that this can {\em not} continue for $\beta$ very close to $\beta_{crit}$ due to the fact that the range we described above,
becomes so small, that $P$ tends to get everywhere a constant value (the value $P_{0}$, which makes
the potential negative). But $P$ must be zero at $\rho =0$ and $\rho \rightarrow \infty$, thus there will be a considerable increase in the $(\partial_{\rho} P)^{2}$
term of the functional, and that makes the benefits of the negative value of the potential to go away, while the trivial solution $P=0$
becomes energetically favorable.
\item{$\beta \leq \beta_{crit}$: }
In that case, the potential term can no longer become negative. It can be either positive or zero and because of the fact that the energy
functional is a sum of five positive terms, it's reasonable to prefer the zero value which, at the same time, minimizes all the terms of the functional.
\end{itemize}

From all the above, one can observe that the crucial difference between the above two cases of $\beta$, is that in the $\beta > \beta_{crit}$ case, 
the energy can have a minimum value (through the potential term) which corresponds to a non-trivial solution for $P$. The existence of negative values 
of the potential are the ``way-through'' that make this possible.

\section{{\normalsize Conclusions}}

We studied a $U(1)_{A}$ model with a GL potential with a cubic term added to it. 
After the numerical analysis we did, we came to the conclusion that for $\beta \leq \beta_{crit}$ we find
no non-trivial solution. For $ 2.13 \leq \beta \leq 2.20$ we get non-trivial solutions
which have the profiles we present in figs.\ref{pa217}-\ref{pa213}. Over $\beta =2.20$ we have no non-trivial solutions.
We analyze and explain our results in these cases.

The form of the potential we have in this search,
can be found  in condensed matter physics as well. 
On the other hand, one could try to make a loop out of the straight string studied above.
We did this, but it was difficult to study mainly due to the fact that there are two instability modes, one having to do with the defect
itself and another which has to do with the loop that tends to shrink due to its tension. 
The former instability is excluded in \cite{toros} for topological reasons. More details about the latter,
can be found in the next chapter.

%===============================================================================================================================
%===============================================================================================================================
%===============================================================================================================================
%===============================================================================================================================
%===============================================================================================================================
%===============================================================================================================================

\chapter{On axially symmetric solitons in Abelian-Higgs models}

\newpage
\section{{\normalsize Introduction}}

In a series of papers \cite{c1, c3} classically stable, metastable quasi-topological domain walls and strings in simple topologically trivial models,
as well as in the two-Higgs Standard Model (2HSM) were studied. They are local minima of the energy functional and can quantum
mechanically tunnel to the vacuum, not being protected by an absolutely conserved quantum number. In \cite{c3} a search for
spherically symmetric particle-like solitons in the 2HSM with a simplified Higgs potential was performed without success.
Although the existence of spherically symmetric particle-like solitons in the 2HSM has not been ruled out, we shall here look, instead,
for axially symmetric solutions in a similar system.

\begin{figure}[t]
\vspace{0.4 in.}
\centering
\includegraphics[scale=0.52]{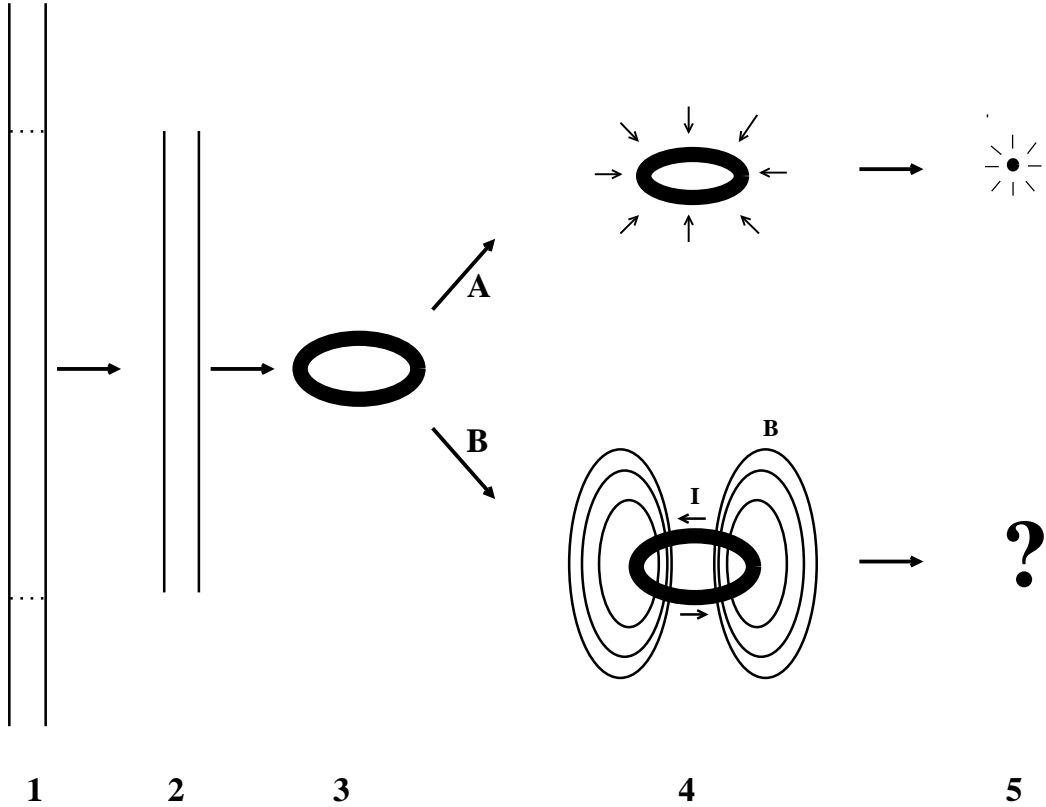}
\caption{\small {\bf $U(1)_{A}\times U(1)_{W}$ model:} Step 1: Infinite straight string. Step 2: Cut a piece of it.
Step 3: Periodically connect its ends together to form a loop. Step 4A: Rings without current shrink due to their tension...
Step 5A: ...and finally collapse. Step 4B: Superconducting vortex rings can be stabilized?... Step 5B: ... or not?\label{plan} }
\end{figure}

\begin{figure}[t]
\vspace{0.4 in.}
\centering
\includegraphics[scale=0.52]{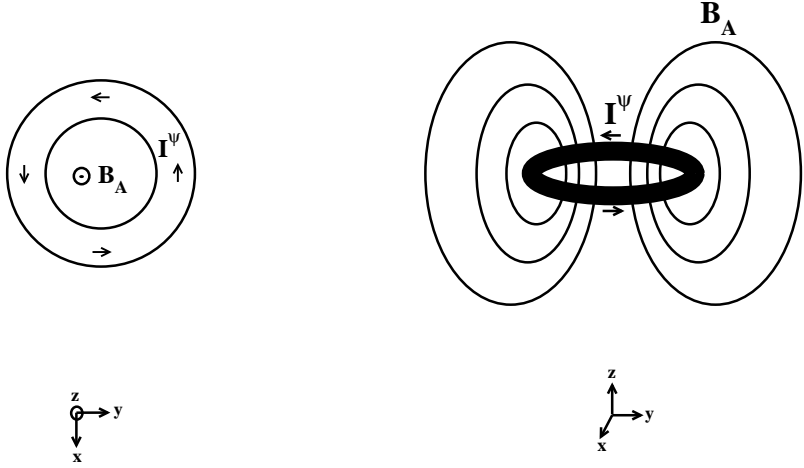}
\caption{\small {\bf $U(1)_{A}\times U(1)_{W}$ model:} A $x-y$ profile of the superconducting ring (left) as well as a $y-z$ profile (right)
where one can view how the mechanism we propose, against shrinking, works.\label{mechh} }
\end{figure}

Consider a model with superconducting strings \cite{c3, c4}. Take a piece of such string, close it to form a donut-shaped loop and let current in it.
A magnetic field due to the supercurrent will be passing through the hole of the donut (fig.\ref{mechh}). The energy of  the loop has, a term
proportional to the length of the string and will tend to shrink the radius of the donut to zero and the ring to extinction. However,
another force opposes this tendency. Namely, as the loop shrinks, the magnetic field lines are squeezed in the hole, since, due to 
the Meissner effect, they cannot leave the loop. They are trapped inside the hole of the donut, oppose further shrinking and might 
even stabilize the string. A brief plan of what we are searching, can be seen in fig.\ref{plan}.

This, as well as other arguments \cite{vort2}-\cite{vort12} are inspiring but not conclusive. The magnetic field will not be trapped inside the loop if the penetration depth
of the superconductor is larger than the thickness of the ring. Also, once the magnetic field gets strong  it can destroy
superconductivity and penetrate \cite{c8}. Further, there is a maximum current a superconductor can support (current quenching).
This sets a limit on the magnetic field one can have through the loop, and this may not be enough to stabilize it.
Thus, the above approach may work at best  in a certain region of the parameter space, depending also on the defect characteristics \cite{c3}.
The purpose of this work is to apply the above straightforward idea to search for string loops in a $U(1)\times U(1)$ gauge model
and to determine the parameter space, if any, for their existence and stability.

Another interesting subject is to have a rotating ring. The rotation
is another extra factor which could help the ring to stabilize. This work was done with success both analytically and numerically in \cite{vort1}
where vortons are exhibited. Another recent example of rotating superconducting electroweak strings can be found in \cite{c8a}, while for
a review on electroweak strings, the reader should also check \cite{c8b}. Finally, a work on static classical vortex rings in
$SU(2)$ non-Abelian Yang-Mills-Higgs model can be found in \cite{c8c}.

\section{{\normalsize The $U(1)_{A}\times U(1)_{W}$ model }}

The  {\bf Lagrangian density} of our model is:
\begin{equation}
\label{laggg}
\mathcal{L}=-\frac{1}{4}F_{\mu\nu}^{2}-\frac{1}{4}W_{\mu\nu}^{2}+|D_{\mu}\psi|^{2}+|\tilde{D}_{\mu}\phi|^{2}-U(|\phi|,|\psi|)
\end{equation}
where the covariant derivatives are $D_{\mu}\psi \equiv \partial_{\mu}\psi+ieA_{\mu}\psi$, $\tilde{D}_{\mu}\phi \equiv \partial_{\mu}\phi+iqW_{\mu}\phi$,
the strength of the fields are $F_{\mu\nu}=\partial_{\mu}A_{\nu}-\partial_{\nu}A_{\mu}$, $W_{\mu\nu}=\partial_{\mu}W_{\nu}-\partial_{\nu}W_{\mu}$, 
while $e$ and $q$ stand as the relevant $U(1)$ charges.
The above Lagrangian (\ref{laggg}), is invariant under the following $U(1)$ gauge transformations
\begin{eqnarray*}
\phi \rightarrow e^{ip(\mathbf{x})} \phi &,& \;\;\; W_{\mu}\rightarrow W_{\mu}-\frac{1}{q}\partial_{\mu}p(\mathbf{x})\\
\psi \rightarrow e^{ib(\mathbf{x})} \psi &,& \;\;\; A_{\mu}\rightarrow A_{\mu}-\frac{1}{e}\partial_{\mu}b(\mathbf{x})\\
\end{eqnarray*}
where $b(\mathbf{x}),\; p(\mathbf{x})$ are position dependent phases.
We choose the potential $U$
\begin{equation}
U(|\phi|,|\psi|)=\frac{g_{1}}{4}\big( |\phi|^{2} -v_{1}^{2}\big)^{2}+\frac{g_{2}}{4}\big( |\psi|^{2}- v_{2}^{2}\big)^{2}+\frac{g_{3}}{2}|\phi|^{2}|\psi|^{2}-\frac{g_{2}}{4}v_{2}^{4}
\end{equation}
The vacuum  $|\phi|= v_{1} \neq 0$, $|\psi|=0$, breaks $U(1)_{W} \times U(1)_{A} \rightarrow U(1)_{A}$,
giving non-zero mass to $W$. The photon field stays massless. There, $U(v_{1},0)=0$. 
The vacuum manifold $\mathcal{M}$ in this theory is a circle $S^{1}$ and the first homotopy group of $\mathcal{M}$
is $\pi_{1}(\mathcal{M})=\pi_{1}(S^{1})=\mathbf{Z}\neq\mathbf{1}$ which signals the existence of strings.
In regions where $|\phi|=0$, the field $|\psi|$ is arranged to be
non-vanishing and $U(1)_{W}\times U(1)_{A}\rightarrow U(1)_{W}$. Thus, $U(1)_{A}\rightarrow \mathbf{1}$ and
one may generate an electric current flowing along regions with vanishing $|\phi|$.
Hence, this theory has superconducting strings \cite{c4}.

In fact, with $|\phi|=F$ and $|\psi|=P$, the extrema of the potential are four:
\begin{itemize}
\item{ $F=0=P$ with $U(0,0)=\frac{1}{4}g_{1}v_{1}^{4}$.}
\item{ $F=0$, $P=v_{2}$ with $U(0,v_{2})=\frac{1}{4} g_{1}v_{1}^{4}-\frac{1}{4}g_{2}v_{2}^{4}$.}
\item{ $F=v_{1}, P=0$ with $U(v_{1},0)=0$.}
\item{ $F=a_{1}, P=a_{2}$ with $U(a_{1},a_{2})= a_{3}-\frac{1}{4}g_{2}v_{2}^{4}$}
\end{itemize}
with 
\begin{eqnarray*}
a_{1}=\Bigg(\frac{g_{2}(g_{1}v_{1}^{2}-g_{3}v_{2}^{2})}{g_{1}g_{2}-g_{3}^{2}}\Bigg)^{1/2},\;\;
a_{2}=\Bigg(\frac{g_{1}(g_{2}v_{2}^{2}-g_{3}v_{1}^{2})}{g_{1}g_{2}-g_{3}^{2}}\Bigg)^{1/2}, \\
a_{3}=\frac{g_{3}(g_{1}g_{3}(g_{3}-g_{1}g_{2})v_{1}^{4}+g_{2}g_{3}(g_{3}-g_{1}g_{2})v_{2}^{4}-2g_{1}g_{2}(g_{3}^{2}-g_{1}g_{2})v_{1}^{2}v_{2}^{2})}{4(g_{1}g_{2}-g_{3}^{2})^{2}}
\end{eqnarray*}
The vacuum that is of interest to us is: $|\phi|=v_{1},\;\;|\psi|=0$ and  leaves unbroken the electromagnetic $U(1)_{A}$.

We can choose $\phi = (f_{1}(\mathbf{x})+v_{1})e^{i\theta_{1}(\mathbf{x})}$ 
and $\psi = f_{2}(\mathbf{x})e^{i\theta_{2}(\mathbf{x})}$ with $<0|f_{1}(\mathbf{x})|0>=0=<0|f_{2}(\mathbf{x})|0>$.
Plugging these in the Lagrangian, we have the masses of the fields:
\begin{equation}
m_{A}=0,\;\;m_{W}=q v_{1},\;\;m_{\phi}^{2}=g_{1}v_{1}^{2},\;\;m_{\psi}^{2}=\frac{1}{2}\big(g_{3}v_{1}^{2}-g_{2}v_{2}^{2}\big)
\end{equation}
When $|\phi| =v_{1}$, the potential becomes:
\begin{equation}
U(v_{1}, |\psi|)= |\psi|^{2}\Bigg(\frac{g_{2}}{4}|\psi|^{2}+k^{2}\Bigg)
\end{equation}
where $k^{2}= \frac{1}{2}(g_{3}v_{1}^{2}-g_{2}v_{2}^{2})$. Searching for the minimum we find:
\begin{equation}
\frac{dU}{d|\psi|}=0\rightarrow |\psi|\Bigg(2k^{2}+g_{2}|\psi|^{2}\Bigg)=0
\end{equation}
When $k^{2}>0$, we have only one minimum at $|\psi|=0$. Otherwise ($k^{2}<0$) we have a minimum at
$|\psi|^{2}=-\frac{2k^{2}}{g_{2}}$ which  means that the photon acquires mass. Thus, we choose the condition $k^{2}>0$ which
reads as the right equation that follows:
\begin{equation}
\label{conditions}
g_{1}>g_{2}\frac{v_{2}^{4}}{v_{1}^{4}},\;\;\; g_{3}>g_{2}\frac{v_{2}^{2}}{v_{1}^{2}}
\end{equation}
The left equation is another condition which comes from the fact that when $|\phi|=0$ and $|\psi|=v_{2}$ the
potential has value 
\begin{equation}
U(0, v_{2})= \frac{g_{1}v_{1}^{4}}{4} - \frac{g_{2}v_{2}^{4}}{4}
\end{equation}
and in order to ensure that this is greater from the value of the potential at the vacuum, which is $0$,
we arrive at the condition above. This makes the value of the vacuum of interest, lower than the other extremum values
as well since
\begin{equation}
U(0,0)\; >\; U(0,v_{2})\; >\; 0=U(0,0)
\end{equation}

\section{{\normalsize The $U(1)_{A}\times U(1)_{W}$ model: Search for superconducting vortex rings}}

Configurations with torus-like shape, representing a piece of a $U(1)_{W}\rightarrow \mathbf{1}$ Nielsen-Olesen string \cite{c9},
closed to form a loop, are of interest in this search. Thus, we will  require $\phi$ to vanish on a circle of radius $a$ (the torus radius)
$\phi (\rho =a, z=0)=0$. At infinity ($\rho\rightarrow \infty, z\rightarrow \infty$), we have the vacuum of the theory. This translates to $|\phi|\rightarrow v_{1}$,
$|\psi| \rightarrow 0$.
The {\bf ansatz} for the fields is:
\begin{eqnarray*}
\phi(\rho,\varphi,z)&=&F(\rho,z)e^{iM\Theta(\rho,z)} \\
\psi(\rho,\varphi,z)&=&P(\rho,z)e^{iN\varphi} \\
\mathbf{A}(\rho,\varphi,z)&=&\frac{A_{\varphi}(\rho,z)}{\rho}\;\hat{\varphi} \\
\mathbf{W}(\rho,\varphi,z)&=&W_{\rho}(\rho,z)\;\hat{\rho}+W_{z}(\rho,z)\;\hat{z} 
\end{eqnarray*}
where $M$, $N$ are the winding numbers of the relevant fields, $\hat{\rho}$, $\hat{\varphi}$, $\hat{z}$ are the
cylindrical unit vectors
and we define 
\begin{equation}
\Theta(\rho,z)=\arctan \Big( \frac{z}{\rho -a}\Big)
\end{equation}
We use cylindrical coordinates $(t,\rho , \varphi , z)$,
with space-time metric $g_{\mu\nu}= diag(1,-1,-\rho^{2},-1)$. We work in the $A^{0}=0=W^{0}$ gauge.
Especially for the gauge fields, we suppose the above form based on the following reasonable thoughts:
The $\mathbf{W}$-field is the one related to the formation of the string thus, it exists in the constant-$\varphi$ plane.
This means that in general its non-vanishing components will be $W_{\rho}$ and $W_{z}$.
The $\mathbf{A}$-field is the one produced by the supercurrent flowing inside the toroidal object. The current is 
in the $\hat{\varphi}$ direction thus, we in general expect the non-vanishing component to be $A_{\varphi}$.
Finally, as it concerns the scalar fields $\phi, \psi$, they follow the geometry of the toroidal defect which has axial
symmetry thus, we expect that their amplitude is independent of $\varphi$.

A more general choice for $\phi$ would be
\begin{equation}
\phi(\rho,\varphi,z)  =  F(\rho,z)e^{iM\Theta(\rho,z) +i\chi(\rho, z)}
\end{equation}
where $\chi(\rho, z)$, an arbitrary function. But gauge invariance allows us to change $\phi \rightarrow \phi e^{ib(\mathbf x)}$, where $b(\mathbf{x})$
an arbitrary space-dependent phase. We can choose $b(\mathbf{x})= - \chi(\rho, z)$ thus, gauge fixing removes the arbitrary function $\chi$.

We proceed by extracting the energy-momentum tensor and by considering the above ansatz we will
derive the energy functional. Couple the system to an external gravitational field $g_{\mu\nu}$ and define
\begin{equation}
T^{\mu\nu}=-\frac{2}{\sqrt{-g}}\frac{\delta S}{\delta g_{\mu\nu}}
\end{equation}
The model coupled to the background gravitational field is
\begin{eqnarray*}
\mathcal{L}&=&\sqrt{-g}\Bigg(-\frac{1}{4}g^{\mu\lambda}g^{\nu\rho}F_{\mu\nu}F_{\lambda\rho}+g^{\mu\nu}(D_{\mu}\psi)^{*}D_{\nu}\psi
-\frac{1}{4}g^{\mu\lambda}g^{\nu\rho}W_{\mu\nu}W_{\lambda\rho}\\
&+&g^{\mu\nu}(\tilde{D}_{\mu}\phi)^{*}\tilde{D}_{\nu}\phi-U\Bigg)
\end{eqnarray*}
and by using 
\begin{equation*}
\delta g^{\mu\nu}=-g^{\mu a} g^{\beta \nu}\delta  g_{a \beta}, \;\;\;\delta \sqrt{-g} = \frac{1}{2} \sqrt{-g} g^{a \beta}\delta g_{a \beta}
\end{equation*}
we get
\begin{equation}
T^{a \beta}=-F^{a}_{\mu}F^{\beta}_{\nu}g^{\mu\nu}+2(D^{a}\psi)^{*}D^{\beta}\psi-W^{a}_{\mu}W^{\beta}_{\nu}g^{\mu\nu}+2(\tilde{D}^{a}\phi)^{*}\tilde{D}^{\beta}\phi
-\frac{g^{a\beta}\mathcal{L}}{\sqrt{-g}}
\end{equation}
We are interested in the energy of the system which is the integral of the $00$ component of the energy-momentum tensor
\begin{equation}
E=\int d^{n}x \sqrt{\bar{g}} \; T^{00}
\end{equation}
where $\sqrt{\bar{g}}$ is the determinant of the spatial part of the metric.
The energy density for the static ansatz above is
\begin{equation}
T^{00}=\frac{1}{4} g^{ik}g^{jl}F_{ij}F_{kl}-g^{ij}(D_{i}\psi)^{*}D_{j}\psi +\frac{1}{4}g^{ik}g^{jl}W_{ij}W_{kl}-g^{ij}(\tilde{D}_{i}\phi)^{*}\tilde{D}_{j}\phi +U
\end{equation}
The terms of the energy density are
\begin{eqnarray*}
\frac{1}{4}g^{ik}g^{jl}F_{ij}F_{kl}&=&\frac{1}{2}\Bigg(\frac{(\partial_{\rho}A_{\varphi})^{2}}{\rho^{2}}+\frac{(\partial_{z}A_{\varphi})^{2}}{\rho^{2}}\Bigg)\\
\frac{1}{4}g^{ik}g^{jl}W_{ij}W_{kl}&=&\frac{1}{2}(\partial_{\rho}W_{z}-\partial_{z}W_{\rho})^{2}\\
-g^{ij}(D_{i}\psi)^{*}D_{j}\psi&=&(\partial_{\rho}P)^{2}+(\partial_{z}P)^{2}+\frac{P^{2}}{\rho^{2}}(eA_{\varphi}+N)^{2}\\
-g^{ij}(\tilde{D}_{i}\phi)^{*}\tilde{D}_{j}\phi&=&(\partial_{\rho}F)^{2}+(\partial_{z}F)^{2}+((qW_{\rho}+M\partial_{\rho}\Theta)^{2}+(qW_{z}+M\partial_{z}\Theta)^{2})F^{2}
\end{eqnarray*}

\begin{figure}
\centering
\includegraphics[scale=0.96]{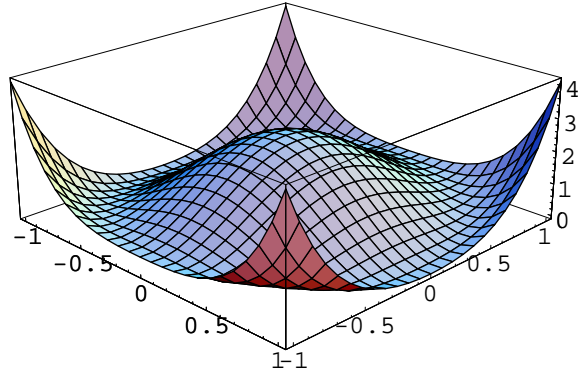}
\caption{\small {\bf $U(1)_{A}\times U(1)_{W}$ model:} A wider view of the ``Mexican hat'' potential (eq.\ref{potu1}) for  $(g_{1},g_{2},u)=(10,8,1)$.\label{mhpos}}
\end{figure}

Thus, with the above ansatz, the {\bf energy functional} takes the form:
\begin{eqnarray}
\label{funcu1}
E&=&2\pi v_{1} \int_{0}^{\infty}\rho d\rho \int_{-\infty}^{\infty}dz \Bigg[ \frac{1}{2\rho^{2}}\Big( (\partial_{\rho}A_{\varphi})^{2}+(\partial_{z}A_{\varphi})^{2}\Big)+{}
                                                   \nonumber\\
{}&+&(\partial_{\rho}P)^{2}+(\partial_{z}P)^{2}+\frac{P^{2}}{\rho^{2}}(eA_{\varphi}+N)^{2}+{}
                                                          \nonumber\\
{}&+&(\partial_{\rho}F)^{2}+(\partial_{z}F)^{2}+\frac{1}{2} (\partial_{\rho}W_{z}-\partial_{z}W_{\rho})^{2}+{}
                                                  \nonumber\\
{}&+&\Big( (qW_{\rho}+M\partial_{\rho}\Theta)^{2}+(qW_{z}+M\partial_{z}\Theta)^{2}\Big)F^{2}+U(F,P) \Bigg]
\end{eqnarray}
and the potential $U$  (fig.\ref{mhpos}) can be written as follows:
\begin{equation}
\label{potu1}
U(F,P)=\frac{g_{1}}{4}\big( F^{2}-1\big)^{2}+\frac{g_{2}}{4}\big( P^{2}-u^{2}\big)^{2}+\frac{g_{3}}{2}F^{2}P^{2}-\frac{g_{2}}{4}u^{4}
\end{equation}
where  $u\equiv v_{2}/v_{1}$. This is the energy functional we  use in our numerical work.
The {\bf conditions} (\ref{conditions}) to be satisfied by the parameters become:
\begin{equation}
g_{1} > g_{2}u^{4}\;\;\;,\;\;\; g_{3} > g_{2}u^{2}
\end{equation}

The gauge fields $\mathbf{A},\;\mathbf{W}$ have  magnetic fields of the following form:
\begin{eqnarray*}
\mathbf{\nabla \times A}=\mathbf{B_{A}}= \frac{1}{\rho}\Bigg(\frac{\partial A_{\varphi}}{\partial \rho}\;\hat{z}-\frac{\partial A_{\varphi}}{\partial z}\;\hat{\rho}\Bigg) \\
\mathbf{\nabla \times W}=\mathbf{B_{W}}= -\Bigg(\frac{\partial W_{z}}{\partial \rho}-\frac{\partial W_{\rho}}{\partial z}\Bigg)\;\hat{\varphi}
\end{eqnarray*}

The {\bf field equations} follow:
\begin{eqnarray*}
\partial_{\rho}^{2}F+\partial_{z}^{2}F+\frac{\partial_{\rho}F}{\rho}-\frac{g_{1}}{2}\Big(F^{2}-1\Big)F-\frac{g_{3}}{2}P^{2}F-{}
 \nonumber\\
{}\Bigg[ \Bigg(qW_{\rho}-M\frac{zcos^{2}\Theta}{(\rho-a)^{2}}\Bigg)^{2}+\Bigg(qW_{z}+M\frac{cos^{2}\Theta}{(\rho-a)}\Bigg)^{2}\Bigg]F=0 \\
\partial_{\rho}^{2}P+\partial_{z}^{2}P+\frac{\partial_{\rho}P}{\rho}-\Big(eA_{\varphi}+N\Big)^{2}\frac{P}{\rho^{2}}-
\frac{g_{2}}{2}\Big(P^{2}-u^{2}\Big)P-\frac{g_{3}}{2}F^{2}P=0 \\
\partial_{\rho}^{2}A_{\varphi}+\partial_{z}^{2}A_{\varphi}-\frac{\partial_{\rho}A_{\varphi}}{\rho}
-2eP^{2}\Big(eA_{\varphi}+N\Big)=0 \\
\partial_{z}^{2}W_{\rho}-\partial_{z}\partial_{\rho}W_{z}-2qF^{2}\Bigg(qW_{\rho}-M\frac{zcos^{2}\Theta}{(\rho-a)^{2}}\Bigg)=0 \\
\partial_{\rho}^{2}W_{z}+\frac{1}{\rho}\Big(\partial_{\rho}W_{z}-\partial_{z}W_{\rho}\Big)-\partial_{\rho}\partial_{z}W_{\rho}
-2qF^{2}\Bigg(qW_{z}+M\frac{cos^{2}\Theta}{(\rho-a)}\Bigg)=0
\end{eqnarray*}

We can also write down the  currents associated with $\phi$ field, namely $j_{\rho}^{\phi}$ and $j_{z}^{\phi}$ and the total current $\mathcal{I}^{\phi}$
out of these as well as the supercurrent $\mathcal{I}^{\psi}$ associated with the $\psi$ field. These are
\begin{equation}
\mathcal{I}^{\phi}= \sqrt{(j_{\rho}^{\phi})^{2}+(j_{z}^{\phi})^{2}}\; ,\;\;\; \mathcal{I}^{\psi}= -\frac{2eP^{2}}{\rho}(eA_{\varphi}+N)
\end{equation}
where
\begin{equation}
j_{\rho}^{\phi}= -2qF^{2}(qW_{\rho}+M\partial_{\rho} \Theta), \;\;\;  j_{z}^{\phi}= -2qF^{2}(qW_{z}+M\partial_{z} \Theta)
\end{equation}

\begin{figure}
\centering
\includegraphics[scale=0.45 ,angle=270]{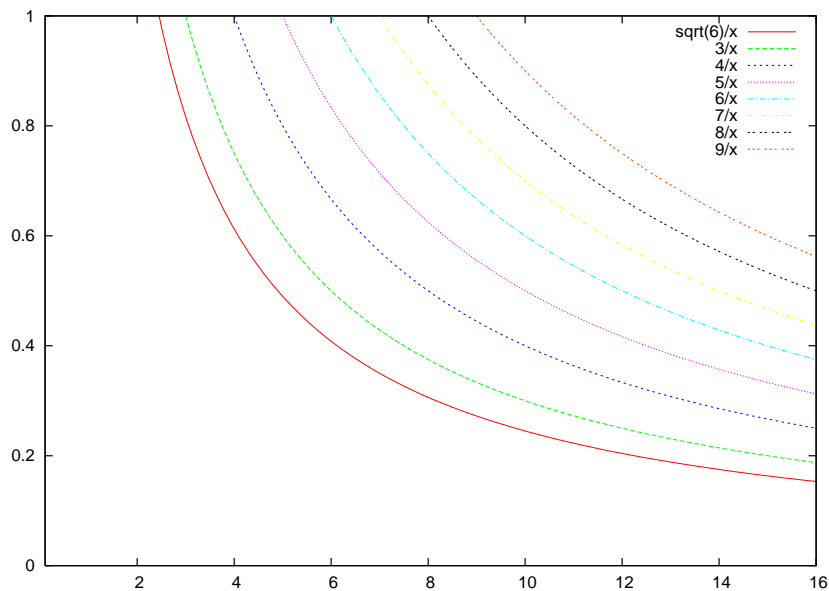}
\caption{\small {\bf $U(1)_{A}\times U(1)_{W}$ model:} For different values of $\sqrt{g_{1}}$, we plot the lower bound over which the condition (\ref{cd}) is satisfied.
The plot is $P_{max}$ vs. $e$ and we plot the function $P_{max}=\frac{\sqrt{g_{1}}}{e}$.\label{param} }
\end{figure}
As explained in the Introduction, Meissner effect is of crucial importance for the stability of the torus-like object. The magnetic fields
produced  by the supercurrent $\mathcal{I}^{\psi}$ penetrate into the toroidal defect in a distance dictated by the penetration depth.
In this theory, the $U(1)_{\mathbf{A}}$ symmetry breaks inside the string and the photon acquires mass $m_{\mathbf{A}}^{2}=e^{2}<P>^{2}$
where $<P>$ is the expectation value of the charge condensate in the vicinity of the string core. There is the superconducting sector of the defect.
The penetration depth is $\lambda=\frac{1}{m_{\mathbf{A}}}=\frac{1}{e<P>}$. But $<P>=P_{max}\leq u$ thus, one can have an estimate for $\lambda$, with
a lower bound for its value. This is $\lambda \geq \frac{1}{eu}$ where the equality holds when $P_{max}=u$.
On the other hand, one can also compute an upper bound for the thickness of the defect. We know that $r_{\phi}=\frac{1}{m_{\phi}}=\frac{1}{\sqrt{g_{1}}}$.
If our concern is to search for stable rings, a reasonable step is to demand the penetration depth to be smaller than the string thickness which means
\begin{equation}
\label{cd}
\lambda < r_{\phi} \Rightarrow \frac{1}{e P_{max}} < \frac{1}{\sqrt{g_{1}}} \Rightarrow e^{2} P_{max}^{2}> g_{1}
\end{equation}
This is the condition  needed in this case. From the above condition, we get the diagram shown in fig.\ref{param} where
one can see the area where it's more possible to find stable solutions if they exist. 
The numerical results we  present  later, are what we found while searching inside this region.

 Below, we analyze a way to derive some virial relations in order to check our results.
 The energy density for a {\em static solution}, which is of interest here, reads:
 \begin{equation}
 T^{00}=\frac{1}{4}F_{ij}F_{ij}+|\tilde{D}_{i}\phi|^{2}+\frac{1}{4}W_{ij}W_{ij}+|D_{i}\psi|^{2}+U(|\phi|,|\psi|)\equiv \varepsilon
 \end{equation}
where $F_{ij}= \partial_{i}A_{j}-\partial_{j}A_{i}$ and $W_{ij}=\partial_{i}W_{j}-\partial_{j}W_{i}$.
 For the solution of the system, we can write:
 \begin{equation}
 \frac{\delta E}{\delta \phi}=\frac{\delta E}{\delta \phi^{*}}=\frac{\delta E}{\delta W_{i}}=
\frac{\delta E}{\delta \psi}=\frac{\delta E}{\delta \psi^{*}}=\frac{\delta E}{\delta A_{i}}=0
 \end{equation}
 Define
 \begin{equation}
 f_{i}\equiv \frac{\delta E}{\delta \phi}\partial_{i}\phi+\frac{\delta E}{\delta \phi^{*}}\partial_{i}\phi^{*}+\frac{\delta E}{\delta W_{l}}\partial_{i}W_{l}+
   \frac{\delta E}{\delta \psi}\partial_{i}\psi+\frac{\delta E}{\delta \psi^{*}}\partial_{i}\psi^{*}+\frac{\delta E}{\delta A_{j}}\partial_{i}A_{j}=0
 \end{equation}
 which together with
 \begin{equation}
 \frac{\delta E}{\delta \Phi}=\frac{\partial T^{00}}{\partial \Phi}
 -\partial_{k}\frac{\partial T^{00}}{\partial (\partial_{k} \Phi)}
 \end{equation}
 can be written in a shorter form $f_{i}=\partial_{j} G_{ij}$ where
 \begin{eqnarray*}
 G_{ij}&=&\varepsilon \delta_{ij}-\frac{\partial \varepsilon}{\partial (\partial_{j}\phi)}\partial_{i} \phi
 -\frac{\partial \varepsilon}{\partial (\partial_{j} \phi^{*})}\partial_{i} \phi^{*} 
 -\frac{\partial \varepsilon}{\partial (\partial_{j} A_{k})}\partial_{i}A_{k}\\
 &-&\frac{\partial \varepsilon}{\partial (\partial_{j}\psi)}\partial_{i} \psi
 -\frac{\partial \varepsilon}{\partial (\partial_{j} \psi^{*})}\partial_{i} \psi^{*} 
 -\frac{\partial \varepsilon}{\partial (\partial_{j} W_{l})}\partial_{i}W_{l}
 \end{eqnarray*}
 This means that any static solution of the field equations satisfies
 \begin{equation}
 \label{eqq3}
 \partial_{j} G_{ij}=0
 \end{equation}
 But, we have 
 \begin{eqnarray*}
\frac{\partial \varepsilon}{\partial(\partial_{j}\phi)} &=& (\tilde{D}_{j}\phi)^{*} \\
\frac{\partial \varepsilon}{\partial(\partial_{j}\phi^{*})} &=& \tilde{D}_{j}\phi \\
\frac{\partial \varepsilon}{\partial(\partial_{j}A_{k})} &=& \partial_{j}A_{k} - \partial_{k}A_{j} = F_{jk}=\epsilon_{ijk}B_{A_{i}}\\
\frac{\partial \varepsilon}{\partial(\partial_{j}\psi)} &=& (D_{j}\psi)^{*} \\
\frac{\partial \varepsilon}{\partial(\partial_{j}\psi^{*})} &=& D_{j}\psi \\
\frac{\partial \varepsilon}{\partial(\partial_{j}W_{l})}&=& \partial_{j}W_{l} - \partial_{l}W_{j} = W_{jl}=\epsilon_{ijl}B_{W_{i}}
\end{eqnarray*}
 which leads to
\begin{eqnarray*}
G_{ij}&=& \varepsilon \delta_{ij}-(\tilde{D}_{j}\phi)^{*}\partial_{i}\phi-\tilde{D}_{j}\phi\partial_{i}\phi^{*}-W_{jl}\partial_{i}W_{l}- \\
&-&(D_{j}\psi)^{*}\partial_{i}\psi-D_{j}\psi\partial_{i}\psi^{*}-F_{jk}\partial_{i}A_{k}= \\
&=&\varepsilon \delta_{ij}-(\tilde{D}_{j}\phi)^{*}\tilde{D}_{i}\phi-\tilde{D}_{j}\phi(\tilde{D}_{i}\phi)^{*}+iqW_{i}\phi(\tilde{D}_{j}\phi)^{*}-\\
&-&iqW_{i}\phi^{*}\tilde{D}_{j}\phi-W_{jl}W_{il}-W_{jl}\partial_{l}W_{i}-\\
&-&(D_{j}\psi)^{*}D_{i}\psi-D_{j}\psi(D_{i}\psi)^{*}+ieA_{i}\psi(D_{j}\psi)^{*}-\\
&-&ieA_{i}\psi^{*}D_{j}\psi-F_{jk}F_{ik}-F_{jk}\partial_{k}A_{i}=\\
&=&\varepsilon \delta_{ij}-(\tilde{D}_{i}\phi)^{*}\tilde{D}_{j}\phi-(\tilde{D}_{j}\phi)^{*}\tilde{D}_{i}\phi-W_{il}W_{jl}-\partial_{l}(W_{jl}W_{i})\\
&+&W_{i}\Big( \partial_{l}W_{jl}-iq(\phi^{*}\tilde{D}_{j}\phi-(\tilde{D}_{j}\phi)^{*}\phi)\Big)-\\
&-&(D_{i}\psi)^{*}D_{j}\psi-(D_{j}\psi)^{*}D_{i}\psi-F_{jk}F_{ik}-\partial_{k}(F_{jk}A_{i})+ \\
&+&A_{i}\Big( \partial_{k}F_{jk}-ie(\psi^{*}D_{j}\psi-(D_{j}\psi)^{*}\psi)\Big)
\end{eqnarray*}
 where in the second equality we just add and subtract the necessary terms in order to have the terms only with covariant derivatives.
The latter happens because all the other terms are gone as we see below:
 \begin{eqnarray*}
 \partial_{j}\partial_{k}(\partial_{j}A_{k}-\partial_{k}A_{j})A_{i}=0&\Rightarrow& (\partial_{j}^{2}\partial_{k}A_{k}-\partial_{k}^{2}\partial_{j}A_{j})A_{i}=0 \\
 \partial_{j}\partial_{l}(\partial_{j}W_{l}-\partial_{l}W_{j})W_{i}=0&\Rightarrow& (\partial_{j}^{2}\partial_{l}W_{l}-\partial_{l}^{2}\partial_{j}W_{j})W_{i}=0 \\
 \end{eqnarray*}
and from the field equations, the following terms are also zero
 \begin{eqnarray*}
 \frac{\delta E}{\delta W_{j}}=0=\frac{\delta T^{00}}{\partial(\partial_{l} W_{j})}=
 -iq(\phi^{*}\tilde{D}_{j}\phi-\phi (\tilde{D}_{j}\phi)^{*})+\partial_{l}W_{jl}=0 \\
\frac{\delta E}{\delta A_{j}}=0=\frac{\delta T^{00}}{\partial(\partial_{k} A_{j})}=
 -ie(\psi^{*}D_{j}\psi-\psi (D_{j}\phi)^{*})+\partial_{k}F_{jk}=0
\end{eqnarray*}
 thus
 \begin{equation}
 G_{ij}=\varepsilon \delta_{ij} -W_{il}W_{jl}-(\tilde{D}_{i}\phi)^{*}\tilde{D}_{j}\phi-(\tilde{D}_{j}\phi)^{*}\tilde{D}_{i}\phi
             -F_{jk}F_{ik}-(D_{i}\psi)^{*}D_{j}\psi-(D_{j}\psi)^{*}D_{i}\psi
 \end{equation}
 but
 \begin{eqnarray*}
 F_{ik}F_{jk}&=& \epsilon_{nik}B_{A_{n}}\epsilon_{mjk}B_{A_{m}}={\mathbf B_{A}^{2}} \delta_{ij}- B_{A_{i}}B_{A_{j}}\\
 W_{il}W_{jl}&=& \epsilon_{pil}B_{W_{p}}\epsilon_{qjl}B_{W_{q}}={\mathbf B_{W}^{2}} \delta_{ij}- B_{W_{i}}B_{W_{j}}\\
 \end{eqnarray*}
 which means that
 \begin{eqnarray*}
  G_{ij}&=&\Bigg(-\frac{{\mathbf B_{A}^{2}}}{2}-\frac{{\mathbf B_{W}^{2}}}{2}+|\tilde{D}\phi|^{2}+|D\psi|^{2}+U\Bigg)\delta_{ij}+B_{W_{i}}B_{W_{j}}+B_{A_{i}}B_{A_{j}}-{}
                            \nonumber\\
 {}&-&(D_{i}\psi)^{*}D_{j}\psi-(D_{j}\psi)^{*}D_{i}\psi-(\tilde{D}_{i}\phi)^{*}\tilde{D}_{j}\phi-(\tilde{D}_{j}\phi)^{*}\tilde{D}_{i}\phi
 \end{eqnarray*}
 From (\ref{eqq3}) one obtains
 \begin{equation}
 \label{eqq1}
 \int d^{3}x G_{ik}=0=\int d^{3}x \partial_{j}(x_{k}G_{ij})=\int dS_{j}x_{k}G_{ij}
 \end{equation}
 The last equality follows from Gauss theorem, while the first is due to the fact that
 \begin{equation}
 \partial_{j}(x_{k}G_{ij})=\frac{\partial x_{k}}{\partial x_{j}}G_{ij} + x_{k}(\partial_{j}G_{ij})=\delta_{kj}G_{ij}=G_{ik}
 \end{equation}
 where we also used (\ref{eqq3}).
 Take the trace of (\ref{eqq1}). We have
\begin{equation}
 \int d^{3}x TrG_{ik}= 2\pi \int \rho d\rho dz\Bigg(-\frac{{\mathbf B_{A}^{2}}}{2}-\frac{{\mathbf B_{W}^{2}}}{2}+|\tilde{D}\phi |^{2}+|D\psi |^{2}+3U\Bigg)
\end{equation}
while the left hand side of (\ref{eqq1}) in the $\hat{\varphi}$ direction of integration of the toroidal object gives
\begin{equation}
2\pi \int dS_{\varphi} G_{\varphi \varphi}=0
\end{equation}
and (\ref{eqq1}) finally ends up to the following:
\begin{equation}
\label{viraplo}
2 \pi \int \rho d\rho dz \Bigg( \frac{1}{2}\Big( B_{W}^{2}+B_{A}^{2}\Big) - U\Bigg)=0
\end{equation}

%====================================================================================================================================

Another way to derive {\bf virial} relations is through Derrick's scaling argument.
The virial relation for the field $\phi$ of our model, must have the constraint $\phi_{\kappa}(\rho=a, z)=\phi (\rho=a, \kappa z) =0$.
Consider the rescalings $\rho\rightarrow \rho$, $z\rightarrow \kappa z$, 
$F_{\kappa}\rightarrow F$, $P_{\kappa}\rightarrow P$, $A_{\varphi_{\kappa}} \rightarrow A_{\varphi}$,
 $W_{\rho,z_{\kappa}} \rightarrow \kappa W_{\rho,z}$. Then, we find the minimum through the relation $\frac{\partial E}{\partial \kappa} =0$ when $\kappa =1 $.
 The virial relation for our model, which is a way to check our numerical results is:
 \begin{eqnarray*}
I_{1}&=& 2\pi v_{1} \int_{0}^{\infty} \rho d\rho \int_{-\infty}^{\infty} dz \Bigg[\frac{1}{2}(\partial_{\rho} W_{z} - \partial_{z} W_{\rho})^{2}+
             \frac{1}{2\rho^{2}} (\partial_{z} A_{\varphi})^{2}+(\partial_{z} P)^{2}+{}
          \nonumber\\
        {}&+&(\partial_{z} F)^{2}+2F^{2}\Bigg( qW_{\rho}(qW_{\rho}+M\partial_{\rho}\Theta )+(qW_{z}+M\partial_{z} \Theta)^{2}\Bigg)
\Bigg] \\
I_{2}&=&  -2\pi v_{1} \int_{0}^{\infty} \rho d\rho \int_{-\infty}^{\infty} dz \Bigg[\frac{1}{2\rho^{2}}(\partial_{\rho}A_{\varphi})^{2}+(\partial_{\rho} F)^{2}+(\partial_{\rho} P)^{2}+{}
        \nonumber\\
{}&+&\frac{P^{2}}{\rho^{2}}(eA_{\varphi}+N)^{2}+(\partial_{z}W_{\rho})(\partial_{\rho}W_{z}- \partial_{z}W_{\rho})+{}
                       \nonumber\\
{}&+&F^{2}\Bigg( (qW_{\rho}+M\partial_{\rho} \Theta)^{2} + (qW_{z}+M\partial_{z} \Theta)^{2}\Bigg)+{}
	                                                                                 \nonumber\\
     {}&+&\frac{g_{1}}{4}(F^{2}-1)^{2} + \frac{g_{2}}{4} (P^{2} -u^{2})^{2} +\frac{g_{3}}{2} F^{2}P^{2}-\frac{g_{2}}{4}u^{4}\Bigg] 
 \end{eqnarray*}	
and we must have $I_{1}+I_{2}=0$. We define the index $V=\frac{||I_{1}|-|I_{2}||}{|I_{1}|+|I_{2}|}$ and we want its value to be as small as possible.
We can derive many other virial relations by assuming generally for a field $\phi$, the ``double'' rescaling $\phi (\vec{x}) \rightarrow  \kappa \phi(\mu \vec{x})$
and then demanding $\frac{\partial E}{\partial \kappa}|_{\kappa = 1 = \mu} =0 =\frac{\partial E}{\partial \mu}|_{\kappa = 1 = \mu}$. For example,
consider the following rescalings $\rho\rightarrow \rho $, $z\rightarrow \mu z$, $F_{\kappa}\rightarrow F$, $P_{\kappa}\rightarrow \kappa P$,
$A_{\varphi_{\kappa}} \rightarrow A_{\varphi}$, $W_{\rho ,z_{\kappa}} \rightarrow  W_{\rho,z}$. 
We have
\begin{eqnarray*}
I_{3}&=& 2\pi v_{1} \int_{0}^{\infty} \rho d\rho \int_{-\infty}^{\infty} dz \Bigg[ \frac{1}{2\rho^{2}} (\partial_{z} A_{\varphi})^{2}+(\partial_{z} P)^{2}+
               (\partial_{z} F)^{2}+{}
			                                          \nonumber\\
     {}&+&2F^{2}\Bigg( M\partial_{z} \Theta (qW_{z} +M\partial_{z} \Theta )\Bigg)+\frac{1}{2}(\partial_{\rho}W_{z}-\partial_{z}W_{\rho})^{2}\Bigg]  \\
I_{4}&=& -2\pi v_{1} \int_{0}^{\infty} \rho d\rho \int_{-\infty}^{\infty} dz \Bigg[\frac{1}{2\rho^{2}} (\partial_{\rho} A_{\varphi})^{2}+(\partial_{\rho} P)^{2}
                                   +(\partial_{\rho} F)^{2}+{}
	                                                                   \nonumber\\
        {}&+&\frac{P^{2}}{\rho^{2}}(eA_{\varphi} +N)^{2}+\partial_{\rho}W_{z}(\partial_{\rho}W_{z}-\partial_{z}W_{\rho})+{}
        \nonumber\\
        {}&+&F^{2}\Bigg( (qW_{\rho}+M\partial_{\rho}\Theta)^{2}+(qW_{z}+M\partial_{z}\Theta)^{2}\Bigg) +{}
		\nonumber\\
        {}&+&\frac{g_{1}}{4}(F^{2}-1)^{2} + \frac{g_{2}}{4} (P^{2} -u^{2})^{2} +\frac{g_{3}}{2} F^{2}P^{2}-\frac{g_{2}}{4}u^{4}\Bigg] 
\end{eqnarray*}
together with
\begin{eqnarray*}
I_{5}&=& 4\pi v_{1} \int_{0}^{\infty} \rho d\rho \int_{-\infty}^{\infty} dz \Bigg[ (\partial_{\rho} P)^{2}+ (\partial_{z} P)^{2}
                  +\frac{P^{2}}{\rho^{2}} (e A_{\varphi} +N)^{2} \Bigg]  \\
I_{6}&=& 2\pi v_{1} \int_{0}^{\infty} \rho d\rho \int_{-\infty}^{\infty} dz \Bigg[g_{2}\Big( P^{2}-u^{2}\Big) P^{2} + g_{3}F^{2}P^{2}\Bigg]
\end{eqnarray*}
where, as above, we must have $I_{3}+I_{4}=0=I_{5}+I_{6}$.

\section{{\normalsize The $U(1)_{A}\times U(1)_{W}$ model: Numerical results}}

A standard minimization algorithm is used to minimize the energy functional of (\ref{funcu1}). The algorithm is 
written in C. One can find details about the algorithm on page 425 of \cite{c6} but, briefly, the basic idea is this:
Given an appropriate initial guess, there are several corrections to it, having as a criterion the minimization of the energy
in every step. When the corrections on the value of the energy are smaller than $\sim 10^{-8}$ the program stops and we get the final results.
A 90$\times$20 grid for every of the five functions is used, that is, 90 points on $\rho$-axis and 20 on $z$.
Here, we begin with fixed torus radius $a$. Then, the configuration with minimum energy for this $a$ is found. Other values of $a$
are chosen as well and the same process goes on until we plot the energy vs. the torus radius $E(a)$. It would be very
interesting to find a non-trivial minimum of the energy (in $a_{min}\neq 0$), which would correspond to stable toroidal defects with radius $a_{min}$.

\begin{figure}
\vspace{0.4 in.}
\centering
\includegraphics[scale=0.45]{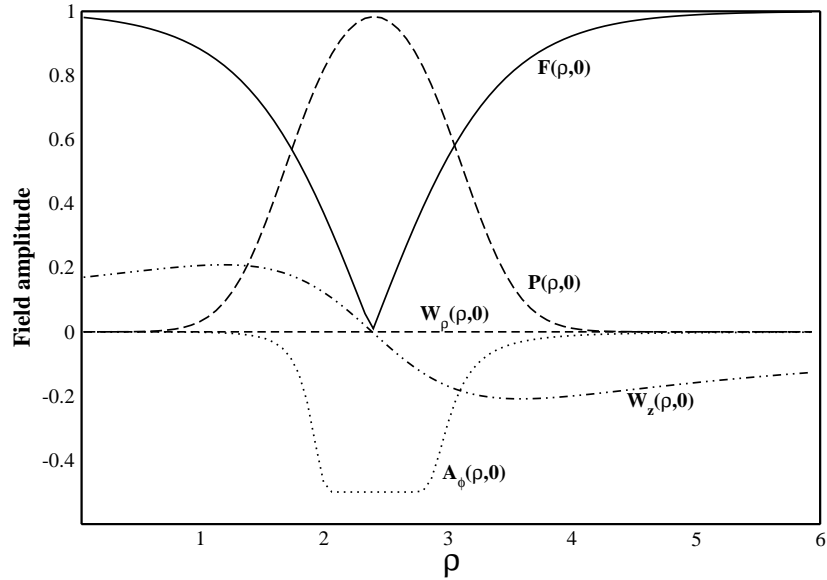}
\caption{\small {\bf $U(1)_{A}\times U(1)_{W}$ model:} A typical plot of the initial guess we use for the five fields, for the lowest winding state $M=1, N=1$
on $z=0$ plane.\label{ig}}
\end{figure}
\begin{figure}
%\vspace{0.4 in.}
\centering
\includegraphics[scale=1.1]{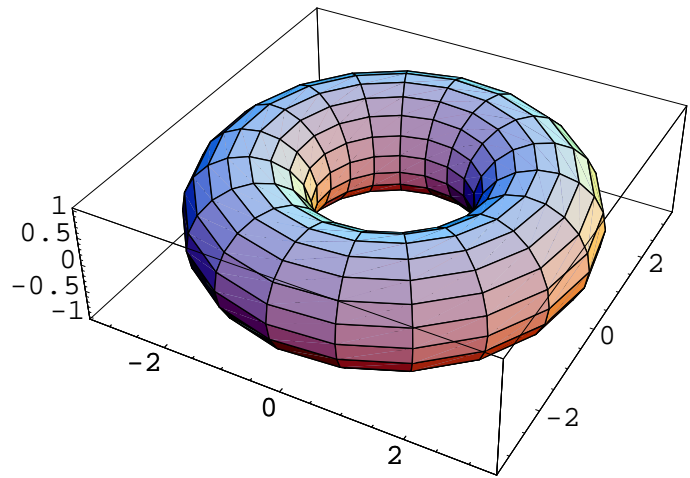}
\caption{\small {\bf $U(1)_{A}\times U(1)_{W}$ model:} This is a typical picture of the toroidal soliton we are after. \label{igt}}
\end{figure}
\begin{figure}
%\vspace{0.4 in.}
\centering
\includegraphics[scale=1.1]{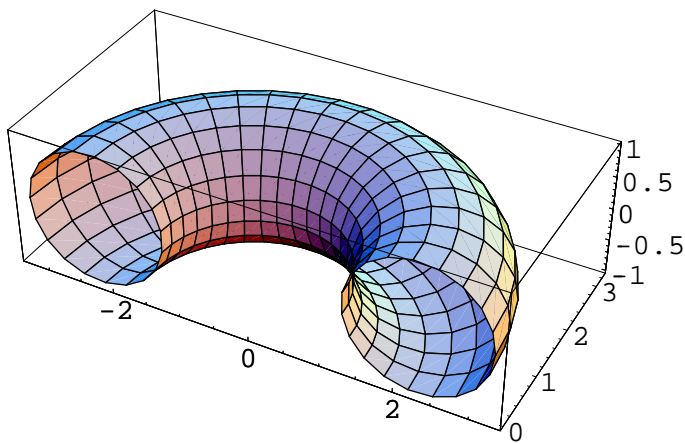}
\caption{\small {\bf $U(1)_{A}\times U(1)_{W}$ model:} The profile of the torus has two identical cycles. The initial configuration of fields 
(plotted in fig.\ref{ig}) is the same in both of them thus, we choose to plot only the right side.  \label{igt2}}
\end{figure}

The {\bf initial guess} (figure \ref{ig}) we use for our computation is:
\begin{eqnarray*}
F(\rho,z)&=& \tanh((\rho-a)^{2}+z^{2})^{M/2} \\
P(\rho,z)&=& \tanh(\rho^{N}) (1-\tanh((\rho -a)^{2}+z^{2}) \\
A_{\varphi}(\rho,z)&=& -\frac{N}{e}\tanh\Bigg(\frac{\xi \rho^{2}}{((\rho-a)^{2}+z^{2})^{2}}\Bigg) \\
W_{\rho}(\rho,z)&=&\frac{Mz\cos^{2}\Theta}{q(\rho-a)^{2}}\Bigg(\frac{(\rho-a)^{2}+z^{2}}{(\rho-a)^{2}+z^{2}+(a^{2}/4)}\Bigg)^{2} \\
W_{z}(\rho,z)&=&-\frac{M\cos^{2}\Theta}{q(\rho-a)}\Bigg(\frac{(\rho-a)^{2}+z^{2}}{(\rho-a)^{2}+z^{2}+(a^{2}/4)}\Bigg)
\end{eqnarray*}
where $\xi$ a constant. Figures \ref{igt}, \ref{igt2} indicate what exactly we plot in fig.\ref{ig} and where.
This initial guess also satisfies the appropriate asymptotics
\begin{itemize}
\item{near $\rho =0$: 
\begin{equation}
F\neq 0 , \;\; P\sim \rho^{N}, \;\; A_{\varphi} \sim \rho^{2}f(z)
\end{equation} }
\item{near $(\rho =a , z=0)$: 
\begin{equation}
F\sim \tilde{\rho}^{M/2}, \;\; W_{\rho}=0=W_{z}
\end{equation} }
\item{at infinity: 
\begin{eqnarray*}
F&\sim& 1-\mathcal{O}(e^{-\sqrt{\tilde{\rho}}}),\;\;\; P\sim \mathcal{O}(e^{-\sqrt{\rho^{2} +z^{2}}}) {}
  \nonumber\\
{}W_{\rho} &\sim& -\frac{M}{q}\partial_{\rho}\Theta|_{\infty} +\mathcal{O}(e^{-\sqrt{\tilde{\rho}}}),\;\;\; 
W_{z} \sim -\frac{M}{q}\partial_{z}\Theta|_{\infty} +\mathcal{O}(e^{-\sqrt{\tilde{\rho}}})
\end{eqnarray*} }
\end{itemize}
where $\tilde{\rho}\equiv (\rho -a)^{2}+z^{2}$.

\begin{figure}[t]
\vspace{0.4 in.}
\centering
\includegraphics[scale=0.45]{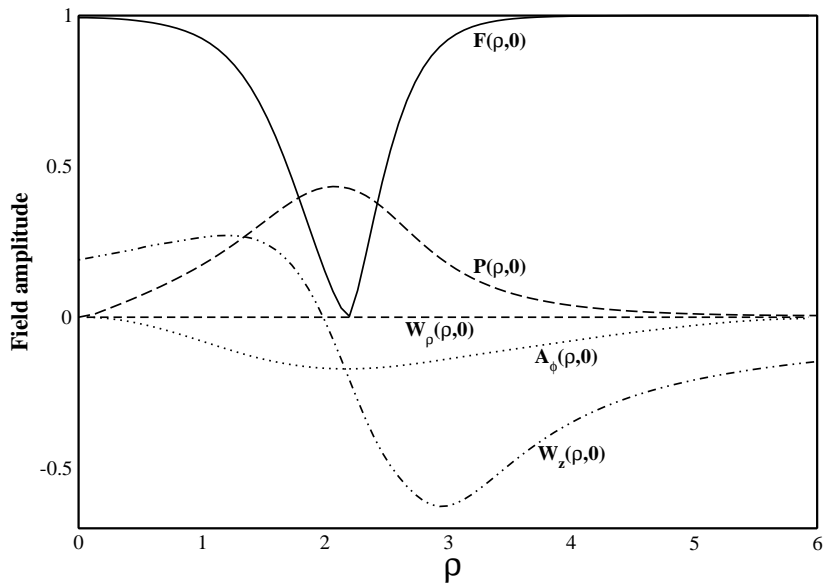}
\caption{\small {\bf $U(1)_{A}\times U(1)_{W}$ model:} Typical plot of the final configuration of fields. Parameters are
$M=1$, $N=1$, $e=5$, $q=m_{W}=2$, $g_{1}=14$, $m_{\phi}=\sqrt{g_{1}}=3.74$, $g_{2}=12$, $g_{3}=14$, $u=1$, 
$m_{\psi}= \sqrt{(g_{3}-g_{2}u^{2})/2}= 1$, $v_{1}=7.5\cdot 10^{-3}$. Energy $E=0.72$ and virial is $1.5\cdot 10^{-3}$.  \label{p1}}
\end{figure}
For fixed torus radius (i.e. here $a\approx 2.2$) we present a typical graph of the final configuration of the
lowest energy (see fig.\ref{p1}). 
We also present the plot of the energy of the system vs. the radius
of the toroidal object which reveals the instability of the system as there is no non-trivial minimum (see fig.\ref{enercond}).

From the results, we can point out a few things. Firstly, the greater the value of $e$ we use, the stronger the supercurrent becomes.
Secondly,  the greater the value of $e$ we use, the lower the radius $a$ where the supercurrent quenches (fig.\ref{jepe}). 
These are expected as the increase of $e$ makes the condition of equation (\ref{cd})
stronger, something which means that the mass of the photon increases and the penetration depth
decreases at the same time. It is also reasonable that a stronger current can ``defend''  the defect, against the magnetic field,
a little longer.

Another observation is that as  $g_{2}$ increases and becomes close to $g_{1}$,
we see that the maximum current increases and the quenching takes place again at lower $a$ (fig.\ref{jepg2}). This is expected as one can see from the potential
in equation (\ref{potu1}) of the energy functional, because the stronger the coupling $g_{2}$ is, the more important the relevant term $g_{2}(P^{2}-u^{2})^{2}/4$ becomes.
The latter has as a consequence, the increase of $P_{max}$  which counteracts the effects from the increasing $g_{2}$ coupling.

The parameter space where we searched, starts from $g_{1} =4 $ ($m_{\phi} =2$). 
In order to search the model, we reached values around $g_{1} = 30$ ($m_{\phi}\approx 5.48$)
over which, $e$ has to be very large in order to respect the condition (\ref{cd}). There is also the
fact that great values of $e$ in general is an unwanted feature since we use a semiclassical approach.
As it concerns the other couplings we have $g_{3}= g_{1}$ and  $g_{2}=g_{1}-k$ with $0.5\leq k \leq 8$, ($u=1$).

\section{{\normalsize Explanation concerning the instability of the vortex ring}}

From the  condition  of equation (\ref{cd}), it is understood that we are enforced to  lower $g_{1}$ as much as possible and/or  increase $e$. But
these steps are not as easy as they might seem. There are some limitations.
The lower bound on the value of $g_{1}$ has
a reasonable explanation. For low values of $g_{1}$, the coupling $g_{2}$ is also low (because $g_{1}>g_{2}u^{4}$).
Now,  when $g_{2}$ is small enough, the changes on the term $g_{2}(P^{2}-u^{2})^{2}/4$ are unimportant for the energy, in comparison
to the term $(\partial_{\rho}P)^{2}$. In that case, the lowering of the last term minimizes the energy, something which means that
$P \rightarrow 0$. Indeed, this is numerically observed.
There is also a lower limit on $e$ which is reasonable because the lowering of $e$ results to an increasing
penetration depth.

\begin{figure}[t]
\centering
\includegraphics[scale=0.45]{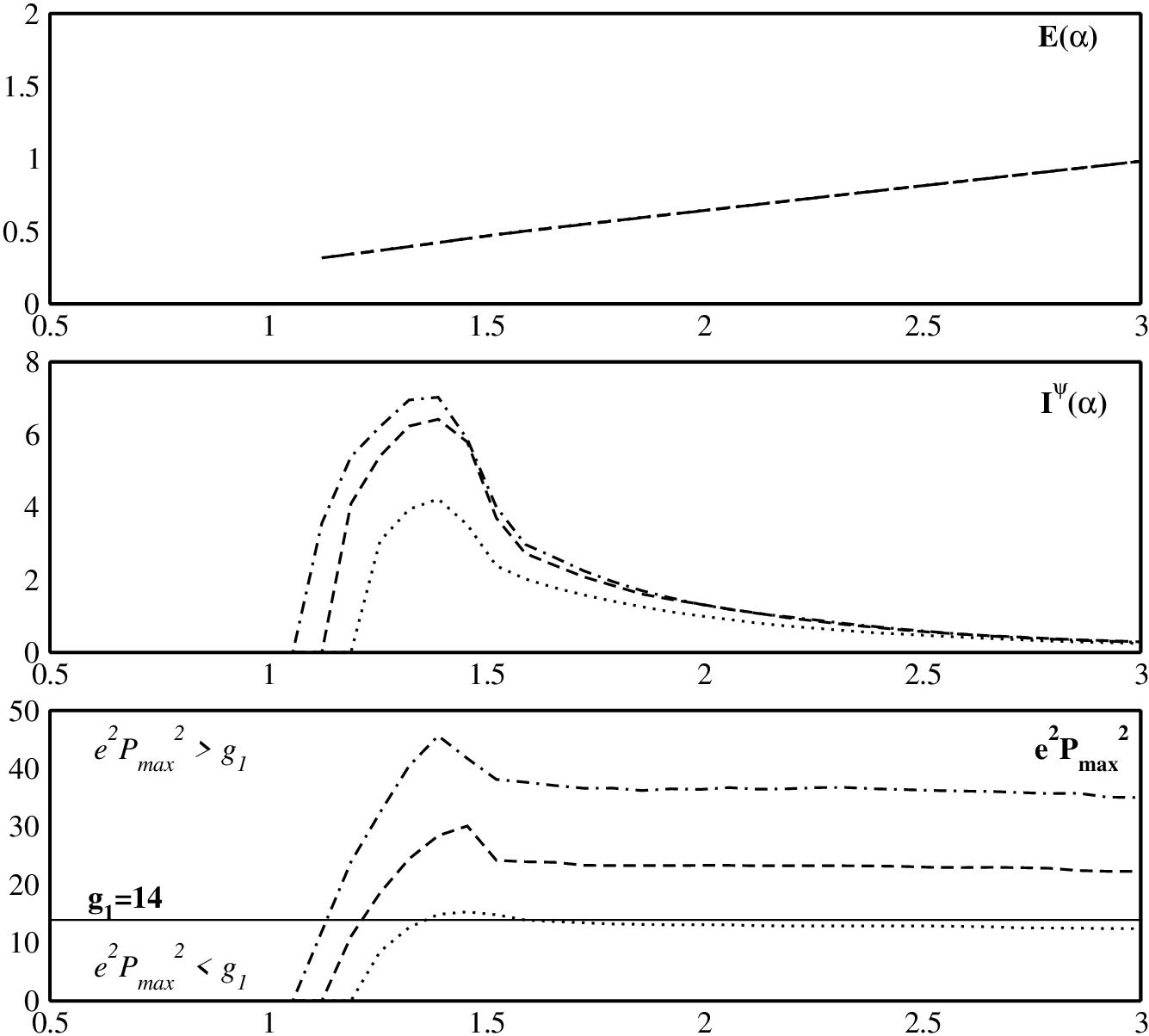}
\caption{\small {\bf $U(1)_{A}\times U(1)_{W}$ model:} The top graph is the energy vs. the torus radius $E(a)$. The middle graph
is the supercurrent $I^{\psi}$ vs. $a$. In that graph one can clearly see current quenching. The bottom graph is the
quantity $e^{2}P_{max}^{2}$ vs. $a$ (or $m_{\mathbf{A}}$ vs. $a$) where one can see the area in which the condition of eq.\ref{cd} holds
(lines over the $g_{1}$-limit line). The resistance from the magnetic field can be seen as a  sharp increase on the supercurrent.
Dotted lines are for $e=6$, dashed for $e=8$ while dashed and dotted for $e=10$. All plots are for $(M,N,u,g_{1},g_{2},g_{3})$=$(1,1,1,14,12.5,14)$.\label{jepe}}
\end{figure}
\begin{figure}[t]
\centering
\includegraphics[scale=0.45]{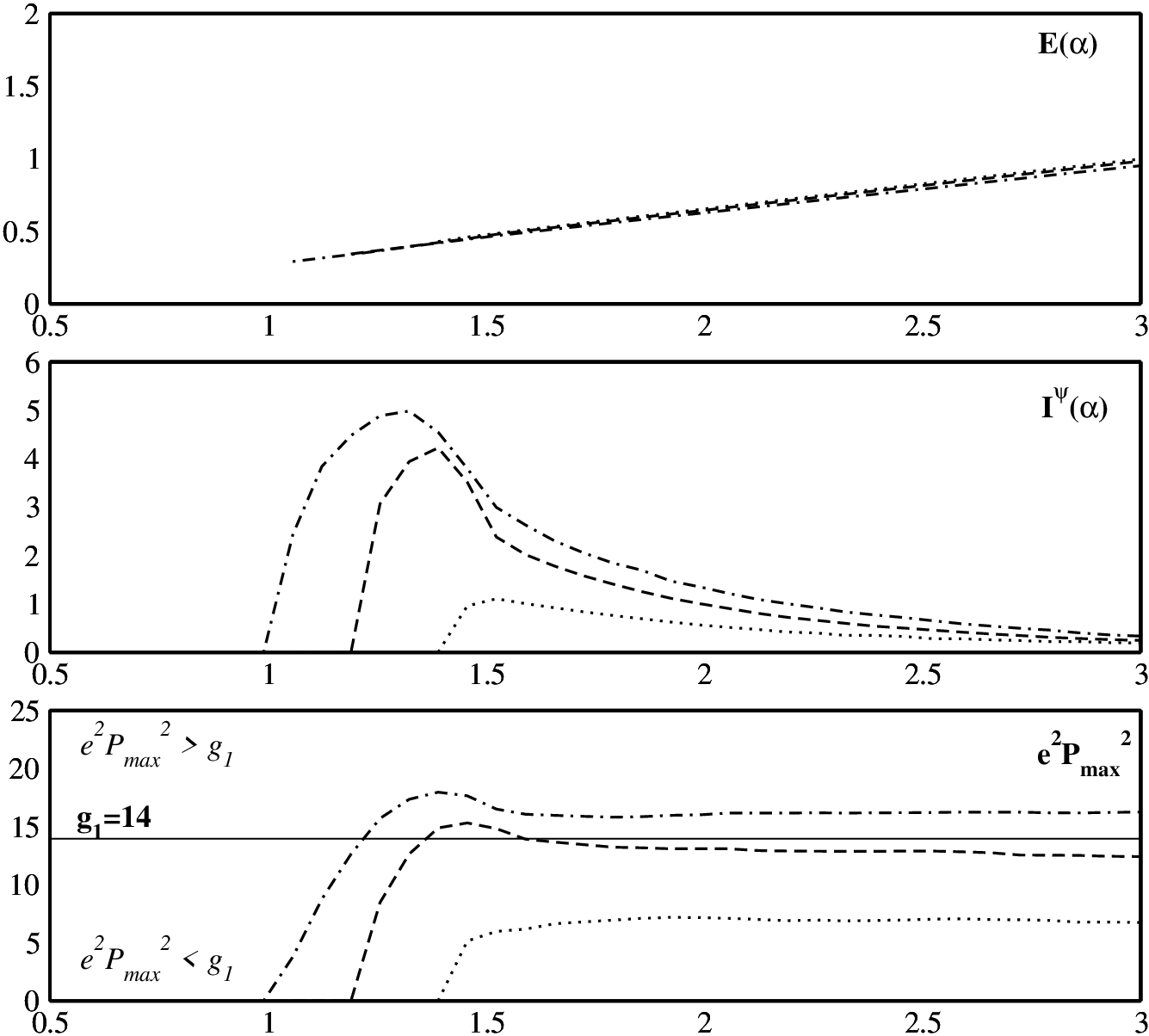}
\caption{\small {\bf $U(1)_{A}\times U(1)_{W}$ model:} The top graph is the energy vs. the torus radius $E(a)$. The middle graph
is the supercurrent $I^{\psi}$ vs. $a$. In that graph one can clearly see current quenching. The bottom graph is the
quantity $e^{2}P_{max}^{2}$ vs. $a$ (or $m_{\mathbf{A}}$ vs. $a$) where one can see the area in which the condition of eq.\ref{cd}  holds
(lines over the $g_{1}$-limit line). The resistance from the magnetic field can be seen as a  sharp increase on the supercurrent. 
Dotted lines are for $g_{2}=12$, dashed for $g_{2}=12.5$ while dashed and dotted for $g_{2}=13$. All plots are for $(M,N,u,e,g_{1},g_{3})$=$(1,1,1,6,14,14)$.\label{jepg2}}
\end{figure}
\begin{figure}[t]
%\vspace{1 in.}
\vspace{0.4 in.}
\centering
\includegraphics[scale=0.45]{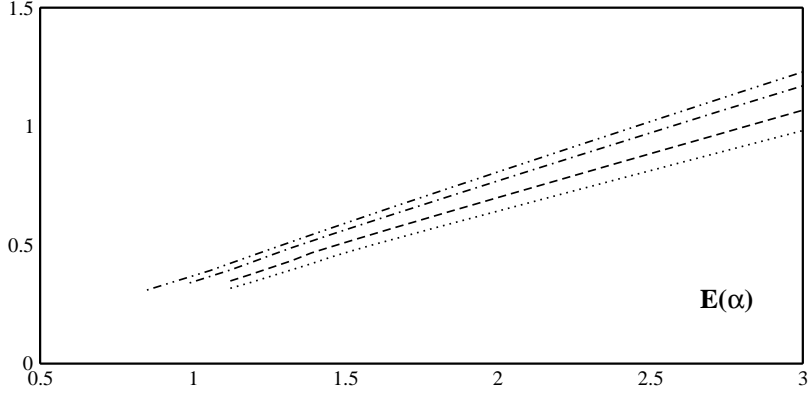}
\caption{\small {\bf $U(1)_{A}\times U(1)_{W}$ model:} The plot of the energy of the system vs. the radius of the torus for four
different sets of parameters. Dotted is for $(g_{1}, g_{2}, g_{3})$= $(14,12.5,14)$, dashed is for $(18,15,18)$, dashed and dotted is for $(25,20,25)$
while dashed with two dots is for $(30,24,30)$. For all data sets we have $(e,M,N,u)$=$(10,1,1,1)$.
As one can observe, there exists no minimum.\label{enercond}}
\end{figure}

We searched for stable objects for values over these  limits described above. The numerical results are exhibited
in figs.\ref{jepe}-\ref{enercond} and as we see, these objects are unstable. We argue that the explanation for the instability is
{\bf current quenching} and that, only for high values of $e$. For lower values of $e$, that is, of the order of $1$, we have,
according to equation (\ref{cd}), that the penetration depth is much bigger than the string thickness thus, stability is out of the question.
The latter is also numerically observed.

Now, we base our aspect about quenching on qualitative as well as quantitative arguments. 
We observe that as the torus shrinks, the supercurrent suddenly drops to zero which signals
the destruction of the defect. Just before the sharp drop, we notice that the supercurrent rises with increasing rate. This must be due to the resistance the torus meets
from the magnetic lines as it shrinks. 
One can observe that as the supercurrent increases and the condition of equation (\ref{cd}) is satisfied at the same time (i.e. see dashed and dotted line of fig.\ref{jepe}),
suddendly the current is lost. This can be explained only through current quenching.
The above phenomenon is not observed when equation (\ref{cd}) is not satisfied (i.e. see dotted line of
fig.\ref{jepg2}). There, as the magnetic lines penetrate the ring, they meet almost no resistance since $\lambda$ is much greater than $r_{\phi}$.

Another observation which supports our quenching argument is that, rough estimations on the maximum current a string can sustain, lead us to the
following formula (see page 129 of \cite{c8} or Appendix I at the end of the chapter) which makes a small overestimation in order to have an upper limit:
\begin{equation}
\label{maxj}
\mathcal{I}^{\psi}_{max} < \sqrt{\sigma} e u
\end{equation}

where $\sigma \equiv \int\int d\rho dz\; P^{2}$. In figs.\ref{jepe}-\ref{jepg2}, the maximum value of the supercurrent is close to the limit
of the estimation of equation (\ref{maxj}). The table below gathers the estimated $\mathcal{I}^{\psi}_{est.}$ (according to equation (\ref{maxj}))
and the computed maximum supercurrent ($\mathcal{I}^{\psi}_{com.}$) for the parameters of these figures.
\newline
\begin{center}
\begin{tabular}{|l|l|l|l||l||l||}
\hline
$e$ & $g_{1}$ & $g_{2}$ & $g_{3}$ &  $\mathcal{I}^{\psi}_{est.}$ & $\mathcal{I}^{\psi}_{com.}$ \\
\hline
\hline
6 & 14 & 12.5  & 14 & 4.6  & 4.3    \\
\hline
6 & 14 & 13    & 14 & 5.2  & 5.0    \\
\hline
8 & 14 & 12.5  & 14 & 6.7  & 6.4    \\
\hline
10 & 14 & 12.5 & 14 & 8.0  & 7.2    \\
\hline
\end{tabular}
\end{center}
%\newline
%\newline
\vspace{0.2 in.}

Finally, one can make an estimation of the value of the supercurrent which would stabilize the ring, namely $\mathcal{I}^{\psi}_{stab.}$.
This can be done as follows. As explained in the introduction, there is the tension of the string which shrinks the loop and
the magnetic field which opposes this tendency. When  the ring is stabilized we have $E_{tension}=E_{magnetic}$. Here, $E_{tension}\sim 1$
and $E_{magnetic}=2\pi v_{1}\int \rho d\rho \int dz (B_{A}^{2}/2)$. Without any calculation, one can point out that, since
the {\em total} energy in the quenching radius is below $E_{tension}=1$, then the $E_{magnetic}$ which is a fraction of it, would
be even smaller. Recall that $B_{A}\propto \mathcal{I}^{\psi}$, which means that we need a $\mathcal{I}^{\psi}_{stab.}$ which is well above
the maximum current we can have inside the defect. Calculations of the magnetic energy are in agreement with the above observation and
place its value around $E_{magnetic}\sim 5\cdot 10^{-3} << E_{tension} \sim 1$. This translates to the following conclusion:
$\mathcal{I}^{\psi}_{stab.} \gtrsim 10\cdot \mathcal{I}^{\psi}_{est.}$.

Thus, we find out that the current needed for stabilization, is at least ten times bigger than the value of the maximum current our
string can sustain. We also observe that our numerical maximum current values are close to the theoretical estimations about quenching.
This means, that we will have current quenching as an ``obstacle'' towards stabilization, since the supercurrent will not be enough
in order to create the magnetic field needed.

\newpage

\begin{figure}
\vspace{0.4 in.}
\centering
\includegraphics[scale=0.45]{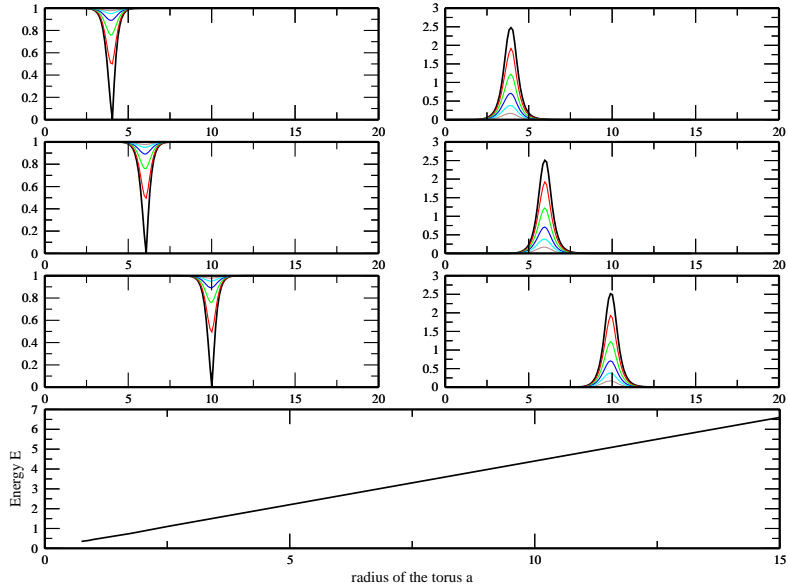}
\caption{\small {\bf $U(1)_{W}$ model:} Profiles for the scalar field $F(\rho,z)$ on the left and for the magnetic field $B_{W}$ on the right side, 
for three random values of $a$. On the bottom graph there is the energy vs. radius of the torus $a$. Parameters are $(M, m_{\phi}, m_{W}, v_{1})$=$(1, 4.47, 2, 0.01)$.
Black solid line is for $z=0$.\label{fw} }
\end{figure}
\section{{\normalsize $U(1)_{W}$ model:  Vortex ring without supercurrent}}

Here, we present a simpler model of a string of toroidal geometry but without the $\psi$ scalar field
responsible for the existence of supercurrent.
In this model, we only have the scalar field $\phi(\rho,z)$ together with its gauge field $W(\rho,z)$. The presence of the latter 
gives us the freedom to choose $\Theta (\rho,z)=\arctan (z/(\rho-a))$, as we did before.
The {\bf initial guess} we use for the fields is the following:
\begin{eqnarray*}
F(\rho,z)&=& \tanh((\rho-a)^{2}+z^{2})^{M/2} \\
W_{\rho}(\rho,z)&=&\frac{Mz\cos^{2}\Theta}{q(\rho-a)^{2}}\Bigg(\frac{(\rho-a)^{2}+z^{2}}{(\rho-a)^{2}+z^{2}+(a^{2}/4)}\Bigg)^{2} \\
W_{z}(\rho,z)&=&-\frac{M\cos^{2}\Theta}{q(\rho-a)}\Bigg(\frac{(\rho-a)^{2}+z^{2}}{(\rho-a)^{2}+z^{2}+(a^{2}/4)}\Bigg)
\end{eqnarray*}

As there is no current flowing inside the string there will be no Meissner effect which means that there is no
reason to prevent the vortex ring from collapsing. Thus, we expect to find no non-trivial minimum on the energy of the system,
something which signals the instability of that object (fig.\ref{fw}).

\section{{\normalsize Conclusions}}

Future experiments in LHC could answer whether metastable particle-like solitons exist in MSSM or 2HSM {\em or} not. 
In \cite{c3} there is a search for spherically symmetric solitons in the frame of the 2HSM  with a simplified potential.
Here we search for axially symmetric solitons which, if stable, will have a mass of the order of TeV \cite{c3}.
We considered the $U(1)_{A}\times U(1)_{W}$ model, where the existence of the vortex is ensured, 
for topological reasons. There, we searched for stable toroidal strings. We present and analyze our observations.
This chapter tries to answer to expectations having to do with observations of stable axially symmetric solitons which would be possible to detect in later experiments 
of LHC. For relatively small values of $e$ ($\sim 1$), which are of interest in that case, the system
seems to have  no stable vortex rings. In fact, this instability is present even in other parameter areas where we searched (i.e. $e\geq 6$ see figs.\ref{jepe},\ref{jepg2}).
We explain why we believe that the main reason of instability is current quenching.

\newpage

%===========================APPENDIX COORDINATE SYSTEMS=====================================
%===========================APPENDIX COORDINATE SYSTEMS=====================================
%===========================APPENDIX COORDINATE SYSTEMS=====================================

\section{{\normalsize APPENDIX}}

\subsection{{\normalsize Part I: Derivation of the formula  $\mathcal{I}^{\psi}_{max} \leq \sqrt{\sigma} e u$}}

To derive the above formula, we follow the steps of \cite{c8} (pages $129$-$130$).
The current density is given by
\begin{equation}
j^{\mu}= ie(\overline{\psi} D^{\mu}\psi - \psi \overline{D^{\mu}\psi})
\end{equation}
where $\psi = P(\rho)e^{i\Phi (\varphi)}$ and the covariant derivative $D^{\mu}= \partial^{\mu}+ieA^{\mu}$.
Substituting above, the absolute value of the current density is $j^{\varphi}= 2eP^{2}D^{\varphi}\Phi$
and the total supercurrent is the integral of the cross section of the current density
\begin{equation}
\mathcal{I}= \int_{0}^{\infty} d\rho \int_{-\infty}^{\infty} dz \;\; j^{\varphi}
\end{equation}
from which we get $\mathcal{I}=2\sigma e D^{\varphi}\Phi$, where we 
denote $\int\int d\rho dz P^{2}$ by $\sigma$.

The term $g_{2}(P^{2}-u^{2})^{2}/4$ of the potential, enforces the expectation value of $P$ to reach $u$.
On the other hand, in the energy functional there is the term $P^{2}(D^{\varphi}\Phi)^{2}$ 
which tries to ``counteract'' the above potential term.
The latter means that $(D^{\varphi}\Phi)^{2}$ acts as a ``negative mass'' squared.
Because of that term, the expectation value of $P$, say $P_{0}$, is not $u$ but
\begin{equation}
P_{0}^{2}\approx u^{2}-\frac{2}{g_{2}}(D^{\varphi}\Phi)^{2}
\end{equation}
The above decrease has an effect on the supercurrent as well. It decreases its value thus, 
\begin{equation}
\mathcal{I}= 2\sigma e (D^{\varphi}\Phi)\Bigg[ 1- 2\sigma \frac{(D^{\varphi}\Phi)^{2}}{u^{2}} \Bigg]
\end{equation}

In the above equation, only $D^{\varphi}\Phi$ has dimensions (of mass) and we rename it as $X$.
We want to find the maximum value of the current:
\begin{equation}
\frac{d\mathcal{I}}{dX}=0\rightarrow X\approx \frac{u}{\sqrt{\sigma}}
\end{equation}
and by substituting, we find out that the maximum value of the supercurrent is
\begin{equation}
\mathcal{I}_{max}\leq \sqrt{\sigma} eu
\end{equation}

\subsection{{\normalsize Part II: Toroidal coordinate system}}

The toroidal coordinates $(\eta, \xi, \varphi)$ are related to the cartesian $(x, y, z)$ through the
following definition:
\begin{equation}
x=\frac{a}{D}\sinh\eta \cos\varphi, \; y=\frac{a}{D}\sinh\eta \sin\varphi, \; z=\frac{a}{D}\sin\xi
\end{equation}
with $a$ a scale factor of the torus, $D=\cosh\eta - \cos\xi$ and $0\leq \eta <\infty $,
$0\leq \xi \leq 2\pi $, $0\leq \varphi \leq 2\pi $. 

In order to find the field equations, we need to relate the derivatives of cartesian coordinates with
those of toroidal coordinates, thus the following formulas can be very useful:

 \begin{eqnarray*}
\frac{\partial }{\partial x } = - \frac{\cos\varphi (\cosh\eta \cos\xi -1)}{a} \frac{\partial }{\partial \eta}
-\frac{\sin\xi \sinh\eta \cos\varphi}{a} \frac{\partial}{\partial \xi} - \frac{D\sin\varphi}
{a \sinh\eta }\frac{\partial}{\partial \varphi} \\
\frac{\partial }{\partial y } = - \frac{\sin\varphi (\cosh\eta \cos\xi -1)}{a} \frac{\partial }{\partial \eta}
-\frac{\sin\xi \sinh\eta \sin\varphi}{a} \frac{\partial }{\partial \xi} + \frac{D\cos\varphi}
{a \sinh\eta }\frac{\partial }{\partial \varphi} \\
\frac{\partial }{\partial z } = - \frac{\sin\xi \sinh\eta }{a}\frac{\partial }{\partial \eta }
+\frac{\cosh\eta \cos\xi -1}{a}\frac{\partial }{\partial \xi } \\
\nabla^{2} \equiv \partial_{x}^{2}+\partial_{y}^{2}+\partial_{z}^{2}=-\frac{D}{a^{2}}
 \Bigg[(\cos\xi -\cosh\eta)\Bigg( \frac{1}{\sinh^{2}\eta}
\frac{\partial^{2}}{\partial \varphi^{2}} + \frac{\partial^{2}}{\partial \xi^{2}} \Bigg)+{}
                                                                                    \nonumber\\
 {}+sin\xi \frac{\partial}{\partial \xi}+(\cos\xi \cosh\eta -1)\frac{1}{\sinh\eta} 
\frac{\partial}{\partial \eta} - D
\frac{\partial^{2}}{\partial \eta^{2}} \Bigg]
 \end{eqnarray*}

The relations between unitary vectors of the toroidal and the cartesian coordinate system follow:
\begin{eqnarray*}
\hat{h}=\frac{1}{D}\Big[\cos\varphi(1 -\cosh\eta \cos\xi)\:\hat{i}
+\sin\varphi(1 -\cosh\eta \cos\xi)\:\hat{j}-\sinh\eta \sin\xi \:\hat{k} \Big] \\
\hat{\xi}=\frac{1}{D}\Big[ -\sinh\eta \sin\xi (\cos\varphi\: \hat{i}
+ \sin\varphi\: \hat{j}) +(\cos\xi \cosh\eta -1)\:\hat{k}\Big] \\
\hat{\varphi}=-\sin\varphi \:\hat{i}+\cos\varphi \:\hat{j}
\end{eqnarray*}

\subsection{{\normalsize Part III: Cylindrical coordinate system}}

The cylindrical coordinates $(\rho,\varphi,z)$ are related to the cartesian $(x,y,z)$ through the
following definition:
\begin{equation}
x=\rho \cos\varphi,\;\;y= \rho \sin\varphi,\;\;z=z
\end{equation}

In order to find the field equations, we need to relate the derivatives of cartesian coordinates with
those of cylindrical coordinates thus, the following formulas can be very useful:

\begin{eqnarray*}
\frac{\partial}{\partial x}=\cos\varphi \frac{\partial}{\partial \rho}-\frac{\sin\varphi}{\rho}\frac{\partial}{\partial \varphi} \\
\frac{\partial}{\partial y}=\sin\varphi \frac{\partial}{\partial \rho}+\frac{\cos\varphi}{\rho}\frac{\partial}{\partial \varphi} \\
\frac{\partial^{2}}{\partial x^{2}}= \cos^{2}\varphi \frac{\partial^{2}}{\partial \rho^{2}}+\frac{\sin2\varphi}{\rho^{2}}\frac{\partial}{\partial \varphi}
-\frac{\sin2\varphi}{\rho} \frac{\partial^{2}}{\partial \rho \partial \varphi}+\frac{\sin^{2}\varphi}{\rho}\frac{\partial}{\partial \rho}
+\frac{\sin^{2}\varphi}{\rho^{2}}\frac{\partial^{2}}{\partial \varphi^{2}} \\
\frac{\partial^{2}}{\partial y^{2}}= \sin^{2}\varphi \frac{\partial^{2}}{\partial \rho^{2}}-\frac{\sin2\varphi}{\rho^{2}}\frac{\partial}{\partial \varphi}
-\frac{\sin2\varphi}{\rho} \frac{\partial^{2}}{\partial \rho \partial \varphi}+\frac{\cos^{2}\varphi}{\rho}\frac{\partial}{\partial \rho}
+\frac{\cos^{2}\varphi}{\rho^{2}}\frac{\partial^{2}}{\partial \varphi^{2}} \\
\frac{\partial^{2}}{\partial x \partial y}= \frac{\sin2\varphi}{2} \frac{\partial^{2}}{\partial \rho^{2}}-\frac{\cos2\varphi}{\rho^{2}}\frac{\partial}{\partial \varphi}
+\frac{\cos2\varphi}{\rho} \frac{\partial^{2}}{\partial \rho \partial \varphi}-\frac{\sin^{2}\varphi}{2\rho}\frac{\partial}{\partial \rho}
-\frac{\sin^{2}\varphi}{2\rho^{2}}\frac{\partial^{2}}{\partial \varphi^{2}} \\
\nabla^{2}=\frac{1}{\rho}\frac{\partial}{\partial \rho}\bigg(\rho\frac{\partial}{\partial \rho}\bigg)+\frac{1}{\rho^{2}}\frac{\partial^{2}}{\partial \varphi^{2}}
+\frac{\partial^{2}}{\partial z^{2}}
\end{eqnarray*}

The relations between unitary vectors of the cylindrical and the cartesian coordinate system follow:
\begin{eqnarray*}
\hat{\rho}=\cos\varphi \;\hat{i} +\sin\varphi\; \hat{j} \\
\hat{\varphi}= -\sin\varphi\; \hat{i} +\cos\varphi\; \hat{j} \\
\hat{z}=\hat{k}
\end{eqnarray*}

\subsection{{\normalsize Part IV: Field profiles in different $z$ levels }}

Here, we present the dependence of the fields of the $U(1)\times U(1)$ model examined above, on $z$ variable.
For that reason, we plot the fields of fig.\ref{p1} for the same parameters, but in different $z$ levels
in order to observe their behavior.

\begin{figure}
\vspace{0.4 in.}
\centering
\includegraphics[scale=0.54]{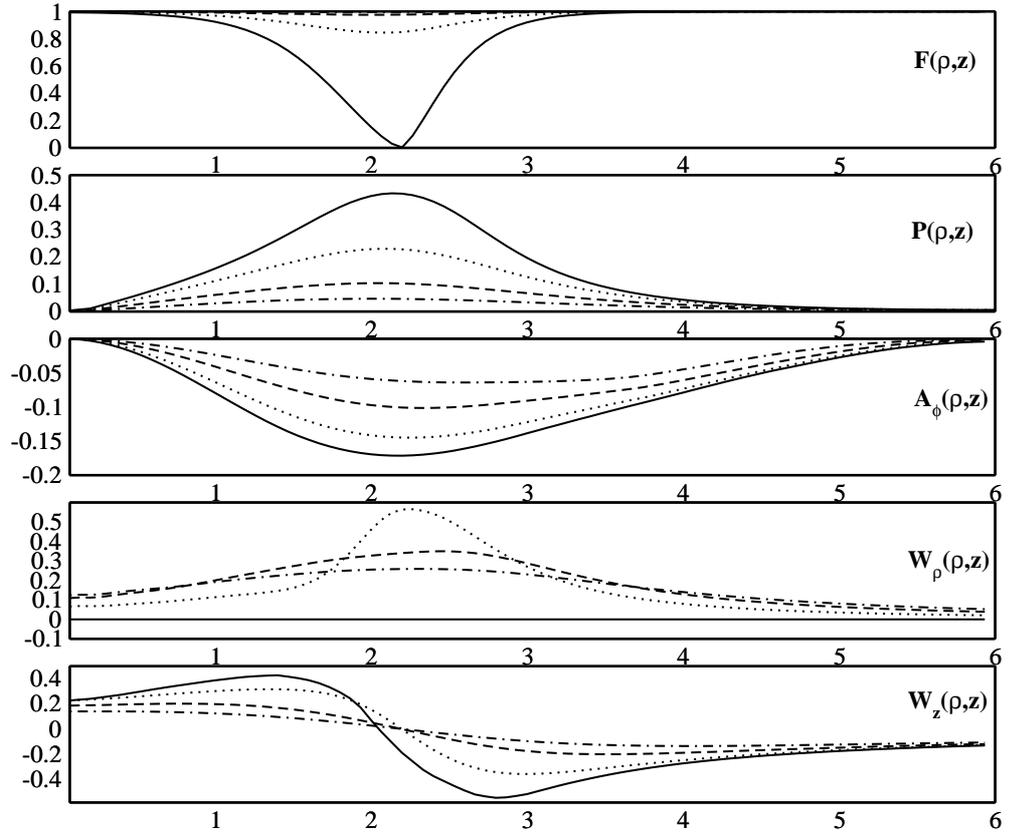}
\caption{\small {\bf $U(1)_{A}\times U(1)_{W}$ model:} Typical plot of the final configuration of fields.
Solid line: $z=0$, dotted line: $z=0.6$, dashed line: $z=1.2$, dashed and dotted line: $z=1.8$.
 Parameters are $M=1$, $N=1$, $e=5$, $q=m_{W}=2$, $g_{1}=14$, $m_{\phi}=\sqrt{g_{1}}=3.74$, $g_{2}=12$, $g_{3}=14$, $u=1$, 
$m_{\psi}= \sqrt{(g_{3}-g_{2}u^{2})/2}= 1$, $v_{1}=7.5\cdot 10^{-3}$. Energy $E=0.72$ and virial is $1.5\cdot 10^{-3}$.  \label{p1z}}
\end{figure}

%=================================================================================================================================
%=================================================================================================================================
%=================================================================================================================================
%=================================================================================================================================
%=================================================================================================================================
%=================================================================================================================================

\chapter{The extended $U(1)\times U(1)$ model for vortex rings}

\newpage

\section{{\normalsize Introduction}}

Stable strings of the previous $U(1)$ model (examined in chapter $4$) could be useful in order to create vortex rings and study
their stability. It could also be relatively helpful numerically, as we would have four fields for minimization
in the energy functional instead of five we had in \cite{toros} (chapter $5$).
This is not as easy as it might seem, since there are two instability modes. The vortex itself
is not necessarily stable while forming a torus, and the latter has the tendency to shrink due to its tension.
The first instability can be avoided in a $U(1)\times U(1)$ model as the one presented in \cite{toros} (chapter $5$),
where a numerical search for bosonic superconducting static vortex rings in a $U(1)_{A}\times U(1)_{W}$ model is examined.
There, the existence of straight strings is ensured for topological reasons.
The superconductivity of the loop though, does not seem to prevent shrinking.
The conclusion there, is that current quenching takes place before stabilization.

Here we deal with an extended version of the previously discussed model of chapter $5$. We
add appropriate higher derivative terms which might help the ring to stabilize.
We present and analyze our results.

\section{{\normalsize The  $U(1)_{A}\times U(1)_{W}$ model}}
The model discussed in \cite{toros} (chapter $5$), is being described by the Lagrangian density
\begin{equation}
\label{el0}
\mathcal{L}_{0}=-\frac{1}{4}F_{\mu\nu}^{2}-\frac{1}{4}W_{\mu\nu}^{2}+|D_{\mu}\psi|^{2}+|\tilde{D}_{\mu}\phi|^{2}-U(|\phi|,|\psi|)
\end{equation}
where the covariant derivatives are $D_{\mu}\psi \equiv \partial_{\mu}\psi+ieA_{\mu}\psi$, $\tilde{D}_{\mu}\phi \equiv \partial_{\mu}\phi+iqW_{\mu}\phi$,
the strength of the fields are $F_{\mu\nu}=\partial_{\mu}A_{\nu}-\partial_{\nu}A_{\mu}$, $W_{\mu\nu}=\partial_{\mu}W_{\nu}-\partial_{\nu}W_{\mu}$, 
while $e$ and $q$ stand as the relevant $U(1)$ charges.
The potential $U$ is
\begin{equation}
U(|\phi|,|\psi|)=\frac{g_{1}}{4}\big( |\phi|^{2} -v_{1}^{2}\big)^{2}+\frac{g_{2}}{4}\big( |\psi|^{2}- v_{2}^{2}\big)^{2}+\frac{g_{3}}{2}|\phi|^{2}|\psi|^{2}-\frac{g_{2}}{4}v_{2}^{4}
\end{equation}
The vacuum of this theory is $|\phi|= v_{1} \neq 0$, $|\psi|=0$ and breaks $U(1)_{W} \times U(1)_{A} \rightarrow U(1)_{A}$,
giving non-zero mass to $W$. The photon field stays massless. There, $U(v_{1},0)=0$. 
The vacuum manifold $\mathcal{M}$ in this theory is a circle $S^{1}$ and the first homotopy group of $\mathcal{M}$
is $\pi_{1}(\mathcal{M})=\pi_{1}(S^{1})=\mathbf{Z}$ which signals the existence of strings.
In regions where $|\phi|=0$, the field $|\psi|$ is arranged to be
non-vanishing and $U(1)_{W}\times U(1)_{A}\rightarrow U(1)_{W}$. Thus, $U(1)_{A}\rightarrow \mathbf{1}$ and
electric current flows along regions with vanishing $|\phi|$.
Hence, this theory has superconducting strings \cite{c4}.
The vacuum of the theory leaves unbroken the electromagnetic $U(1)_{A}$.
For $g_{3}v_{1}^{2} > g_{2}v_{2}^{2}$ this vacuum is stable, while $g_{1}v_{1}^{4}>g_{2}v_{2}^{4}$
ensures that it is the global minimum of the potential.
The mass spectrum is 
\begin{equation}
m_{A}=0,\;\;m_{W}=q v_{1},\;\;m_{\phi}^{2}=g_{1}v_{1}^{2},\;\;m_{\psi}^{2}=\frac{1}{2}\big(g_{3}v_{1}^{2}-g_{2}v_{2}^{2}\big)
\end{equation}

\section{{\normalsize The extended $U(1)_{A}\times U(1)_{W}$ model}}

We intend to modify the above model by adding higher derivative terms of the fields $\phi$ and $\psi$ and 
find out whether such changes can stabilize the ring. By following Derrick's scaling argument \cite{der}, one can argue that
terms such as $|D_{\mu}\psi|^{4}$ or $|\tilde{D}_{\mu}\phi|^{4}$ or $|\tilde{D}_{\mu}\phi|^{2}|D_{\mu}\psi|^{2}$ could be helpful. 
Also, in an investigation of a similar model \cite{hiet},
the conclusions lead to the same path, in order to search for possibilities of stabilizing such solitons against radial shrinking.
Thus, we have the following {\bf Lagrangian density}:
\begin{equation}
\label{lgex}
\mathcal{L}=\mathcal{L}_{0}+c_{\phi}|\tilde{D}_{\mu}\phi|^{4}+c_{\psi}|D_{\mu}\psi|^{4}+c_{\phi\psi}|\tilde{D}_{\mu}\phi|^{2}|D_{\mu}\psi|^{2}
\end{equation}
where $c_{\phi}$, $c_{\psi}$, $c_{\phi\psi}$ constants.
This Lagrangian (\ref{lgex}) with the extra terms, exhibits the symmetries of the original Lagrangian $\mathcal{L}_{0}$ (\ref{el0}).

Configurations with torus-like shape, representing a piece of a $U(1)_{W}\rightarrow \mathbf{1}$ Nielsen-Olesen string,
closed to form a loop, are of interest in this search. Thus, we will  require $\phi$ to vanish on a circle of radius $a$ (the torus radius)
$\phi (\rho =a, z=0)=0$. At infinity ($\rho\rightarrow \infty, z\rightarrow \infty$), we have the vacuum of the theory. This translates to $|\phi|\rightarrow v_{1}$,
$|\psi| \rightarrow 0$.
The  {\bf ansatz} for the fields is:
\begin{eqnarray*}
\phi(\rho,\varphi,z)&=&F(\rho,z)e^{iM\Theta(\rho,z)} \\
\psi(\rho,\varphi,z)&=&P(\rho,z)e^{iN\varphi} \\
\mathbf{A}(\rho,\varphi,z)&=&\frac{A_{\varphi}(\rho,z)}{\rho}\;\hat{\varphi} \\
\mathbf{W}(\rho,\varphi,z)&=&W_{\rho}(\rho,z)\;\hat{\rho}+W_{z}(\rho,z)\;\hat{z} 
\end{eqnarray*}
where $M$, $N$ are the winding numbers of the relevant fields, $\hat{\rho}$, $\hat{\varphi}$, $\hat{z}$ are the cylindrical unit vectors and the phase 
\begin{equation}
\Theta(\rho,z) \equiv \arctan \Big(\frac{z}{\rho-a}\Big)
\end{equation}
We use cylindrical coordinates $(t,\rho , \varphi , z)$,
with space-time metric that has the form $g_{\mu\nu}$=$diag(1,-1,-\rho^{2},-1)$. We work in the $A^{0}=0=W^{0}$ gauge.
We follow the ansatz of \cite{toros}.

With the above ansatz, the  {\bf energy functional} to be minimized takes the form:
\begin{eqnarray}
\label{funcu11}
E&=&2\pi v_{1} \int_{0}^{\infty}\rho d\rho \int_{-\infty}^{\infty}dz \Bigg[ \frac{1}{2\rho^{2}}\Big( (\partial_{\rho}A_{\varphi})^{2}+(\partial_{z}A_{\varphi})^{2}\Big)
+\frac{1}{2} (\partial_{\rho}W_{z}-\partial_{z}W_{\rho})^{2}+{}
                                                   \nonumber\\
{}&+&(\partial_{\rho}P)^{2}+(\partial_{z}P)^{2}+(\partial_{\rho}F)^{2}+(\partial_{z}F)^{2}+\frac{P^{2}}{\rho^{2}}(eA_{\varphi}+N)^{2}+{}
                                                          \nonumber\\
{}&+&\Big( (qW_{\rho}+M\partial_{\rho}\Theta)^{2}+(qW_{z}+M\partial_{z}\Theta)^{2}\Big)F^{2}+{}
                                                  \nonumber\\
{}&+&c_{\phi}\Bigg\{ (\partial_{\rho}F)^{2}+(\partial_{z}F)^{2}+ \Big((qW_{\rho}+M\partial_{\rho}\Theta)^{2}+(qW_{z}+M\partial_{z}\Theta)^{2}\Big)F^{2}\Bigg\}^{2}+{}
                 \nonumber\\
{}&+&c_{\psi}\Bigg\{  (\partial_{\rho}P)^{2}+(\partial_{z}P)^{2}+\frac{P^{2}}{\rho^{2}}(eA_{\varphi}+N)^{2}\Bigg\}^{2}+{}
                   \nonumber\\
{}&+&c_{\phi\psi}\Bigg\{\Bigg((\partial_{\rho}F)^{2}+(\partial_{z}F)^{2}+\Big( (qW_{\rho}+M\partial_{\rho}\Theta)^{2}+(qW_{z}+M\partial_{z}\Theta)^{2}\Big)F^{2}\Bigg){}
                 \nonumber\\
{}&\cdot&\Bigg( (\partial_{\rho}P)^{2}+(\partial_{z}P)^{2}+\frac{P^{2}}{\rho^{2}}(eA_{\varphi}+N)^{2}\Bigg)\Bigg\}+U(F,P) \Bigg]
\end{eqnarray}
and the potential $U$ follows:
\begin{equation}
\label{potu11}
U(F,P)=\frac{g_{1}}{4}\big( F^{2}-1\big)^{2}+\frac{g_{2}}{4}\big( P^{2}-u^{2}\big)^{2}+\frac{g_{3}}{2}F^{2}P^{2}-\frac{g_{2}}{4}u^{4}
\end{equation}
where  $u\equiv v_{2}/v_{1}$. This is the energy functional we  use for our numerical analysis.
The  conditions to be satisfied by the parameters become:
\begin{equation}
g_{1} > g_{2}u^{4}\;\;\;,\;\;\; g_{3} > g_{2}u^{2}
\end{equation}

The  magnetic fields are
\begin{eqnarray*}
\mathbf{\nabla \times A}=\mathbf{B_{A}}= \frac{1}{\rho}\Bigg(\frac{\partial A_{\varphi}}{\partial \rho}\;\hat{z}-\frac{\partial A_{\varphi}}{\partial z}\;\hat{\rho}\Bigg) \\
\mathbf{\nabla \times W}=\mathbf{B_{W}}= -\Bigg(\frac{\partial W_{z}}{\partial \rho}-\frac{\partial W_{\rho}}{\partial z}\Bigg)\;\hat{\varphi}
\end{eqnarray*}
while the currents associated with $\phi$ field, namely $j_{\rho}^{\phi}$ and $j_{z}^{\phi}$ and the total current $\mathcal{I}^{\phi}$
out of these as well as the supercurrent $\mathcal{I}^{\psi}$ associated with the $\psi$ field are
\begin{equation}
\mathcal{I}^{\phi}= \sqrt{(j_{\rho}^{\phi})^{2}+(j_{z}^{\phi})^{2}}\; ,\;\;\; \mathcal{I}^{\psi}= -\frac{2eP^{2}}{\rho}(eA_{\varphi}+N)
\end{equation}
where
\begin{equation}
j_{\rho}^{\phi}= -2qF^{2}(qW_{\rho}+M\partial_{\rho} \Theta), \;\;\;  j_{z}^{\phi}= -2qF^{2}(qW_{z}+M\partial_{z} \Theta)
\end{equation}

Finally, in order to check our numerical results, we can derive {\bf virial} relations through Derrick's scaling argument.
Below we present the virial relations we use in our search.
Consider the rescalings $\rho\rightarrow \rho$, $z\rightarrow \kappa z$, 
$F_{\kappa}\rightarrow F$, $P_{\kappa}\rightarrow P$, $A_{\varphi_{\kappa}} \rightarrow A_{\varphi}$,
 $W_{\rho,z_{\kappa}} \rightarrow \kappa W_{\rho,z}$. By demanding $\frac{\partial E}{\partial \kappa} =0$ when $\kappa =1 $
and if we define
 \begin{eqnarray*}
I_{1}&=& 2\pi v_{1} \int_{0}^{\infty} \rho d\rho \int_{-\infty}^{\infty} dz \Bigg[\frac{1}{2}(\partial_{\rho} W_{z} - \partial_{z} W_{\rho})^{2}+
             \frac{1}{2\rho^{2}} (\partial_{z} A_{\varphi})^{2}+(\partial_{z} P)^{2}+{}
          \nonumber\\
        {}&+&(\partial_{z} F)^{2}+2F^{2}\Bigg( qW_{\rho}(qW_{\rho}+M\partial_{\rho}\Theta )+(qW_{z}+M\partial_{z} \Theta)^{2}\Bigg)+{}
            \nonumber\\
        {}&+&c_{\phi}\Bigg\{4F^{2}(\partial_{z}F)^{2}\Bigg( (qW_{\rho}+M\partial_{\rho} \Theta)^{2} + (qW_{z}+M\partial_{z} \Theta)^{2}\Bigg)+{}
                \nonumber\\
        {}&+&3(\partial_{z}F)^{4}+2(\partial_{\rho}F)^{2}(\partial_{z}F)^{2}+{}
        \nonumber\\
        {}&+&\Bigg(4F^{4}\Bigg( (qW_{\rho}+M\partial_{\rho} \Theta)^{2} + (qW_{z}+M\partial_{z} \Theta)^{2}\Bigg){}
         \nonumber\\
        &\cdot&\Big( qW_{\rho}(qW_{\rho}+M\partial_{\rho} \Theta)+ (qW_{z}+M\partial_{z} \Theta)^{2}\Big)\Bigg)+{}
        \nonumber\\
        {}&+&4F^{2}\Big( (\partial_{\rho}F)^{2}+(\partial_{z}F)^{2}\Big) \Big( qW_{\rho}(qW_{\rho}+M\partial_{\rho} \Theta) + (qW_{z}+M\partial_{z} \Theta)^{2}\Big)\Bigg\}+{}
             \nonumber\\
        {}&+&c_{\psi}\Bigg\{ 3(\partial_{z}P)^{4}+2(\partial_{\rho}P)^{2}(\partial_{z}P)^{2}+\frac{2P^{2}}{\rho^{2}}(\partial_{z}P)^{2}(eA_{\varphi}+N)^{2} \Bigg\}+{}
        \nonumber\\
 \end{eqnarray*}
 \begin{eqnarray*}
         &+&c_{\phi\psi}\Bigg\{ (\partial_{z}P)^{2}(\partial_{\rho}F)^{2}+(\partial_{\rho}P)^{2}(\partial_{z}F)^{2}+3(\partial_{z}P)^{2}(\partial_{z}F)^{2}+{}
          \nonumber\\
        {}&+&\frac{P^{2}}{\rho^{2}}(eA_{\varphi}+N)^{2}(\partial_{z}F)^{2}+\Bigg( \frac{2P^{2}F^{2}}{\rho^{2}}(eA_{\varphi}+N)^{2} +2F^{2}\Big((\partial_{\rho}P)^{2}
        +(\partial_{z}P)^{2}\Big) \Bigg){}
        \nonumber\\
        {}&\cdot&\Big( qW_{\rho}(qW_{\rho}+M\partial_{\rho}\Theta)+(qW_{z}+M\partial_{z}\Theta)^{2}\Big)+{}
        \nonumber\\
        {}&+&(\partial_{z}P)^{2}F^{2}\Bigg( (qW_{\rho}+M\partial_{\rho} \Theta)^{2} + (qW_{z}+M\partial_{z} \Theta)^{2}\Bigg)\Bigg\}\Bigg] 
\end{eqnarray*}
\begin{eqnarray*}
I_{2}&=&  -2\pi v_{1} \int_{0}^{\infty} \rho d\rho \int_{-\infty}^{\infty} dz \Bigg[\frac{1}{2\rho^{2}}(\partial_{\rho}A_{\varphi})^{2}+(\partial_{\rho} F)^{2}+(\partial_{\rho} P)^{2}+{}
        \nonumber\\
        {}&+&\frac{P^{2}}{\rho^{2}}(eA_{\varphi}+N)^{2}+(\partial_{z}W_{\rho})(\partial_{\rho}W_{z}- \partial_{z}W_{\rho})+{}
        \nonumber\\
        {}&+&F^{2}\Bigg( (qW_{\rho}+M\partial_{\rho} \Theta)^{2} + (qW_{z}+M\partial_{z} \Theta)^{2}\Bigg)+{}
     \nonumber\\
     {}&+&c_{\phi}\Bigg\{ (\partial_{\rho}F)^{4}+F^{4}\Bigg( (qW_{\rho}+M\partial_{\rho} \Theta)^{2} + (qW_{z}+M\partial_{z} \Theta)^{2}\Bigg)^{2} +{}
     \nonumber \\
     {}&+& 2F^{2}\Big( (\partial_{\rho}F)^{2}+(\partial_{z}F)^{2}\Big)\Bigg( (qW_{\rho}+M\partial_{\rho} \Theta)^{2} + (qW_{z}+M\partial_{z} \Theta)^{2}\Bigg)\Bigg\}+{}
     \nonumber\\
     {}&+& c_{\psi}\Bigg\{ (\partial_{\rho}P)^{4}+\frac{P^{4}}{\rho^{4}}(eA_{\varphi}+N)^{4}+\frac{2P^{2}}{\rho^{2}}(\partial_{\rho}P)^{2}(eA_{\varphi}+N)^{2}\Bigg\}+{}
     \nonumber\\
     {}&+& c_{\phi\psi}\Bigg\{ (\partial_{\rho}P)^{2}(\partial_{\rho}F)^{2} +\frac{P^{2}}{\rho^{2}}(eA_{\varphi}+N)^{2}(\partial_{\rho}F)^{2}+{}
     \nonumber\\
     {}&+&\Bigg( \frac{P^{2}F^{2}}{\rho^{2}}(eA_{\varphi}+N)^{2} +F^{2}(\partial_{\rho}P)^{2}\Bigg){}
     \nonumber\\
     {}&\cdot&\Bigg( (qW_{\rho}+M\partial_{\rho} \Theta)^{2} + (qW_{z}+M\partial_{z} \Theta)^{2}\Bigg)\Bigg\}+{}
                                  \nonumber\\
     {}&+&\frac{g_{1}}{4}(F^{2}-1)^{2} + \frac{g_{2}}{4} (P^{2} -u^{2})^{2} +\frac{g_{3}}{2} F^{2}P^{2}-\frac{g_{2}}{4}u^{4}\Bigg] 
 \end{eqnarray*}	
we must have $I_{1}+I_{2}=0$. We define the index $V=\frac{||I_{1}|-|I_{2}||}{|I_{1}|+|I_{2}|}$ and we want its value to be as small as possible.
We can have many other virial relations by assuming generally for a field $\phi$, the ``double'' rescaling $\phi (\vec{x}) \rightarrow  \kappa \phi(\mu \vec{x})$
and then demand $\frac{\partial E}{\partial \kappa}|_{\kappa = 1 = \mu} =0 =\frac{\partial E}{\partial \mu}|_{\kappa = 1 = \mu}$. For example,
we check our results through the following relations as well.
Consider the following rescalings $\rho\rightarrow \rho $, $z\rightarrow \mu z$, $F_{\kappa}\rightarrow F$, $P_{\kappa}\rightarrow \kappa P$,
$A_{\varphi_{\kappa}} \rightarrow A_{\varphi}$, $W_{\rho ,z_{\kappa}} \rightarrow  W_{\rho,z}$. 
We define
\begin{eqnarray*}
I_{3}&=& 2\pi v_{1} \int_{0}^{\infty} \rho d\rho \int_{-\infty}^{\infty} dz \Bigg[ \frac{1}{2\rho^{2}} (\partial_{z} A_{\varphi})^{2}+(\partial_{z} P)^{2}+
               (\partial_{z} F)^{2}+{}
			                                          \nonumber\\
     {}&+&\frac{1}{2}(\partial_{\rho}W_{z}-\partial_{z}W_{\rho})^{2}+2F^{2}\Bigg( M\partial_{z} \Theta (qW_{z} +M\partial_{z} \Theta )\Bigg)+{}
     \nonumber\\
     {}&+&c_{\phi}\Bigg\{ 3(\partial_{z} F)^{4}+2(\partial_{\rho}F)^{2}(\partial_{z}F)^{2}+{}
     \nonumber\\
     {}&+&4F^{4}\Bigg( (qW_{\rho}+M\partial_{\rho} \Theta)^{2} + (qW_{z}+M\partial_{z} \Theta)^{2}\Bigg)\Big( M\partial_{z}\Theta (qW_{z}+M\partial_{z}\Theta )\Big)+{}
     \nonumber\\
     {}&+&4F^{2}(\partial_{z}F)^{2}\Bigg( (qW_{\rho}+M\partial_{\rho} \Theta)^{2} + (qW_{z}+M\partial_{z} \Theta)^{2}\Bigg)+{}
     \nonumber\\
     {}&+&4F^{2}\Big( (\partial_{\rho}F)^{2}+(\partial_{z}F)^{2}\Big)\Big( M\partial_{z}\Theta (qW_{z}+M\partial_{z}\Theta )\Big)\Bigg\}+{}
     \nonumber\\
     {}&+&c_{\psi}\Bigg\{ 3(\partial_{z}P)^{4} +2(\partial_{\rho}P)^{2}(\partial_{z}P)^{2}+\frac{2P^{2}}{\rho^{2}}(\partial_{z}P)^{2}(eA_{\varphi}+N)^{2}\Bigg\}+{}
     \nonumber\\
     {}&+&c_{\phi\psi} \Bigg\{ (\partial_{\rho}P)^{2}(\partial_{z}F)^{2}+(\partial_{z}P)^{2}(\partial_{\rho}F)^{2}+3(\partial_{z}P)^{2}(\partial_{z}F)^{2}+{}
     \nonumber\\
     {}&+&\frac{P^{2}}{\rho^{2}}(eA_{\varphi}+N)^{2}(\partial_{z}F)^{2}+{}
     \nonumber\\
     {}&+&2F^{2}\Bigg( (\partial_{\rho}P)^{2} +(\partial_{z}P)^{2} + \frac{P^{2}}{\rho^{2}}(eA_{\varphi}+N)^{2}\Bigg) (M\partial_{z} \Theta (qW_{z}+M\partial_{z}\Theta))+{}
     \nonumber\\
     {}&+&(\partial_{z}P)^{2}F^{2}\Bigg( (qW_{\rho}+M\partial_{\rho} \Theta)^{2} + (qW_{z}+M\partial_{z} \Theta)^{2}\Bigg)\Bigg\}\Bigg] 
\end{eqnarray*}
\begin{eqnarray*}
I_{4}&=& -2\pi v_{1} \int_{0}^{\infty} \rho d\rho \int_{-\infty}^{\infty} dz \Bigg[\frac{1}{2\rho^{2}} (\partial_{\rho} A_{\varphi})^{2}+(\partial_{\rho} P)^{2}
                                   +(\partial_{\rho} F)^{2}+{}
	                                                                   \nonumber\\
        {}&+&\frac{P^{2}}{\rho^{2}}(eA_{\varphi} +N)^{2}+\partial_{\rho}W_{z}(\partial_{\rho}W_{z}-\partial_{z}W_{\rho})+{}
        \nonumber\\
        {}&+&F^{2}\Bigg( (qW_{\rho}+M\partial_{\rho}\Theta)^{2}+(qW_{z}+M\partial_{z}\Theta)^{2}\Bigg) +{}
        \nonumber\\
        {}&+&c_{\phi}\Bigg\{ (\partial_{\rho}F)^{4} + F^{4}\Bigg( (qW_{\rho}+M\partial_{\rho} \Theta)^{2} + (qW_{z}+M\partial_{z} \Theta)^{2}\Bigg)^{2}+{}
        \nonumber\\
        {}&+&2F^{2}\Big( (\partial_{\rho}F)^{2}+(\partial_{z}F)^{2}\Big)\Bigg( (qW_{\rho}+M\partial_{\rho} \Theta)^{2} + (qW_{z}+M\partial_{z} \Theta)^{2}\Bigg)\Bigg \} +{}
        \nonumber\\
        {}&+&c_{\psi}\Bigg\{ (\partial_{\rho}P)^{4}+\frac{P^{4}}{\rho^{4}}(eA_{\varphi}+N)^{4} +\frac{2P^{2}}{\rho^{2}}(\partial_{z}P)^{2}(eA_{\varphi}+N)^{2}\Bigg\}+{}
        \nonumber\\
        {}&+&c_{\phi\psi}\Bigg\{ (\partial_{\rho}P)^{2}(\partial_{\rho}F)^{2}+\frac{P^{2}}{\rho^{2}}(eA_{\varphi}+N)^{2}(\partial_{\rho}F)^{2}+{}
        \nonumber\\
        {}&+&\Bigg(\frac{P^{2}F^{2}}{\rho^{2}}(eA_{\varphi}+N)^{2}+(\partial_{\rho}P)^{2}F^{2}\Bigg){}
        \nonumber\\
        {}&\cdot&\Bigg( (qW_{\rho}+M\partial_{\rho} \Theta)^{2} + (qW_{z}+M\partial_{z} \Theta)^{2}\Bigg)\Bigg\}+{}
        \nonumber\\
        {}&+&\frac{g_{1}}{4}(F^{2}-1)^{2} + \frac{g_{2}}{4} (P^{2} -u^{2})^{2} +\frac{g_{3}}{2} F^{2}P^{2}-\frac{g_{2}}{4}u^{4}\Bigg] 
\end{eqnarray*}
together with
\begin{eqnarray*}
I_{5}&=& 4\pi v_{1} \int_{0}^{\infty} \rho d\rho \int_{-\infty}^{\infty} dz \Bigg[ (\partial_{\rho} P)^{2}+ (\partial_{z} P)^{2}
                  +\frac{P^{2}}{\rho^{2}} (e A_{\varphi} +N)^{2} +{}
                  \nonumber\\
                  {}&+&c_{\psi}\Bigg\{ 2(\partial_{\rho}P)^{4}+2(\partial_{z}P)^{4}+\frac{2P^{4}}{\rho^{4}}(eA_{\varphi}+N)^{4}+4(\partial_{\rho}P)^{2}(\partial_{z}P)^{2}+{}
                  \nonumber\\
                  {}&+&\frac{4P^{2}}{\rho^{2}}\Big( (\partial_{\rho}P)^{2}+(\partial_{z}P)^{2}\Big)(eA_{\varphi}+N)^{2}\Bigg\}+{}
                  \nonumber\\
                  {}&+&c_{\phi\psi}\Bigg\{ (\partial_{\rho}P)^{2}(\partial_{\rho}F)^{2}+(\partial_{\rho}P)^{2}(\partial_{z}F)^{2}+
                  (\partial_{z}P)^{2}(\partial_{\rho}F)^{2}+(\partial_{z}P)^{2}(\partial_{z}F)^{2}+{}
                  \nonumber\\
                  {}&+&F^{2}\Bigg( (\partial_{\rho}P)^{2}+(\partial_{z}P)^{2} +\frac{P^{2}}{\rho^{2}}(eA_{\varphi}+N)^{2}\Bigg){}
                  \nonumber\\
                  {}&\cdot&\Bigg( (qW_{\rho}+M\partial_{\rho} \Theta)^{2} + (qW_{z}+M\partial_{z} \Theta)^{2}\Bigg)+{}
                  \nonumber\\
                  {}&+&\frac{P^{2}}{\rho^{2}}(eA_{\varphi}+N)^{2}\Big( (\partial_{\rho}P)^{2}+(\partial_{z}P)^{2}\Big)\Bigg\}\Bigg] \\
I_{6}&=& 2\pi v_{1} \int_{0}^{\infty} \rho d\rho \int_{-\infty}^{\infty} dz \Bigg[g_{2}\Big( P^{2}-u^{2}\Big) P^{2} + g_{3}F^{2}P^{2}\Bigg]
\end{eqnarray*}
where, as above, we must have $I_{3}+I_{4}=0=I_{5}+I_{6}$.

In the energy functional (\ref{funcu11}), the terms that come from the $|\tilde{D}_{i}\phi|^{4}$ extra term, are multiplied with $c_{\phi}$.
In fact, these terms are proportional to $\partial_{\rho}F$ and $\partial_{z}F$. Thus, if one chooses $(c_{\phi},c_{\psi},c_{\phi\psi})=(1,0,0)$, then
there are more $F$-derivative terms in the functional. Energy minimization lowers these terms, something which one expects to lead to a thicker
string. This is a wanted feature in order to stabilize the ring. This would have another consequence. The extra ``$F$-terms'', enforce the
$F$ field to stay away from its vacuum expectation value within a larger area. This means that $F\approx 0$ inside a bigger area.
There, the potential becomes
\begin{equation}
U(0,P)= \frac{g_{1}}{4}-\frac{g_{2}u^{4}}{4}+\frac{g_{2}}{4}\Big(P^{2}-u^{2}\Big)^{2}
\end{equation}
and its value increases because of the term $g_{1}(F^{2}-1)^{2}/4$, as $F\rightarrow 0$. 
Minimization tends to make $P\rightarrow u$, which tries to compensate for that increase. The latter means that $P$ will 
increase and this is another wanted feature since the supercurrent $\mathcal{I}^{\psi} \propto P^{2}$.

On the other hand, in (\ref{funcu1}), the terms that come from the $|D_{i}\psi|^{4}$ extra term, are multiplied with $c_{\psi}$.
In fact, these terms are proportional to $\partial_{\rho}P$ and $\partial_{z}P$. Thus, if one chooses $(c_{\phi},c_{\psi},c_{\phi\psi})=(0,1,0)$, then
the  $P$-derivative terms become more. The minimization of them is expected to decrease the charge condensate $P$. This decrease
is an unwanted feature.

Finally, the terms that come from the $|\tilde{D}_{i}\phi|^{2}|D_{i}\psi|^{2}$ term, are multiplied with $c_{\phi\psi}$. The consequences
of the addition of this term can be seen if we observe that the extra terms are of the form
\begin{equation}
(\partial P)^{2}\Bigg[ (\partial_{\rho}F)^{2}+(\partial_{z}F)^{2}+\cdots\Bigg]
\end{equation}
where the dots represent the rest of the terms which are positive as well.
We expect a stronger decrease of the charge condensate $P$. This is because $\partial P < 1$ which means that $(\partial P)^{2} > (\partial P)^{4}$.
Thus, the need for minimizing the derivative terms of $P$ becomes stronger than in the case of $(c_{\phi},c_{\psi},c_{\phi\psi})=(0,1,0)$.
Apart from this, it is also the fact that, since $\partial F \sim 1$, the weight of the $P$-derivative terms is now greater than unity and this is another factor
which would tend to make $P\rightarrow 0$ or, at least, smaller than in the case  $(c_{\phi},c_{\psi},c_{\phi\psi})=(0,1,0)$.

Theoretically, the fact that the terms $|D_{i}\psi|^{4}$ and $|\tilde{D}_{i}\phi|^{2}|D_{i}\psi|^{2}$ tend to shrink the charge condensate $P$,
can also be seen from the virial relation $I_{5}+I_{6}=0$ above. The integral $I_{6}$ is negative (since $P<u$) and the addition of extra terms leaves it
unchanged. On the other hand, $I_{5}$ is a sum of positive terms and the above two extra derivative terms rise the value of $I_{5}$.
Thus, the only way to satisfy that virial relation is either to increase $g_{2}$ and/or to decrease $P$. If $I_{5}$ is large enough, then
$P$ will be enforced to become zero in order to satisfy the $I_{5}+I_{6}=0$ relation.

\section{{\normalsize The extended $U(1)_{A}\times U(1)_{W}$ model: Numerical results }}

We use the same minimization algorithm, as in \cite{toros}, to minimize the energy functional presented in (\ref{funcu11}).
A $90\times 20$ grid for every of the five functions is used, that is, $90$ points on $\rho$-axis and $20$ on $z$.
We begin with fixed torus radius $a$. Then, the configuration with minimum energy for this $a$ is found. Other values of $a$
are chosen as well and the same process goes on until we plot the energy vs. the torus radius $E(a)$. It would be very
interesting to find a non-trivial minimum of the energy (in $a_{min}\neq 0$), which would correspond to stable toroidal defects with radius $a_{min}$.
One crucial check of our results is done through  virial relations.
\begin{figure}
\vspace{0.4 in.}
\centering
\includegraphics[scale=0.45]{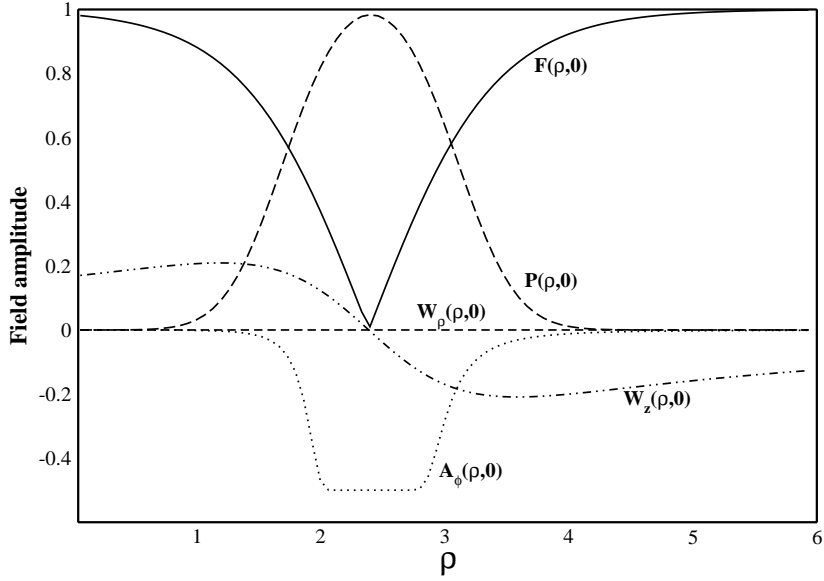}
\caption{\small {\bf Extended $U(1)_{A}\times U(1)_{W}$ model:} A typical plot of the initial guess we use, for the lowest winding state $M=1, N=1$
on $z=0$ plane.\label{ig1}}
\end{figure}

The {\bf initial guess} (fig.\ref{ig1}) we use for our computation is:
\begin{eqnarray*}
F(\rho,z)= \tanh((\rho-a)^{2}+z^{2})^{M/2} \\
P(\rho,z)= \tanh(\rho^{N}) (1-\tanh((\rho -a)^{2}+z^{2}) \\
A_{\varphi}(\rho,z)= -\frac{N}{e}\tanh\Bigg(\frac{\rho^{2}}{((\rho-a)^{2}+z^{2})^{2}}\Bigg) \\
W_{\rho}(\rho,z)=\frac{Mz\cos^{2}\Theta}{q(\rho-a)^{2}}\Bigg(\frac{(\rho-a)^{2}+z^{2}}{(\rho-a)^{2}+z^{2}+(a^{2}/4)}\Bigg)^{2} \\
W_{z}(\rho,z)=-\frac{M\cos^{2}\Theta}{q(\rho-a)}\Bigg(\frac{(\rho-a)^{2}+z^{2}}{(\rho-a)^{2}+z^{2}+(a^{2}/4)}\Bigg)
\end{eqnarray*}
This initial guess also satisfies the appropriate asymptotics
\begin{itemize}
\item{near $\rho =0$: 
\begin{equation}
F\neq 0 , \;\; P\sim \rho^{N}, \;\; A_{\varphi} \sim \rho^{2}f(z)
\end{equation} }
\item{near $(\rho =a , z=0)$: 
\begin{equation}
F\sim \tilde{\rho}^{M/2}, \;\; W_{\rho}=0=W_{z}
\end{equation} }
\item{at infinity: 
\begin{eqnarray}
F\sim 1-\mathcal{O}(e^{-\sqrt{\tilde{\rho}}}),\;\;\; P\sim \mathcal{O}(e^{-\sqrt{\rho^{2} +z^{2}}}) {}
  \nonumber\\
{}W_{\rho} \sim -\frac{M}{q}\partial_{\rho}\Theta|_{\infty} +\mathcal{O}(e^{-\sqrt{\tilde{\rho}}}),\;\;\; 
W_{z} \sim -\frac{M}{q}\partial_{z}\Theta|_{\infty} +\mathcal{O}(e^{-\sqrt{\tilde{\rho}}})
\end{eqnarray} }
\end{itemize}
where $\tilde{\rho}\equiv (\rho -a)^{2}+z^{2}$.

Based on \cite{toros} (chapter $5$), we search on parameter areas where $e$ acquires relatively large values, but they
are interesting concerning the possible stability of the loop. The reason was analyzed in that paper (chapter) and stems
from the need to have string thickness greater than the penetration depth as well as strong supercurrent.

\begin{figure}
\vspace{0.4 in.}
\centering
\includegraphics[scale=0.45]{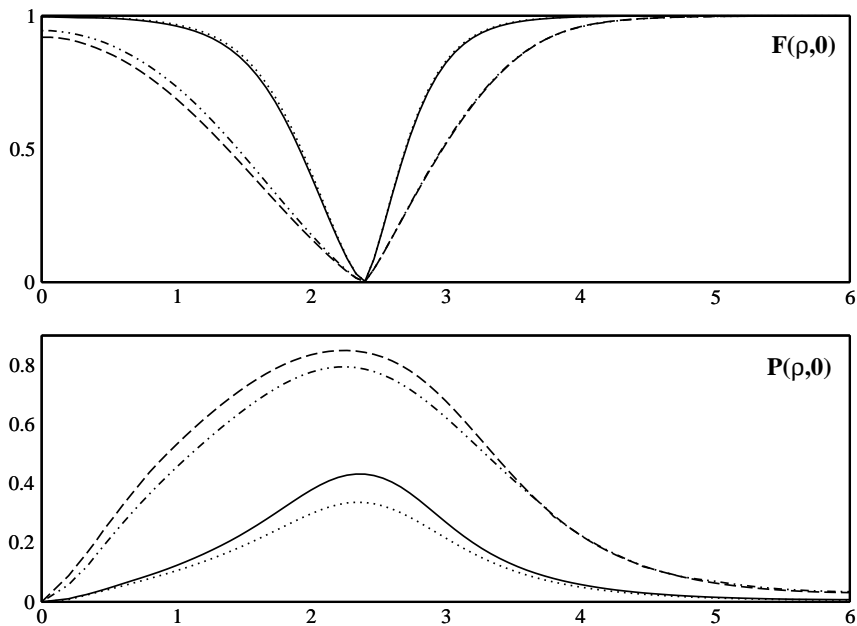}
\caption{\small {\bf Extended $U(1)_{A}\times U(1)_{W}$ model:}
 A typical graph which exhibits the effects of the higher derivative terms on the scalar fields. Solid line is for $(c_{\phi},c_{\psi},c_{\phi\psi})$=$(0,0,0)$,
dotted line for  $(c_{\phi},c_{\psi},c_{\phi\psi})$=$(0,1,0)$, dashed for  $(c_{\phi},c_{\psi},c_{\phi\psi})$=$(1,0,0)$ while dashed and dotted
for $(c_{\phi},c_{\psi},c_{\phi\psi})$=$(1,1,0)$. In the case $(c_{\phi},c_{\psi},c_{\phi\psi})$=$(0,0,1)$, $P$ is trivial. 
Parameters in this figure are $(g_{1},g_{2},g_{3},e,q,u,v_{1},M,N)$= $(14,12,14,6,2,1,7.5\cdot 10^{-3},1,1)$.\label{extraFP}}
\end{figure}
\begin{figure}
\vspace{0.4 in.}
\centering
\includegraphics[scale=0.45]{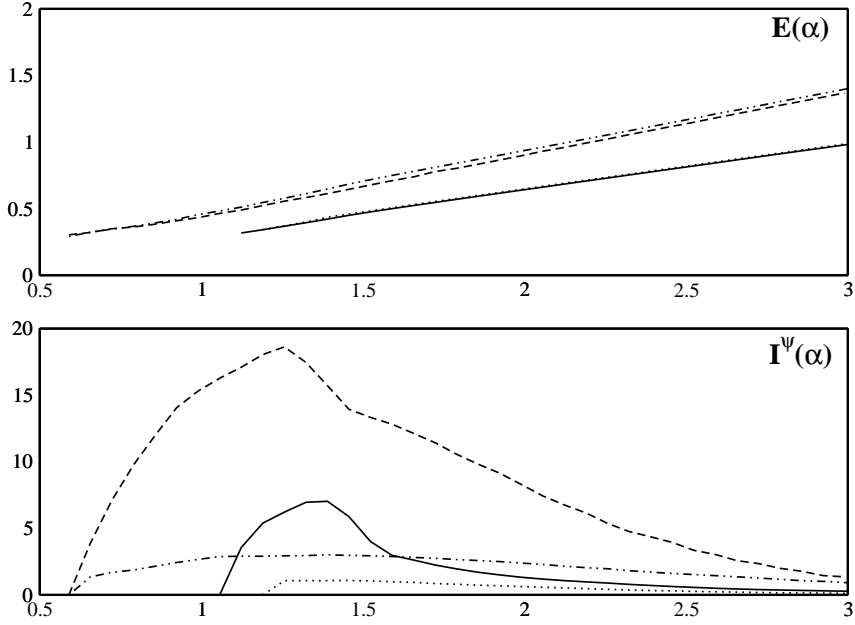}
\caption{\small {\bf Extended $U(1)_{A}\times U(1)_{W}$ model:}
 The effect of the higher derivative terms on the energy and the supercurrent when these are plotted vs. the
radius of the torus. Solid line is for $(c_{\phi},c_{\psi},c_{\phi\psi})$=$(0,0,0)$,
dotted line for  $(c_{\phi},c_{\psi},c_{\phi\psi})$=$(0,1,0)$, dashed for  $(c_{\phi},c_{\psi},c_{\phi\psi})$=$(1,0,0)$
while dashed and dotted for $(c_{\phi},c_{\psi},c_{\phi\psi})$=$(1,1,0)$.
For $(c_{\phi}, c_{\psi},c_{\phi\psi})$=$(0,0,1)$, the supercurrent as well as the charge condensate, are trivial.
Parameters in this figure are $(g_{1},g_{2},g_{3},e,q,u,v_{1},M,N)$=$(14,12.5,14,10,2,1,7.5\cdot 10^{-3},1,1)$.\label{extraEI}}
\end{figure}

The results confirm our expectations stated previously. For example, in fig.\ref{extraFP}, one can 
observe that when the extra higher derivative term is $|D_{i}\psi|^{4}$, then $F$ exhibits no change while
the charge condensate $P$  decreases (dotted line in fig.\ref{extraFP}).
In this case, the  consequence is the reduction of the supercurrent (dotted line in fig.\ref{extraEI})
when compared to the case of the original model without extra terms ($(c_{\phi},c_{\psi},c_{\phi\psi})=(0,0,0)$, see solid line in fig.\ref{extraEI}).
On the other hand, when the extra term is $|\tilde{D}_{i}\phi|^{4}$, then $F$ widens and this leads to the broadening of $P$ as well.
The latter increases (compare dashed and solid lines of fig.\ref{extraFP}) and the supercurrent increases too (dashed line of fig.\ref{extraEI}).
Finally, when the extra term is  $|\tilde{D}_{i}\phi|^{2}|D_{i}\psi|^{2}$, then $P=0$ for the values of $g_{2}$ we use. 
In general, the charge condensate can be non-trivial for higher $g_{2}$.
This happens due to the term $g_{2}(P^{2}-u^{2})^{2}/4$ of the potential. When $g_{2}$ grows, $P$ tends to reach $u$. This is also numerically observed.

One can combine two extra terms to see what happens. In example, we add both $|\tilde{D}_{i}\phi|^{4}$ and $|D_{i}\psi|^{4}$, (case $(c_{\phi},c_{\psi},c_{\phi\psi})$=$(1,1,0)$
in fig.\ref{extraFP}). This results to the addition of the ``favorable'' $F$-derivative terms, but also $P$-derivative terms would be present.
This translates to the growth of $P$ but not as much as in the case  $(c_{\phi},c_{\psi},c_{\phi\psi})$=$(1,0,0)$. We also observed that 
the combination of either $|\tilde{D}_{i}\phi|^{4}$ 
or $|D_{i}\psi|^{4}$ or even both, with the $|\tilde{D}_{i}\phi|^{2}|D_{i}\psi|^{2}$ term, leads to shrinking of $P$ because of the strong effect of the last term.
After the theoretical and numerical analysis, we conclude that the most ``interesting'' extra term is $|\tilde{D}_{i}\phi|^{4}$.

It is clear that current quenching is present here as well (fig.\ref{extraEI}). The extra term $|\tilde{D}_{i}\phi|^{4}$, can increase the supercurrent and can make
the penetration of the magnetic field more difficult, as the string increases its diameter, but this increase is not enough in order
for the ring to stabilize. For the shake of research, we also tried higher values of $c_{\phi}$ in order to make the favorable term more significant.
We also tried higher  values of $e$, but the ring could not stabilize in a non-trivial radius.

\section{{\normalsize Discussion}}

The most crucial terms of the energy functional which could provide for the stability of the ring in a non-zero radius $a$,
are: 
\begin{eqnarray*}
A&=& \int_{0}^{\infty}\rho d\rho \int_{-\infty}^{\infty} dz \;\; \frac{B_{A}^{2}}{2}= 
\int_{0}^{\infty}\rho d\rho \int_{-\infty}^{\infty} dz \;\; \frac{1}{2\rho^{2}}\Big( (\partial_{\rho}A_{\varphi})^{2}+(\partial_{z}A_{\varphi})^{2}\Big)\\
B&=& \int_{0}^{\infty}\rho d\rho \int_{-\infty}^{\infty} dz \;\; \frac{P^{2}}{\rho^{2}}(eA_{\varphi}+N)^{2}
\end{eqnarray*}
These two terms have an explicit total $1/\rho$ behavior which helps them to increase as the torus radius decreases.
The problem is that they are not increasing at a satisfactory rate in order to overcome all the rest terms of the energy which are
decreasing with $\rho$. This is also numerically observed.
Under some radius $a$, the ideal would be to have a strongly increasing charge condensate $P$. Then, as the radius decreases and
at the same time $P$ increases, the above two terms would start to increase with sufficient rate in order to lift the energy of the system.
The magnetic term would increase because as $P^{2}/\rho^{2}$ increases, $|A_{\varphi}|\rightarrow |A_{\varphi_{\mathbf{max}}}|\rightarrow N/e$.
The latter would make the derivatives of $A_{\varphi}$ (and $B_{A}$ as well) to increase as $\rho$ decreases.

This is what we tried to do here, especially with the help of the $|\tilde{D}_{i}\phi|^{4}$ extra term. The charge condensate became more
robust but that was not enough.
This supports the conclusion of \cite{toros} which states that very high values of supercurrent are needed for stabilization.
It seems that such highly increasing currents can not be produced, despite the help of extra terms. 
In fact, numerical details reveal that the rate of increase of the terms $A, B$ above, is $\Delta (A+B) \sim 10^{-3}$,
while the rate of decrease of all the rest terms is $\Delta (E-(A+B)) \sim 3\cdot 10^{-2}$. This means that the rate of increase
of $A,B$ should be $\sim 30$ times bigger. That case would 
require $P\geq u$ which is something that does not seem to satisfy $I_{5}+I_{6}=0$ virial relation.
But even if that was possible, current quenching would be another ``obstacle''.

\section{{\normalsize Conclusions}}

We are based on the model of \cite{toros} (the previous chapter) and analyze an extended version of it, by adding
higher derivative terms in order to check whether they can stabilize the superconducting ring or not. 
Although the $|\tilde{D}_{i}\phi|^{4}$ term is helpful on that direction, it turns out to be insufficient and current quenching prevails.
Finally, we discuss what one would need for a stable ring. This discussion in combination with the results of \cite{toros},
seems to exclude the possibility of existence of such vortex rings in this model.

%=============================================================================================================================
%=============================================================================================================================
%=============================================================================================================================
%=============================================================================================================================
%=============================================================================================================================
%=============================================================================================================================

\chapter{Conclusions}

\newpage

\section{{\normalsize Conclusions}}

Bosonic superconducting cosmic strings and solitons were the main subject of research in this Thesis.
Firstly, we examined a Goldstone model for antiperiodic solitons on $S^{1}$. An analytical study
as well as stability analysis are performed. Classically stable solitons were identified. Such
model can be connected and can give us the experience to deal with realistic particle physics models in our search for
possible metastable localized solitons.

Then, a detailed numerical search for bosonic superconducting straight strings in a $U(1)$ model with
Ginzburg-Landau potential with a cubic term added to it, is done. Such strings exist in a small, numerically
determined region. We fully analyze and explain the reasons of stability there, as well as the reasons of instability
in the rest of the parameter space. Such models can be found in Condensed Matter Physics as well. On the other hand,
this model can provide the basis for searching torus-shaped solitons in the framework of High-Energy Physics.

The stable strings of the above $U(1)$ model though, are not necessarily stable while forming a torus. Thus, we
worked in a $U(1)\times U(1)$ model which admits straight bosonic superconducting strings due to topological reasons.
We cut a piece out of these strings in order to create a torus and examine its stability. We performed a detailed
numerical study which showed that such solitons seem to be unstable. An explanation concerning this instability
is given with the help of theoretical predictions about current quenching as well. There are strong indications
that current quenching is the main reason of instability.

After the above conclusion, we searched further the model by extending it. We add higher derivative terms
which could be in favor of stability of this vortex ring. We analyze and explain our expectations.
Despite the fact that the results are improved, we again have non-stable objects. We discuss the reason
for this and explain what it would be necessary for stability. Out of this discussion, the conclusion
which states that these objects are unstable, is strongly supported. Current quenching seems to be the obstacle,
while there are indications that the vortex rings of this theory, need some specific extreme conditions in order to be stable.
Conditions which can {\em not} be satisfied by the $U(1)\times U(1)$ model we examined, even if current quenching phenomenon was absent.

LHC experiments, which are about to start in early 2008, can answer whether metastable particle-like solitons exist 
in MSSM or 2HSM. In the framework of 2HSM, the existence of spherically symmetric solitons was examined but the 
result was negative. The next step was to search for the possibility of the existence of axially symmetric solitons
with mass of the order of TeV. The confirmation or rejection of the existence of such solitons, would be a very interesting,
important as well as urgent task, since LHC will start working in a few months in a very promising energy range.
The work appeared in this Thesis, gives an answer and completes the above mentioned search on solitons with
axial symmetry.

Apart from this, search for such solitons can be done in other models with higher symmetries or extended models
\cite{c8a}-\cite{toki}. Experimental condensed matter physics is another branch where these gauge theories can apply.
Some examples can be found in \cite{condmat2}-\cite{nl2} and \cite{p1, paramos}.
On the other hand, such theories can give detailed and useful hints for cosmological observation of strings
in general and of loops in particular \cite{cosmr1}-\cite{cosmr9}, \cite{s6}-\cite{CSL}, \cite{loopconst}, \cite{cmb1}-\cite{cc12}.
Finally, another interest comes from the recent developments in Superstring theory. In the framework of large
extra dimensions, long superstrings may be stable and appear at the same energy scale as GUT scale cosmic strings. More
can be found in \cite{css1,css2}, \cite{r1}-\cite{acu}.

%=============================================================================================================================
%=============================================================================================================================
%=============================================================================================================================
%=============================================================================================================================
%=============================================================================================================================
%=============================================================================================================================

\chapter*{PART III}

\section*{ NUMERICAL METHODS}
\addcontentsline{toc}{chapter}{PART III:  NUMERICAL METHODS  }

In this part, we present details of the numerical methods used in models of Part I.
Then, we present full details of the algorithm we finally chose for our research in Part II.
We tested the algorithms on models with known solutions such as, the Nielsen-Olesen vortex and
the straight superconducting string.
Newton-Raphson and energy minimization were comparatively easier to handle, while the latter was faster and 
quite helpful as it concerns the physics of the model and for that reason we chose it for our research.

\chapter{Summary of the algorithms used}

\newpage
\section{{\normalsize Relaxation}}

 \begin{itemize}
\item{\em How it works}
\end{itemize}
  In relaxation methods, ODE's (Ordinary Differential Equations) are replaced by approximate FDE's
 (Finite-Difference Equations) on a grid or mesh of points that fills the domain of interest. The
 general rule is:
 \begin{equation}
 \frac{dy}{dx}\rightarrow \frac{y_{k}-y_{k-1}}{h}, \;\; (x,y)\rightarrow \Big(\frac{x_{k}+x_{k-1}}{2},
 \frac{y_{k}+y_{k-1}}{2}\Big)
 \end{equation}
 where $h=(x_{max}-x_{min})/N$ and $N$ stands for the number of points used to divide the region we
 are interested in. In other words it is the step on $x$-axis. The information the algorithm needs as input in order 
 to start computing the final configuration of fields are, the initial and final conditions as well as an 
 initial guess for the solution. It is important to have an, as good as possible, initial guess. This can be done
 by using the results of an asymptotical analysis on the behavior of the differential equations around $x_{min}$ and $x_{max}$.
 An iterative process is used and the initial guess improves step by step, while every next step uses the
 improved configuration produced by the previous one. Finally, the result is said to relax to the true solution.

\begin{itemize}
\item{\em  Concluding remarks for Relaxation}
\end{itemize}
 The satisfactory output for Nielsen-Olesen was created after many difficulties since the algorithm was very sensitive
 on very small changes of the boundary conditions.
 This had effect on the model of superconducting bosonic string where the method did not led us to a satisfactory result. 
 In general this algorithm was not easy to use and it's not efficient \cite{c6} especially in problems which have solutions
 with oscillatory parts. We continue by working with Runge-Kutta. 

\section{{\normalsize Runge-Kutta}}

\begin{itemize}
\item{\em How it works}
\end{itemize}
 This method advances a solution from $x_{n}$ to $x_{n+1}\equiv x_{n}+h$ through the formula
 $y_{n+1}=y_{n}+h\;f(x_{n},y_{n})$. User supplies the code with the initial boundary conditions only.
 This means that the values of $y_{x_{min}}$ and its derivative at that point are necessary. The "core"
 of the code of a fourth-order Runge-Kutta method follows:
 \begin{eqnarray}
 k_{1}&=&h\;f(x_{n},y_{n}) \\
 k_{2}&=&h\;f\Big(x_{n}+\frac{h}{2},y_{n}+\frac{k_{1}}{2}\Big) \\
 k_{3}&=&h\;f\Big(x_{n}+\frac{h}{2},y_{n}+\frac{k_{2}}{2}\Big) \\
 k_{4}&=&h\;f(x_{n}+h,y_{n}+k_{3}) \\
 y_{n+1}&=&y_{n}+\frac{k_{1}}{6}+\frac{k_{2}}{3}+\frac{k_{3}}{3}+\frac{k_{4}}{6}+\mathcal{O}(h^{5})
 \end{eqnarray}
 which means that every step $h$ involves four evaluations of the right hand side as one can see.
 Apart from that, we also used the fifth-order Runge-Kutta method in order to achieve greater accuracy 
 although our problems involve smooth functions, thus one could support that it might not be necessary.
 The "core" of the fifth-order Runge-Kutta method is presented below:
 \begin{eqnarray}
 k_{1}&=&h\;f(x_{n},y_{n}) \\
 k_{2}&=&h\;f(x_{n}+a_{2}h,y_{n}+b_{21}k_{1}) \\
 k_{3}&=&h\;f(x_{n}+a_{3}h,y_{n}+b_{31}k_{1}+b_{32}k_{2}) \\
 k_{4}&=&h\;f(x_{n}+a_{4}h,y_{n}+b_{41}k_{1}+b_{42}k_{2}+b_{43}k_{3}) \\
 k_{5}&=&h\;f(x_{n}+a_{5}h,y_{n}+b_{51}k_{1}+b_{52}k_{2}+b_{53}k_{3}+b_{54}k_{4}) \\
 k_{6}&=&h\;f(x_{n}+a_{6}h,y_{n}+b_{61}k_{1}+b_{62}k_{2}+b_{63}k_{3}+b_{64}k_{4}+b_{65}k_{5}) \\
 y_{n+1}&=&y_{n}+c_{1}k_{1}+c_{2}k_{2}+c_{3}k_{3}+c_{4}k_{4}+c_{5}k_{5}+c_{6}k_{6}+\mathcal{O}(h^{6})
 \end{eqnarray}
 where $a_{i},\;b_{ij},\;c_{i}$ are the Cash-Karp parameters and can be found on page $717$ of \cite{c6}.

\begin{itemize}
\item{\em Concluding remarks for Runge-Kutta}
\end{itemize}
Runge-Kutta was easier to use in comparison to Relaxation but the results were poor
in the  ``test''-models we tried to use it. Thus, we proceed on Newton-Raphson and energy minimization algorithms.

\newpage
\section{{\normalsize Newton-Raphson}}

\begin{itemize}
\item{\em How it works}
\end{itemize}
 The algorithm that follows, was created in 2004, in collaboration with Dr. E.D.M.Kavoussanaki.

 Suppose a differential equation written in the form of matrices: $A\cdot x=B$, where $A$ is a $N\times N$
 matrix, $x$ is a $N\times 1$ matrix of the unknown function and $B$ is a $N\times 1$ matrix as well. In
 that formulation, the solution of a differential equation involves is the root of the function $F=A\cdot x - B$.
 This means $F(x_{new})=0$, where $new$ means the matrix $x$ after an iteration. Thus $F(x_{new})=$
 $F(x_{old})+F'(x_{old})(x_{new}-x_{old})+\cdots$, which is in fact a Taylor expansion where terms beyond
 linear are considered unimportant. Solving in terms of $\delta = x_{new}-x_{old}$, we have:
 \begin{equation}
 \delta =-\frac{F(x_{old})}{F'(x_{old})}
 \end{equation}
 where the prime denotes differentiation with respect to $x$. The above equation is the correction to our
 initial guess $x$ and iteratively leads us to the solution. 
 Practically, one does the following steps:
 \begin{enumerate}
 
 \item {\bf Write the equation}. In our case we have second-order differential equations. 
 The derivatives are replaced in the following way:
 \begin{equation}
 x'\rightarrow \frac{x_{k+1}-x_{k-1}}{2h},\;\;\;\;x''\rightarrow \frac{x_{k+1}+x_{k-1}-2x_{k}}{h^{2}}
 \end{equation}
 
 \item {\bf Write matrices $A$ and $B$}. In this case, matrix $A$ is tri-diagonal. 
 $B$ is initially zero unless the equation is non-homogeneous with respect to $x$. 
 Otherwise, it contains these non-homogeneous terms. If the equation is non-linear,
 then $A=A_{L}+A_{NL}$, where $L$ and $NL$ stands for Linear an Non-Linear respectively.
 
 \item {\bf Compute the inverse of the Jacobian of $A$}. A general formula for the calculation of the
 Jacobian of a matrix is:
 \begin{equation}
 J_{ij}=\frac{\partial F_{i}}{\partial x_{j}}
 \end{equation}
 The Jacobian, and its inversion can also be done at once with 
 {\em clapack \footnote{More information at www.netlib.org}} 
routine package in order to achieve greater efficiency
\footnote{In fact it was observed that by using this package, the program needed at most 1/3 of the time it needed before in 
order to run. The difference was greater by increasing the grid points.}. 
 
 \item {\bf Compute correction}. The above step does the correction and this is added to our initial
 guess in the first iteration. Thus, our initial guess is improved in every iteration.
 
 \item {\bf Find solution}. When the correction $\delta$ becomes smaller than a user-defined small value,
 the program stops and we get the numerical solution.
 
 \end{enumerate}

\begin{itemize}
 \item{\em Concluding remarks for Newton-Raphson}
\end{itemize}
 This method, as one can observe in Part I,
 is one of the two  methods which gives satisfactory results on both models described there. A result is satisfactory
 when we have proofs
 \footnote{In Part I, one can find in detail, explanation of ways of how to check if a final configuration
 is the solution of a system of differential equations, with the help of virial relations.}
 that we found the solution, such as the appropriate asymptotics, the satisfaction
 of the equations of motion/field equations and of virial theorem.
 Apart from this, other advantages of this algorithm is the relatively short time it took us to get the results
 shown for both "test-models"
 and it is also the fact that it is relatively easy (compared with Relaxation and Runge-Kutta) to use.

\section{\normalsize Energy minimization}

\begin{itemize}
\item{\em How it works}
\end{itemize}
The general idea is this. Suppose the energy functional of a model of interest. This can involve two or more functions of one or more variables.
One can start with an initial guess for these functions approximating the true solution. A standard minimization algorithm such as the one we used,
adds a correction to the fields but having as a criterion that every such step will reduce the energy functional value. Thus, it is called energy 
minimization. The output is a field configuration for which the energy functional value is minimum. The algorithm needs as input 
the function as well as the derivative of that function at every point.

Details of the specific algorithm can be found on page 428 of \cite{c6}. There are used two subroutines, the first of them called DFPMIN which has as
target to find the so-called ``Newton'' direction which is a direction that makes the value of the energy smaller. Then, after this direction 
is calculated, subroutine LNSRCH is called and finds an appropriate step of correction for every point of every function on the ``Newton'' direction.
When the correction is done, DFPMIN calculates a new ``Newton'' direction based on the new corrected data. When the corrections are smaller than the
machine tolerance ($\approx 10^{-8}$) or the derivative of our function is very small then we are on a minimum.

To determine the next iteration point in this algorithm we do the following:
Consider finding a minimum by using Newton's method to search for a zero on the gradient of the function. Near the current point ${\mathbf x_{i}}$
we have to second order:
\begin{equation}
f({\mathbf x}) = f({\mathbf x_{i}}) + ({\mathbf x}-{\mathbf x_{i}})\cdot \nabla f({\mathbf x_{i}}) + 
\frac{1}{2}({\mathbf x}-{\mathbf x_{i}})\cdot {\mathbf A} \cdot ({\mathbf x}-{\mathbf x_{i}})
\end{equation}
so
\begin{equation}
\nabla f({\mathbf x}) = \nabla f({\mathbf x_{i}}) +  {\mathbf A} \cdot ({\mathbf x}-{\mathbf x_{i}})
\end{equation}
In Newton's method we set $\nabla f({\mathbf x}) =0$ to determine the next iteration step and finally we  have:
\begin{equation}
{\mathbf x_{i+1}} = {\mathbf x_{i}} + {\mathbf A^{-1}} \cdot ( \nabla f({\mathbf x_{i+1}}) -\nabla f({\mathbf x_{i}}))
\end{equation}
The DFPMIN subroutine provides the appropriate formulas for calculating ${\mathbf A^{-1}}$. One can find more details on p.428 of \cite{c6}.

\begin{itemize}
\item{\em Numerical details}
\end{itemize}
A point we need to mention here is that
one has to be very careful when inserting the derivatives (below denoted as $g_{i}$) 
of functions (below denoted as $p_{i}$) such as the one we have in our model. One must write down explicitly the first two or three
terms of the double sum which constitutes our energy functional, as well as the last two in order to see the sequence of appearance of $p_{i}$
in order to compute the relevant $g_{i}$. Then, we choose the variable of differentiation and gather the terms
which include that variable in order to find the derivative. In numerical language, the notion of the term ``variable'' is not necessarily the same as in mathematical
language.

In this model, as we have five functions of two variables, we firstly have to divide the $\rho$-coordinate which has $n$ points, into five
parts representing the functions: $1\rightarrow k_{1}$ for $F$, $k_{1}+1\rightarrow k_{2}$ for $P$, $k_{2}+1 \rightarrow k_{3}$ for $A_{\varphi}$,
$k_{3}+1 \rightarrow k_{4}$ for $W_{\rho}$ and $k_{4}+1 \rightarrow n$ for $W_{z}$. We also have to divide the $z$-coordinate in, say, $n_{z}$ points.
Because of the fact that now we have derivatives of these functions also in the $z$-direction we have to find a way to connect a point of a function
in $z_{ii}$ level with the same point in the $z_{ii_{+}}$ level. Well, according to our formulation, this happens when  we add $n$ to the index of a point.
Thus, if $i$ is for $\rho$ and $ii$ is for $z$, then we can represent a derivative of a function $f$ in $z$ direction as 
\begin{equation}
\frac{df}{dz}=\frac{f[i+n]-f[i]}{h_{z}}
\end{equation}
with
\begin{equation}
h_{z}=\frac{z_{max}-z_{min}}{n_{z}}
\end{equation}
 Also, differentiation of $f$ with respect to $\rho$ is  
\begin{equation}
\frac{df}{d\rho}=\frac{f[i+1]-f[i]}{h_{\rho}}
\end{equation}
 with 
\begin{equation}
h_{\rho}=\frac{\rho_{max}-\rho_{min}}{n}
\end{equation}
In the formulation below, for simplicity, we represent with $ii_{+}$ the next $z$ level and with $ii_{-}$ the previous $z$ level. For example if we are at $z=d$ level
then $ii=d\cdot n$ and $ii_{+}=(d+1)\cdot n$, while $ii_{-}=(d-1)\cdot n$, where $d$ an integer. The energy functional in numerical language, follows:
\begin{eqnarray*}
{\bf E_{i:\:1\rightarrow k_{1}-1,\;\; ii:\:0\rightarrow n(z_{max}-1) }} =2\pi v_{1} h_{\rho}h_{z}\rho_{i}
\Bigg[ \frac{1}{2\rho_{i}^{2}}\Bigg(\Bigg(\frac{p_{k_{2}+i+1+ii}-p_{k_{2}+i+ii}}{h_{\rho}}\Bigg)^{2}+{}
                                                                \nonumber\\
{}+\Bigg(\frac{p_{k_{2}+i+ii_{+}}-p_{k_{2}+i+ii}}{h_{z}}\Bigg)^{2}\Bigg)+\frac{1}{2}\Bigg(\frac{p_{k_{4}+i+1+ii}-p_{k_{4}+i+ii}}{h_{\rho}}-{}
           \nonumber\\
{}-\frac{p_{k_{3}+i+ii_{+}}-p_{k_{3}+i+ii}}{h_{z}}\Bigg)^{2}+{}
          \nonumber\\
{}+\Bigg(\frac{p_{k_{1}+i+1+ii}-p_{k_{1}+i+ii}}{h_{\rho}}\Bigg)^{2}+\Bigg(\frac{p_{k_{1}+i+ii_{+}}-p_{k_{1}+i+ii}}{h_{z}}\Bigg)^{2}+{}
                                                      \nonumber\\
{}+\Bigg(\frac{p_{i+1+ii}-p_{i+ii}}{h_{\rho}}\Bigg)^{2}+\Bigg(\frac{p_{i+ii_{+}}-p_{i+ii}}{h_{z}}\Bigg)^{2}+\frac{p_{k_{1}+i+ii}^{2}}{\rho_{i}^{2}}\Big(ep_{k_{2}+i+ii}+N\Big)^{2}+{}
\nonumber\\
{}+\Bigg[ \Bigg(qp_{k_{3}+i+ii}-M\frac{z_{ii}cos^{2}\Theta}{(\rho_{i}-a)^{2}}\Bigg)^{2}+\Bigg(qp_{k_{4}+i+ii}+M\frac{cos^{2}\Theta}{\rho_{i}-a}\Bigg)^{2}\Bigg] p_{i+ii}^{2}+{}
                                                                                                                       \nonumber\\
{}+\frac{g_{1}}{4}\Big(p_{i+ii}^{2}-1\Big)^{2}+\frac{g_{2}}{4}\Big(p_{k_{1}+i+ii}^{2}-u^{2}\Big)^{2}+\frac{g_{3}}{2}p_{i+ii}^{2}p_{k_{1}+i+ii}^{2} \Bigg]
\end{eqnarray*}
%==========================================================================================================================================
%==========================================================================================================================================
%==========================================================================================================================================
Below we write down the derivatives of the above energy functional for every ``variable''. We start with the $i=1$ level for all functions and all $z$-levels except
for the first and the last one. In fact $i=1$ means $\rho=0$.
\begin{eqnarray*}
{\bf g_{1+ii} }=2\pi v_{1}h_{\rho}h_{z} \rho_{1}\Bigg[ \frac{2}{h_{z}^{2}}(p_{1+ii}-p_{1+ii_{-}})-\frac{2}{h_{\rho}^{2}}(p_{2+ii}-p_{1+ii})-\frac{2}{h_{z}^{2}}(p_{1+ii_{+}}-p_{1+ii})+{}                                                                                                          \nonumber\\
{}+2p_{1+ii}\Bigg[ \Bigg( qp_{k_{3}+1+ii}-M\frac{z_{ii}cos^{2}\Theta}{(\rho_{1}-a)^{2}}\Bigg)^{2}+\Bigg( qp_{k_{4}+1+ii}+M\frac{cos^{2}\Theta}{\rho_{1}-a}\Bigg)^{2}\Bigg]+{}
                                                                                                              \nonumber\\
{}+g_{1}(p_{1+ii}^{2}-1)p_{1+ii}+g_{3}p_{1+ii}p_{k_{1}+1+ii}^{2}\Bigg] \\
{\bf g_{k_{1}+1+ii} }= 2\pi v_{1}h_{\rho}h_{z}\rho_{1}\Bigg[ \frac{2}{h_{z}^{2}}(p_{k_{1}+1+ii}-p_{k_{1}+1+ii_{-}})-\frac{2}{h_{\rho}^{2}}(p_{k_{1}+2+ii}-p_{k_{1}+1+ii})-{}
             \nonumber\\
{}-\frac{2}{h_{z}^{2}}(p_{k_{1}+1+ii_{+}}-p_{k_{1}+1+ii})+{}
                                                                                   \nonumber\\
{}+\frac{2}{\rho_{1}^{2}}p_{k_{1}+1+ii}(ep_{k_{2}+1+ii}+N)^{2}+g_{2}\Big(p_{k_{1}+1+ii}^{2}-u^{2}\Big)p_{k_{1}+1+ii}+g_{3}p_{1+ii}^{2}p_{k1+1+ii}\Bigg]\\
{\bf g_{k_{2}+1+ii} }= 2\pi v_{1}h_{\rho}h_{z}\rho_{1}\Bigg[ \frac{1}{2\rho_{1}^{2}}\Bigg( -\frac{2}{h_{\rho}^{2}}(p_{k_{2}+2+ii}-p_{k_{2}+1+ii})-{}
  \nonumber\\
{}-\frac{2}{h_{z}^{2}}(p_{k_{2}+1+ii_{+}}-p_{k_{2}+1+ii})+{}
             \nonumber\\
{}+\frac{2}{h_{z}^{2}}(p_{k_{2}+1+ii}-p_{k_{2}+1+ii_{-}})\Bigg)+\frac{2ep_{k_{1}+1+ii}^{2}}{\rho_{1}^{2}}(ep_{k_{2}+1+ii}+N)\Bigg] \\
{\bf g_{k_{3}+1+ii}}= 
2\pi v_{1}h_{\rho}h_{z} \rho_{1}\Bigg[ \frac{1}{h_{z}}\Bigg(\frac{p_{k_{4}+2+ii}-p_{k_{4}+1+ii}}{h_{\rho}}-\frac{p_{k_{3}+1+ii_{+}}-p_{k_{3}+1+ii}}{h_{z}}\Bigg)-{}
                                     \nonumber\\
{}-\frac{1}{h_{z}}\Bigg(\frac{p_{k_{4}+2+ii_{-}}-p_{k_{4}+1+ii_{-}}}{h_{\rho}}-\frac{p_{k_{3}+1+ii}-p_{k_{3}+1+ii_{-}}}{h_{z}}\Bigg)+{}
              \nonumber\\
{}+2qp_{1+ii}^{2}\Bigg(qp_{k_{3}+1+ii}-M\frac{z_{ii}cos^{2}\Theta}{(\rho_{1}-a)^{2}}\Bigg)\Bigg] \\
{\bf g_{k_{4}+1+ii} }= 2\pi v_{1}h_{\rho}h_{z} \rho_{1} \Bigg[ -\frac{1}{h_{\rho}}\Bigg(\frac{p_{k_{4}+2+ii}-p_{k_{4}+1+ii}}{h_{\rho}}-\frac{p_{k_{3}+1+ii_{+}}-p_{k_{3}+1+ii}}{h_{z}}\Bigg)+{}
                \nonumber\\
{}+2qp_{1+ii}^{2}\Bigg(qp_{k_{4}+1+ii}+M\frac{cos^{2}\Theta}{\rho_{1}-a}\Bigg)\Bigg] 
\end{eqnarray*}
The derivatives of all functions on all the $\rho$ and $z$ levels apart from the first and the last level follow.
\begin{eqnarray*}
{\bf g_{(i:\;2\rightarrow k_{1}-1)\;\; i+ii} }= 2\pi v_{1}h_{\rho}h_{z} \rho_{i-1}\Bigg[ \frac{2}{h_{\rho}^{2}}(p_{i+ii}-p_{i-1+ii})\Bigg]+{}
                                                                   \nonumber\\
{}+2\pi v_{1}h_{\rho}h_{z} \rho_{i}\Bigg[ \frac{2}{h_{z}^{2}}(p_{i+ii}-p_{i+ii_{-}})-\frac{2}{h_{\rho}^{2}}(p_{i+1+ii}-p_{i+ii})-\frac{2}{h_{z}^{2}}(p_{i+ii_{+}}-p_{i+ii})+{}
                                                                                                          \nonumber\\
{}+2p_{i+ii}\Bigg[ \Bigg( qp_{k_{3}+i+ii}-M\frac{z_{ii}cos^{2}\Theta}{(\rho_{i}-a)^{2}}\Bigg)^{2}+\Bigg( qp_{k_{4}+i+ii}
+M\frac{cos^{2}\Theta}{\rho_{i}-a}\Bigg)^{2}\Bigg]+{}
                                                                                                              \nonumber\\
{}+g_{1}(p_{i+ii}^{2}-1)p_{i+ii}+g_{3}p_{i+ii}p_{k_{1}+i+ii}^{2}\Bigg] \\
{\bf g_{(i:\;2\rightarrow k_{1}-1),\;\; k_{1}+i+ii} }= 2\pi v_{1}h_{\rho}h_{z}\rho_{i-1}\Bigg[ \frac{2}{h_{\rho}^{2}}(p_{k_{1}+i+ii}-p_{k_{1}+i-1+ii})\Bigg]+{}
                                                                               \nonumber\\
{}+2\pi v_{1}h_{\rho}h_{z}\rho_{i}\Bigg[ \frac{2}{h_{z}^{2}}(p_{k_{1}+i+ii}-p_{k_{1}+i+ii_{-}})-\frac{2}{h_{\rho}^{2}}(p_{k_{1}+i+1+ii}-p_{k_{1}+i+ii})-{}
                   \nonumber\\
{}-\frac{2}{h_{z}^{2}}(p_{k_{1}+i+ii_{+}}-p_{k_{1}+i+ii})+{}
                                                                                   \nonumber\\
{}+\frac{2}{\rho_{i}^{2}}p_{k_{1}+i+ii}(ep_{k_{2}+i+ii}+N)^{2}+g_{2}\Big(p_{k_{1}+i+ii}^{2}-u^{2}\Big)p_{k_{1}+i+ii}+g_{3}p_{i+ii}^{2}p_{k_{1}+i+ii}\Bigg]\\
{\bf g_{(i:\;2\rightarrow k_{1}-1),\;\;k_{2}+i+ii} }= 
 2\pi h_{\rho}h_{z}\rho_{i-1}\Bigg[\frac{1}{2\rho_{i-1}^{2}}\Bigg( \frac{2}{h_{\rho}^{2}}(p_{k_{2}+i+ii}-p_{k_{2}+i-1+ii})\Bigg)\Bigg]+{}
                                                                                                    \nonumber\\
{}+ 2\pi v_{1}h_{\rho}h_{z}\rho_{i}\Bigg[ \frac{1}{2\rho_{i}^{2}}\Bigg( -\frac{2}{h_{\rho}^{2}}(p_{k_{2}+i+1+ii}-p_{k_{2}+i+ii})
-\frac{2}{h_{z}^{2}}(p_{k_{2}+i+ii_{+}}-p_{k_{2}+i+ii})+{}
              \nonumber\\
{}+\frac{2}{h_{z}^{2}}(p_{k_{2}+i+ii}-p_{k_{2}+i+ii_{-}})\Bigg)+\frac{2ep_{k_{1}+i+ii}^{2}}{\rho_{i}^{2}}(ep_{k_{2}+i+ii}+N)\Bigg] 
\end{eqnarray*}
\begin{eqnarray*}
{\bf g_{(i:\;2\rightarrow k_{1}-1),\;\;k_{3}+i+ii} }= 2\pi v_{1}h_{\rho}h_{z} \rho_{i}\Bigg[ \frac{1}{h_{z}}\Bigg(\frac{p_{k_{4}+i+1+ii}-p_{k_{4}+i+ii}}{h_{\rho}}-{}
            \nonumber\\
{}-\frac{p_{k_{3}+i+ii_{+}}-p_{k_{3}+i+ii}}{h_{z}}\Bigg)-{}
                                     \nonumber\\
{}-\frac{1}{h_{z}}\Bigg(\frac{p_{k_{4}+i+1+ii_{-}}-p_{k_{4}+i+ii_{-}}}{h_{\rho}}-\frac{p_{k_{3}+i+ii}-p_{k_{3}+i+ii_{-}}}{h_{z}}\Bigg)+{}
              \nonumber\\
{}+2qp_{i+ii}^{2}\Bigg( qp_{k_{3}+i+ii}-M\frac{z_{ii}cos^{2}\Theta}{(\rho_{i}-a)^{2}}\Bigg)\Bigg] \\
{\bf g_{(i:\;2\rightarrow k_{1}-1),\;\;k_{4}+i+ii}}=  2\pi v_{1}h_{z} \rho_{i-1} \Bigg[\Bigg(\frac{p_{k_{4}+i+ii}-p_{k_{4}+i-1+ii}}{h_{\rho}}-{}
      \nonumber\\
{}-\frac{p_{k_{3}+i-1+ii_{+}}-p_{k_{3}+i-1+ii}}{h_{z}}\Bigg)\Bigg]-{}
                \nonumber\\
{}2\pi v_{1}h_{\rho}h_{z} \rho_{i} \Bigg[\frac{1}{h_{\rho}}\Bigg(\frac{p_{k_{4}+i+1+ii}-p_{k_{4}+i+ii}}{h_{\rho}}-\frac{p_{k_{3}+i+ii_{+}}-p_{k_{3}+i+ii}}{h_{z}}\Bigg)+{}
                \nonumber\\
{}+2qp_{i+ii}^{2}\Bigg(qp_{k_{4}+i+ii}+M\frac{cos^{2}\Theta}{\rho_{i}-a}\Bigg)\Bigg] 
\end{eqnarray*}
The derivatives of all functions on the last $\rho$ level and all $z$ levels apart from the first and the last level follow.
\begin{eqnarray*}
{\bf g_{k_{1}+ii} }=  2\pi v_{1}h_{\rho}h_{z} \rho_{k_{1}-1}\Bigg[\frac{2}{h_{\rho}^{2}}(p_{k_{1}+ii}-p_{k_{1}-1+ii})\Bigg] \\
{\bf g_{k_{2}+ii} }= 2\pi v_{1}h_{\rho}h_{z} \rho_{k_{1}-1}\Bigg[ \frac{2}{h_{\rho}^{2}}(p_{k_{2}+ii}-p_{k_{2}-1+ii})\Bigg] \\
{\bf g_{k_{3}+ii} }=  2\pi v_{1} h_{\rho}h_{z}\rho_{k_{1}-1}\Bigg[\frac{1}{2\rho_{k_{1}-1}^{2}}\Bigg( \frac{2}{h_{\rho}^{2}}(p_{k_{3}+ii}-p_{k_{3}-1+ii})\Bigg)\Bigg] \\
{\bf g_{k_{4}+ii} }=0 \\
{\bf g_{n+ii} }= 2\pi v_{1}h_{z}\rho_{k_{1}-1}\Bigg[\Bigg(\frac{p_{n+ii}-p_{n-1+ii}}{h_{\rho}}-\frac{p_{k_{4}-1+ii_{+}}-p_{k_{4}-1+ii}}{h_{z}}\Bigg)\Bigg] \\
\end{eqnarray*}
The derivatives of all functions on all the $\rho$  levels apart from the first and the last level and the $ii=0$ level follow. The $ii=0$ level is in fact the $z=-z_{0}$.
\begin{eqnarray*}
{\bf g_{(i:\;2\rightarrow k_{1}-1),\;\; i+0} } = 2 \pi v_{1}h_{\rho}h_{z}\rho_{i-1}\Bigg[ \frac{2}{h_{\rho}^{2}}(p_{i+0}-p_{i-1+0})\Bigg]+{}
                                         \nonumber\\
{}+ 2 \pi v_{1}h_{\rho}h_{z}\rho_{i}\Bigg[-\frac{2}{h_{z}^{2}}(p_{i+ii_{+}}-p_{i+0})-\frac{2}{h_{\rho}^{2}}(p_{i+1+0}-p_{i+0})+{}
                         \nonumber\\
{}+2p_{i+0}\Bigg(\Bigg(qp_{k_{3}+i+0} -\frac{Mz\cos^{2} \Theta}{(\rho_{i}-a)^{2}}\Bigg)^{2}
+\Bigg(qp_{k_{4}+i+0} -\frac{M\cos^{2} \Theta}{\rho_{i}-a}\Bigg)^{2}\Bigg)+{}
                                \nonumber\\
{}+g_{1}p_{i+0}\Big(p_{i+0}^{2}-1\Big)+g_{3}p_{i+0}p_{k_{1}+i+0}^{2}\Bigg] \\
{\bf g_{(i:\;2\rightarrow k_{1}-1),\;\; k_{1}+i+0} } = 2 \pi v_{1} h_{\rho} h_{z}\rho_{i-1} \Bigg[ \frac{2}{h_{\rho}^{2}}(p_{k_{1}+i+0}-p_{k_{1}+i-1+0})\Bigg]+{}
                         \nonumber\\
{}+ 2 \pi v_{1}h_{\rho}h_{z}\rho_{i} \Bigg[ -\frac{2}{h_{\rho}^{2}}(p_{k_{1}+i+1+0}-p_{k_{1}+i+0})-\frac{2}{h_{z}^{2}}(p_{k_{1}+i+ii_{+}}-p_{k_{1}+i+0})+{}
                       \nonumber\\
{}+\frac{2p_{k_{1}+i+0}}{\rho_{i}^{2}}(ep_{k_{2}+i+0}+N)^{2}+g_{2}p_{k_{1}+i+0}\Big(p_{k_{1}+i+0}^{2}-u^{2}\Big)+g_{3}p_{i+0}^{2}p_{k_{1}+i+0}\Bigg] \\
{\bf g_{(i:\;2\rightarrow k_{1}-1),\;\; k_{2}+i+0} } = 2 \pi v_{1}h_{\rho}h_{z}\rho_{i-1} 
\Bigg[ \frac{1}{2\rho_{i-1}^{2}}\Bigg(\frac{2}{h_{\rho}^{2}}(p_{k_{2}+i+0}-p_{k_{2}+i-1+0})\Bigg)\Bigg]+{}
             \nonumber\\
{}+ 2 \pi v_{1}h_{\rho}h_{z}\rho_{i}\Bigg[\frac{1}{2\rho_{i}^{2}}\Bigg(-\frac{2}{h_{\rho}^{2}}(p_{k_{2}+i+1+0}-p_{k_{2}+i+0})
-\frac{2}{h_{z}^{2}}(p_{k_{2}+i+ii_{+}}-p_{k_{2}+i+0})\Bigg)+{}
              \nonumber\\
{}+\frac{2ep_{k_{1}+i+0}^{2}}{\rho_{i}^{2}}(ep_{k_{2}+i+0}+N)\Bigg] \\
{\bf g_{(i:\;2\rightarrow k_{1}-1),\;\; k_{3}+i+0} } = 2 \pi v_{1}h_{\rho}h_{z}\rho_{i}\Bigg[\frac{1}{h_{z}}\Bigg(\frac{p_{k_{4}+i+1+0}-p_{k_{4}+i+0}}{h_{\rho}}-{}
                \nonumber\\
{}-\frac{p_{k_{3}+i+ii_{+}}-p_{k_{3}+i+0}}{h_{z}}\Bigg)+2qp_{i+0}^{2}\Bigg(qp_{k_{3}+i+0}-\frac{Mz\cos^{2}\Theta}{(\rho_{i}-a)^{2}}\Bigg)\Bigg] 
\end{eqnarray*}
\begin{eqnarray*}
{\bf g_{(i:\;2\rightarrow k_{1}-1),\;\; k_{4}+i+0} } = 2 \pi v_{1}h_{\rho}h_{z}\rho_{i-1}\Bigg[ \frac{1}{h_{\rho}}\Bigg(\frac{p_{k_{4}+i+0}-p_{k_{4}+i-1+0}}{h_{\rho}}-{}
       \nonumber\\
{}-\frac{p_{k_{3}+i-1+ii_{+}}-p_{k_{3}+i-1+0}}{h_{z}}\Bigg)\Bigg]{}
                  \nonumber\\
{}+2 \pi v_{1}h_{\rho}h_{z}\rho_{i}\Bigg[ -\frac{1}{h_{\rho}}\Bigg(\frac{p_{k_{4}+i+1+0}-p_{k_{4}+i+0}}{h_{\rho}}-\frac{p_{k_{3}+i+ii_{+}}-p_{k_{3}+i+0}}{h_{z}}\Bigg)+{}
                \nonumber\\
{}+2qp_{i+0}^{2}\Bigg(qp_{k_{4}+i+0}+\frac{M\cos^{2}\Theta}{\rho_{i}-a}\Bigg)\Bigg] \\
\end{eqnarray*}
The derivatives of all functions on all the $\rho$ levels except for the last level and the last $z$ level follow. The last $z$ level $ii_{max}$ is in fact the $z=z_{0}$.
\begin{eqnarray*}
{\bf g_{(i:\;1\rightarrow k_{1}-1),\;\; i+ii_{max}} } = 2 \pi v_{1}h_{\rho}h_{z}\rho_{i}\Bigg[\frac{2}{h_{z}^{2}}(p_{i+ii_{max}}-p_{i+ii_{max-1}})\Bigg] \\
{\bf g_{(i:\;1\rightarrow k_{1}-1),\;\; k_{1}+i+ii_{max}} } = 2 \pi v_{1}h_{\rho}h_{z}\rho_{i}\Bigg[\frac{2}{h_{z}^{2}}(p_{k_{1}+i+ii_{max}}-p_{k_{1}+i+ii_{max-1}})\Bigg] \\
{\bf g_{(i:\;1\rightarrow k_{1}-1),\;\; k_{2}+i+ii_{max}} } = 2 \pi v_{1}h_{\rho}h_{z}\rho_{i}\Bigg[\frac{1}{2\rho_{i}^{2}}\Bigg(
\frac{2}{h_{z}^{2}}(p_{k_{1}+i+ii_{max}}-p_{k_{1}+i+ii_{max-1}})\Bigg)\Bigg] \\
{\bf g_{(i:\;1\rightarrow k_{1}-1),\;\; k_{3}+i+ii_{max}} } = 2 \pi v_{1}h_{\rho}h_{z}\rho_{i}\Bigg[-\frac{1}{h_{z}}
\Bigg(\frac{p_{k_{4}+i+1+ii_{max-1}}-p_{k_{4}+i+ii_{max-1}}}{h_{\rho}}-{}
                  \nonumber\\
{}-\frac{p_{k_{3}+i+ii_{max}}-p_{k_{3}+i+ii_{max-1}}}{h_{z}}\Bigg)\Bigg] \\
{\bf g_{(i:\;1\rightarrow k_{1}-1),\;\; k_{4}+i+ii_{max}} } = 0
\end{eqnarray*}

\setlength{\unitlength}{1mm}
\begin{picture}(80,80)
\thicklines
%\put(-5,40){\vector(1,0){160}}
\put(-5,40){\vector(1,0){130}}
\put(-5,5){\vector(0,1){70}}

\put(20,5){\line(0,1){70}}
%\put(55,5){\line(0,1){70}}
\put(45,5){\line(0,1){70}}
%\put(85,5){\line(0,1){70}}
\put(70,5){\line(0,1){70}}
%\put(115,5){\line(0,1){70}}
\put(95,5){\line(0,1){70}}
%\put(145,5){\line(0,1){70}}
\put(120,5){\line(0,1){70}}

\thinlines
\put(0,12){\line(0,1){56}}
\put(5,12){\line(0,1){56}}
\put(10,12){\line(0,1){56}}
\put(15,12){\line(0,1){56}}
\put(20,12){\line(0,1){56}}
\put(25,12){\line(0,1){56}}
\put(30,12){\line(0,1){56}}
\put(35,12){\line(0,1){56}}
\put(40,12){\line(0,1){56}}
\put(45,12){\line(0,1){56}}
\put(50,12){\line(0,1){56}}
\put(55,12){\line(0,1){56}}
\put(60,12){\line(0,1){56}}
\put(65,12){\line(0,1){56}}
\put(70,12){\line(0,1){56}}
\put(75,12){\line(0,1){56}}
\put(80,12){\line(0,1){56}}
\put(85,12){\line(0,1){56}}
\put(90,12){\line(0,1){56}}
\put(95,12){\line(0,1){56}}
\put(100,12){\line(0,1){56}}
\put(105,12){\line(0,1){56}}
\put(110,12){\line(0,1){56}}
\put(115,12){\line(0,1){56}}
\put(120,12){\line(0,1){56}}
%\put(125,12){\line(0,1){56}}
%\put(130,12){\line(0,1){56}}
%\put(135,12){\line(0,1){56}}
%\put(140,12){\line(0,1){56}}
%\put(145,12){\line(0,1){56}}

%\put(-5,12){\line(1,0){150}}
%\put(-5,16){\line(1,0){150}}
%\put(-5,20){\line(1,0){150}}
%\put(-5,24){\line(1,0){150}}
%\put(-5,28){\line(1,0){150}}
%\put(-5,32){\line(1,0){150}}
%\put(-5,36){\line(1,0){150}}
%\put(-5,40){\line(1,0){150}}
%\put(-5,44){\line(1,0){150}}
%\put(-5,48){\line(1,0){150}}
%\put(-5,52){\line(1,0){150}}
%\put(-5,56){\line(1,0){150}}
%\put(-5,60){\line(1,0){150}}
%\put(-5,64){\line(1,0){150}}
%\put(-5,68){\line(1,0){150}}

\put(-5,12){\line(1,0){125}}
\put(-5,16){\line(1,0){125}}
\put(-5,20){\line(1,0){125}}
\put(-5,24){\line(1,0){125}}
\put(-5,28){\line(1,0){125}}
\put(-5,32){\line(1,0){125}}
\put(-5,36){\line(1,0){125}}
\put(-5,40){\line(1,0){125}}
\put(-5,44){\line(1,0){125}}
\put(-5,48){\line(1,0){125}}
\put(-5,52){\line(1,0){125}}
\put(-5,56){\line(1,0){125}}
\put(-5,60){\line(1,0){125}}
\put(-5,64){\line(1,0){125}}
\put(-5,68){\line(1,0){125}}

\put(-4,69){$1$}
%\put(20,69){$k_{1}$}
%\put(26,69){$k_{1}+1$}
\put(15,69){$k_{1}$}
\put(21,69){$k_{1}+1$}
%\put(50,69){$k_{2}$}
%\put(56,69){$k_{2}+1$}
\put(40,69){$k_{2}$}
\put(46,69){$k_{2}+1$}
%\put(80,69){$k_{3}$}
%\put(86,69){$k_{3}+1$}
\put(65,69){$k_{3}$}
\put(71,69){$k_{3}+1$}
%\put(110,69){$k_{4}$}
%\put(116,69){$k_{4}+1$}
\put(90,69){$k_{4}$}
\put(96,69){$k_{4}+1$}
%\put(142,69){$n$}
\put(117,69){$n$}
\thicklines
%\put(65,77){\vector(1,0){25}}
%\put(76,79){$i$}

\put(55,77){\vector(1,0){25}}
\put(66,79){$i$}

%\put(146,68){$z=z_{0}$}
%\put(146,12){$z=-z_{0}$}
%\put(148,16){\vector(0,1){20}}
%\put(150,25){$ii$}
%\put(156,39){$z=0$}

\put(121,68){$z=z_{0}$}
\put(121,12){$z=-z_{0}$}
\put(130,16){\vector(0,1){20}}
\put(132,25){$ii$}
\put(126,39){$z=0$}

%\put(-4.5,8.5){$\rho_{min}$}
%\put(16.5,8.5){$\rho_{max}$}
%\put(25.5,8.5){$\rho_{min}$}
%\put(46.5,8.5){$\rho_{max}$}
%\put(55.5,8.5){$\rho_{min}$}
%\put(76.5,8.5){$\rho_{max}$}
%\put(85.5,8.5){$\rho_{min}$}
%\put(106.5,8.5){$\rho_{max}$}
%\put(115.5,8.5){$\rho_{min}$}
%\put(136.5,8.5){$\rho_{max}$}

\put(-4.5,8.5){$\rho_{min}$}
\put(11.5,8.5){$\rho_{max}$}
\put(20.5,8.5){$\rho_{min}$}
\put(36.5,8.5){$\rho_{max}$}
\put(45.5,8.5){$\rho_{min}$}
\put(61.5,8.5){$\rho_{max}$}
\put(70.5,8.5){$\rho_{min}$}
\put(86.5,8.5){$\rho_{max}$}
\put(95.5,8.5){$\rho_{min}$}
\put(111.5,8.5){$\rho_{max}$}

%\put(4,4){${\mathbf F(\rho,z)}$}
%\put(34,4){${\mathbf P(\rho,z)}$}
%\put(64,4){${\mathbf A_{\varphi}(\rho,z)}$}
%\put(94,4){${\mathbf W_{\rho}(\rho,z)}$}
%\put(124,4){${\mathbf W_{z}(\rho,z)}$}

\put(2.5,4){${\mathbf F(\rho,z)}$}
\put(26.5,4){${\mathbf P(\rho,z)}$}
\put(50.5,4){${\mathbf A_{\varphi}(\rho,z)}$}
\put(75,4){${\mathbf W_{\rho}(\rho,z)}$}
\put(100.5,4){${\mathbf W_{z}(\rho,z)}$}

\put(-5,40){\circle*{1}}
\put(-5,44){\circle*{1}}
\put(-5,48){\circle*{1}}
\put(-5,52){\circle*{1}}
\put(-5,56){\circle*{1}}
\put(-5,60){\circle*{1}}
\put(-5,64){\circle*{1}}
\put(-5,68){\circle*{1}}
\put(-5,36){\circle*{1}}
\put(-5,32){\circle*{1}}
\put(-5,28){\circle*{1}}
\put(-5,24){\circle*{1}}
\put(-5,20){\circle*{1}}
\put(-5,16){\circle*{1}}
\put(-5,12){\circle*{1}}
%======================================
\put(0,40){\circle*{1}}
\put(0,44){\circle*{1}}
\put(0,48){\circle*{1}}
\put(0,52){\circle*{1}}
\put(0,56){\circle*{1}}
\put(0,60){\circle*{1}}
\put(0,64){\circle*{1}}
\put(0,68){\circle*{1}}
\put(0,36){\circle*{1}}
\put(0,32){\circle*{1}}
\put(0,28){\circle*{1}}
\put(0,24){\circle*{1}}
\put(0,20){\circle*{1}}
\put(0,16){\circle*{1}}
\put(0,12){\circle*{1}}
%====================================
\put(5,40){\circle*{1}}
\put(5,44){\circle*{1}}
\put(5,48){\circle*{1}}
\put(5,52){\circle*{1}}
\put(5,56){\circle*{1}}
\put(5,60){\circle*{1}}
\put(5,64){\circle*{1}}
\put(5,68){\circle*{1}}
\put(5,36){\circle*{1}}
\put(5,32){\circle*{1}}
\put(5,28){\circle*{1}}
\put(5,24){\circle*{1}}
\put(5,20){\circle*{1}}
\put(5,16){\circle*{1}}
\put(5,12){\circle*{1}}
%===========================================
\put(10,40){\circle*{1}}
\put(10,44){\circle*{1}}
\put(10,48){\circle*{1}}
\put(10,52){\circle*{1}}
\put(10,56){\circle*{1}}
\put(10,60){\circle*{1}}
\put(10,64){\circle*{1}}
\put(10,68){\circle*{1}}
\put(10,36){\circle*{1}}
\put(10,32){\circle*{1}}
\put(10,28){\circle*{1}}
\put(10,24){\circle*{1}}
\put(10,20){\circle*{1}}
\put(10,16){\circle*{1}}
\put(10,12){\circle*{1}}
%===========================================
\put(15,40){\circle*{1}}
\put(15,44){\circle*{1}}
\put(15,48){\circle*{1}}
\put(15,52){\circle*{1}}
\put(15,56){\circle*{1}}
\put(15,60){\circle*{1}}
\put(15,64){\circle*{1}}
\put(15,68){\circle*{1}}
\put(15,36){\circle*{1}}
\put(15,32){\circle*{1}}
\put(15,28){\circle*{1}}
\put(15,24){\circle*{1}}
\put(15,20){\circle*{1}}
\put(15,16){\circle*{1}}
\put(15,12){\circle*{1}}
%===========================================
\put(20,40){\circle*{1}}
\put(20,44){\circle*{1}}
\put(20,48){\circle*{1}}
\put(20,52){\circle*{1}}
\put(20,56){\circle*{1}}
\put(20,60){\circle*{1}}
\put(20,64){\circle*{1}}
\put(20,68){\circle*{1}}
\put(20,36){\circle*{1}}
\put(20,32){\circle*{1}}
\put(20,28){\circle*{1}}
\put(20,24){\circle*{1}}
\put(20,20){\circle*{1}}
\put(20,16){\circle*{1}}
\put(20,12){\circle*{1}}
%===========================================
\put(25,40){\circle*{1}}
\put(25,44){\circle*{1}}
\put(25,48){\circle*{1}}
\put(25,52){\circle*{1}}
\put(25,56){\circle*{1}}
\put(25,60){\circle*{1}}
\put(25,64){\circle*{1}}
\put(25,68){\circle*{1}}
\put(25,36){\circle*{1}}
\put(25,32){\circle*{1}}
\put(25,28){\circle*{1}}
\put(25,24){\circle*{1}}
\put(25,20){\circle*{1}}
\put(25,16){\circle*{1}}
\put(25,12){\circle*{1}}
%===========================================
\put(30,40){\circle*{1}}
\put(30,44){\circle*{1}}
\put(30,48){\circle*{1}}
\put(30,52){\circle*{1}}
\put(30,56){\circle*{1}}
\put(30,60){\circle*{1}}
\put(30,64){\circle*{1}}
\put(30,68){\circle*{1}}
\put(30,36){\circle*{1}}
\put(30,32){\circle*{1}}
\put(30,28){\circle*{1}}
\put(30,24){\circle*{1}}
\put(30,20){\circle*{1}}
\put(30,16){\circle*{1}}
\put(30,12){\circle*{1}}
%===========================================
\put(35,40){\circle*{1}}
\put(35,44){\circle*{1}}
\put(35,48){\circle*{1}}
\put(35,52){\circle*{1}}
\put(35,56){\circle*{1}}
\put(35,60){\circle*{1}}
\put(35,64){\circle*{1}}
\put(35,68){\circle*{1}}
\put(35,36){\circle*{1}}
\put(35,32){\circle*{1}}
\put(35,28){\circle*{1}}
\put(35,24){\circle*{1}}
\put(35,20){\circle*{1}}
\put(35,16){\circle*{1}}
\put(35,12){\circle*{1}}
%===========================================
\put(40,40){\circle*{1}}
\put(40,44){\circle*{1}}
\put(40,48){\circle*{1}}
\put(40,52){\circle*{1}}
\put(40,56){\circle*{1}}
\put(40,60){\circle*{1}}
\put(40,64){\circle*{1}}
\put(40,68){\circle*{1}}
\put(40,36){\circle*{1}}
\put(40,32){\circle*{1}}
\put(40,28){\circle*{1}}
\put(40,24){\circle*{1}}
\put(40,20){\circle*{1}}
\put(40,16){\circle*{1}}
\put(40,12){\circle*{1}}
%===========================================
\put(45,40){\circle*{1}}
\put(45,44){\circle*{1}}
\put(45,48){\circle*{1}}
\put(45,52){\circle*{1}}
\put(45,56){\circle*{1}}
\put(45,60){\circle*{1}}
\put(45,64){\circle*{1}}
\put(45,68){\circle*{1}}
\put(45,36){\circle*{1}}
\put(45,32){\circle*{1}}
\put(45,28){\circle*{1}}
\put(45,24){\circle*{1}}
\put(45,20){\circle*{1}}
\put(45,16){\circle*{1}}
\put(45,12){\circle*{1}}
%===========================================
\put(50,40){\circle*{1}}
\put(50,44){\circle*{1}}
\put(50,48){\circle*{1}}
\put(50,52){\circle*{1}}
\put(50,56){\circle*{1}}
\put(50,60){\circle*{1}}
\put(50,64){\circle*{1}}
\put(50,68){\circle*{1}}
\put(50,36){\circle*{1}}
\put(50,32){\circle*{1}}
\put(50,28){\circle*{1}}
\put(50,24){\circle*{1}}
\put(50,20){\circle*{1}}
\put(50,16){\circle*{1}}
\put(50,12){\circle*{1}}
%===========================================
\put(55,40){\circle*{1}}
\put(55,44){\circle*{1}}
\put(55,48){\circle*{1}}
\put(55,52){\circle*{1}}
\put(55,56){\circle*{1}}
\put(55,60){\circle*{1}}
\put(55,64){\circle*{1}}
\put(55,68){\circle*{1}}
\put(55,36){\circle*{1}}
\put(55,32){\circle*{1}}
\put(55,28){\circle*{1}}
\put(55,24){\circle*{1}}
\put(55,20){\circle*{1}}
\put(55,16){\circle*{1}}
\put(55,12){\circle*{1}}
%===========================================
\put(60,40){\circle*{1}}
\put(60,44){\circle*{1}}
\put(60,48){\circle*{1}}
\put(60,52){\circle*{1}}
\put(60,56){\circle*{1}}
\put(60,60){\circle*{1}}
\put(60,64){\circle*{1}}
\put(60,68){\circle*{1}}
\put(60,36){\circle*{1}}
\put(60,32){\circle*{1}}
\put(60,28){\circle*{1}}
\put(60,24){\circle*{1}}
\put(60,20){\circle*{1}}
\put(60,16){\circle*{1}}
\put(60,12){\circle*{1}}
%===========================================
\put(65,40){\circle*{1}}
\put(65,44){\circle*{1}}
\put(65,48){\circle*{1}}
\put(65,52){\circle*{1}}
\put(65,56){\circle*{1}}
\put(65,60){\circle*{1}}
\put(65,64){\circle*{1}}
\put(65,68){\circle*{1}}
\put(65,36){\circle*{1}}
\put(65,32){\circle*{1}}
\put(65,28){\circle*{1}}
\put(65,24){\circle*{1}}
\put(65,20){\circle*{1}}
\put(65,16){\circle*{1}}
\put(65,12){\circle*{1}}
%===========================================
\put(70,40){\circle*{1}}
\put(70,44){\circle*{1}}
\put(70,48){\circle*{1}}
\put(70,52){\circle*{1}}
\put(70,56){\circle*{1}}
\put(70,60){\circle*{1}}
\put(70,64){\circle*{1}}
\put(70,68){\circle*{1}}
\put(70,36){\circle*{1}}
\put(70,32){\circle*{1}}
\put(70,28){\circle*{1}}
\put(70,24){\circle*{1}}
\put(70,20){\circle*{1}}
\put(70,16){\circle*{1}}
\put(70,12){\circle*{1}}
%===========================================
\put(75,40){\circle*{1}}
\put(75,44){\circle*{1}}
\put(75,48){\circle*{1}}
\put(75,52){\circle*{1}}
\put(75,56){\circle*{1}}
\put(75,60){\circle*{1}}
\put(75,64){\circle*{1}}
\put(75,68){\circle*{1}}
\put(75,36){\circle*{1}}
\put(75,32){\circle*{1}}
\put(75,28){\circle*{1}}
\put(75,24){\circle*{1}}
\put(75,20){\circle*{1}}
\put(75,16){\circle*{1}}
\put(75,12){\circle*{1}}
%===========================================
\put(80,40){\circle*{1}}
\put(80,44){\circle*{1}}
\put(80,48){\circle*{1}}
\put(80,52){\circle*{1}}
\put(80,56){\circle*{1}}
\put(80,60){\circle*{1}}
\put(80,64){\circle*{1}}
\put(80,68){\circle*{1}}
\put(80,36){\circle*{1}}
\put(80,32){\circle*{1}}
\put(80,28){\circle*{1}}
\put(80,24){\circle*{1}}
\put(80,20){\circle*{1}}
\put(80,16){\circle*{1}}
\put(80,12){\circle*{1}}
%===========================================
\put(85,40){\circle*{1}}
\put(85,44){\circle*{1}}
\put(85,48){\circle*{1}}
\put(85,52){\circle*{1}}
\put(85,56){\circle*{1}}
\put(85,60){\circle*{1}}
\put(85,64){\circle*{1}}
\put(85,68){\circle*{1}}
\put(85,36){\circle*{1}}
\put(85,32){\circle*{1}}
\put(85,28){\circle*{1}}
\put(85,24){\circle*{1}}
\put(85,20){\circle*{1}}
\put(85,16){\circle*{1}}
\put(85,12){\circle*{1}}
%===========================================
\put(90,40){\circle*{1}}
\put(90,44){\circle*{1}}
\put(90,48){\circle*{1}}
\put(90,52){\circle*{1}}
\put(90,56){\circle*{1}}
\put(90,60){\circle*{1}}
\put(90,64){\circle*{1}}
\put(90,68){\circle*{1}}
\put(90,36){\circle*{1}}
\put(90,32){\circle*{1}}
\put(90,28){\circle*{1}}
\put(90,24){\circle*{1}}
\put(90,20){\circle*{1}}
\put(90,16){\circle*{1}}
\put(90,12){\circle*{1}}
%===========================================
\put(95,40){\circle*{1}}
\put(95,44){\circle*{1}}
\put(95,48){\circle*{1}}
\put(95,52){\circle*{1}}
\put(95,56){\circle*{1}}
\put(95,60){\circle*{1}}
\put(95,64){\circle*{1}}
\put(95,68){\circle*{1}}
\put(95,36){\circle*{1}}
\put(95,32){\circle*{1}}
\put(95,28){\circle*{1}}
\put(95,24){\circle*{1}}
\put(95,20){\circle*{1}}
\put(95,16){\circle*{1}}
\put(95,12){\circle*{1}}
%===========================================
\put(100,40){\circle*{1}}
\put(100,44){\circle*{1}}
\put(100,48){\circle*{1}}
\put(100,52){\circle*{1}}
\put(100,56){\circle*{1}}
\put(100,60){\circle*{1}}
\put(100,64){\circle*{1}}
\put(100,68){\circle*{1}}
\put(100,36){\circle*{1}}
\put(100,32){\circle*{1}}
\put(100,28){\circle*{1}}
\put(100,24){\circle*{1}}
\put(100,20){\circle*{1}}
\put(100,16){\circle*{1}}
\put(100,12){\circle*{1}}
%===========================================
\put(105,40){\circle*{1}}
\put(105,44){\circle*{1}}
\put(105,48){\circle*{1}}
\put(105,52){\circle*{1}}
\put(105,56){\circle*{1}}
\put(105,60){\circle*{1}}
\put(105,64){\circle*{1}}
\put(105,68){\circle*{1}}
\put(105,36){\circle*{1}}
\put(105,32){\circle*{1}}
\put(105,28){\circle*{1}}
\put(105,24){\circle*{1}}
\put(105,20){\circle*{1}}
\put(105,16){\circle*{1}}
\put(105,12){\circle*{1}}
%===========================================
\put(110,40){\circle*{1}}
\put(110,44){\circle*{1}}
\put(110,48){\circle*{1}}
\put(110,52){\circle*{1}}
\put(110,56){\circle*{1}}
\put(110,60){\circle*{1}}
\put(110,64){\circle*{1}}
\put(110,68){\circle*{1}}
\put(110,36){\circle*{1}}
\put(110,32){\circle*{1}}
\put(110,28){\circle*{1}}
\put(110,24){\circle*{1}}
\put(110,20){\circle*{1}}
\put(110,16){\circle*{1}}
\put(110,12){\circle*{1}}
%===========================================
\put(115,40){\circle*{1}}
\put(115,44){\circle*{1}}
\put(115,48){\circle*{1}}
\put(115,52){\circle*{1}}
\put(115,56){\circle*{1}}
\put(115,60){\circle*{1}}
\put(115,64){\circle*{1}}
\put(115,68){\circle*{1}}
\put(115,36){\circle*{1}}
\put(115,32){\circle*{1}}
\put(115,28){\circle*{1}}
\put(115,24){\circle*{1}}
\put(115,20){\circle*{1}}
\put(115,16){\circle*{1}}
\put(115,12){\circle*{1}}
%===========================================
\put(120,40){\circle*{1}}
\put(120,44){\circle*{1}}
\put(120,48){\circle*{1}}
\put(120,52){\circle*{1}}
\put(120,56){\circle*{1}}
\put(120,60){\circle*{1}}
\put(120,64){\circle*{1}}
\put(120,68){\circle*{1}}
\put(120,36){\circle*{1}}
\put(120,32){\circle*{1}}
\put(120,28){\circle*{1}}
\put(120,24){\circle*{1}}
\put(120,20){\circle*{1}}
\put(120,16){\circle*{1}}
\put(120,12){\circle*{1}}

\end{picture}

Above one can see how a grid, like the one we use, looks like. The five areas where the data points of every function are stored, is clear as well
as the $\rho$ and $z$ levels. 
A brief diagram of the program we use follows. Subroutines DFPMIN and LNSRCH are provided by \cite{c6}. In the following diagram ${\mathbf x}$ is the matrix
that has all the points of the five functions and is the matrix on which the corrections are applied until we have the final position of these points,
the final configuration for the fields and the value of the total energy.
\newline
\newline

\setlength{\unitlength}{1mm}
\begin{picture}(150,150)
\thicklines
\put(-5,150){\line(1,0){110}}
\put(-5,150){\line(0,-1){40}}
\put(-5,110){\line(1,0){110}}
\put(105,150){\line(0,-1){40}}
\put(-2,145){{\bf 1. MAIN PROGRAM}}
\thinlines
\put(-5,143){\line(1,0){110}}
\put(-2,138){$\blacktriangleright$ Insert parameters, initial guess}
\put(-2,133){$\blacktriangleright$ Compute initial $E$ and $\nabla E$}
\put(-2,128){$\blacktriangleright$ CALL DFPMIN--Compute Newton direction ${\mathbf p}$}
\put(90,129){\line(1,0){40}}
\put(130,129){\line(0,-1){32}}
\put(130,97){\vector(-1,0){84}}
\put(-2,123){$\blacktriangleright$ Get final configuration from DFPMIN}
\put(-2,118){$\blacktriangleright$ Compute virial, Produce plots etc.}
\put(-2,113){$\blacktriangleright$ END PROGRAM}
%=========================================================
\thicklines
\put(-5,108){\line(1,0){110}}
\put(-5,108){\line(0,-1){45}}
\put(-5,63){\line(1,0){110}}
\put(105,108){\line(0,-1){45}}
\put(-2,103){{\bf 2. DFPMIN}}
\thinlines
\put(-5,101){\line(1,0){110}}
\put(-2,96){$\blacktriangleright$ Insert data from main}
\put(-2,91){$\blacktriangleright$ Compute Newton direction ${\mathbf p}$ on initial guess}
\put(-2,86){$\blacktriangleright$ CALL LNSRCH--Compute step on ${\mathbf p}$ and correct ${\mathbf x}$}
\put(97,87){\line(1,0){33}}
\put(130,87){\line(0,-1){37}}
\put(130,50){\vector(-1,0){74}}
\put(-2,81){$\blacktriangleright$ Get corrected ${\mathbf x}$ from LNSRCH}
\put(-2,76){$\blacktriangleright$ IF $\Delta {\mathbf x}$ OR $\nabla E$ $\leq 5\cdot 10^{-8}$ RETURN SOLUTION}
\thicklines
\put(94,77){\line(1,0){54}}
\put(148,77){\line(0,1){47}}
\put(148,124){\vector(-1,0){75}}
\put(-2,71){$\blacktriangleright$ ELSE Compute new Newton direction ${\mathbf p}$}
\thinlines
\put(77,72){\line(1,0){23}}
\put(100,72){\line(0,1){18}}
\put(100,90){\line(-1,0){11}}
\put(89,90){\vector(0,-1){2}}
\put(-2,66){$\blacktriangleright$ END DFPMIN}
%=========================================================
\thicklines
\put(-5,61){\line(1,0){110}}
\put(-5,61){\line(0,-1){70}}
\put(-5,-9){\line(1,0){110}}
\put(105,61){\line(0,-1){70}}
\put(-2,56){{\bf 3. LNSRCH}}
\thinlines
\put(-5,54){\line(1,0){110}}
\put(-2,49){$\blacktriangleright$ Insert data from DFPMIN}
\put(-2,44){$\blacktriangleright$ Start with full Newton step $\lambda =1$}
\put(-2,39){$\blacktriangleright$ Correction: ${\mathbf x_{new}}={\mathbf x_{old}} +\lambda {\mathbf p}$}
\put(-2,34){$\blacktriangleright$ Compute $E_{new}({\mathbf x_{new}})$}
\put(-2,29){$\blacktriangleright$ IF $\lambda \leq 5\cdot 10^{-8}$ OR $E_{new}\leq E_{old} +\lambda a g^{'}$ RETURN}
\put(96,30){\line(1,0){39}}
\put(135,30){\line(0,1){52}}
\put(135,82){\vector(-1,0){73}}
\put(-2,24){$\blacktriangleright$ ELSE IF $\lambda =1$ THEN $\lambda_{new}=-g^{'}/2(E_{new}-E_{old}-g^{'})$}
\put(-2,19){$\blacktriangleright$ ELSE IF $b^{2}-3ag^{'}(0)<0$ THEN $\lambda_{new}=\frac{\lambda}{2}$, ($a,b$ given)}
\put(-2,14){$\blacktriangleright$ ELSE $\lambda_{new}=(-b+\sqrt{b^{2}-3ag^{'}(0)})/3a$}
\put(-2,9){$\blacktriangleright$ IF $\lambda_{new}>0.5\lambda$ THEN $\lambda_{new}=0.5\lambda$}
\put(-2,4){$\blacktriangleright$ IF $\lambda_{new}<0.1\lambda$ THEN $\lambda_{new}=0.1\lambda$}
\put(-2,-1){$\blacktriangleright$ $\lambda =\lambda_{new}$, $E_{old}=E_{new}$}
\put(45,-1){\line(1,0){58}}
\put(103,-1){\line(0,1){41}}
\put(103,40){\vector(-1,0){45}}
\put(-2,-6){$\blacktriangleright$ END LNSRCH}
%================================================================
\thicklines
\put(-5,-11){\line(1,0){110}}
\put(-5,-11){\line(0,-1){25}}
\put(-5,-36){\line(1,0){110}}
\put(105,-11){\line(0,-1){25}}
\put(-2,-16){{\bf 4. FUNC}}
\thinlines
\put(-5,-18){\line(1,0){110}}
\put(-2,-23){$\blacktriangleright$ Compute $E$}
\put(-2,-28){$\blacktriangleright$ RETURN the value of $E$}
\put(-2,-33){$\blacktriangleright$ END FUNC}
%================================================================
\thicklines
\put(-5,-38){\line(1,0){110}}
\put(-5,-38){\line(0,-1){25}}
\put(-5,-63){\line(1,0){110}}
\put(105,-38){\line(0,-1){25}}
\put(-2,-43){{\bf 5. DFUNC}}
\thinlines
\put(-5,-45){\line(1,0){110}}
\put(-2,-50){$\blacktriangleright$ Compute the derivatives of $E$}
\put(-2,-55){$\blacktriangleright$ RETURN the values of the derivatives of $E$}
\put(-2,-60){$\blacktriangleright$ END DFUNC}
\end{picture}

\newpage
\begin{itemize}
 \item{\em Concluding remarks for Energy minimization algorithm}
\end{itemize}
It was an algorithm which successfully generated the expected results for our test models in the  shortest time and the output for every model is exhibited 
in the appropriate section. It is the easiest to handle, even when compared to Newton-Raphson, and the results again were multi-tested as one can see in detail. 
Thus, it is used in our research models and provides the solutions we found there.

\clearpage
\newpage

\listoffigures

%******************************************************************************************
%=====================-------------------------============================================
%=====================|     END OF THESIS     |============================================
%=====================|=======================|============================================
%=====================|  BIBLIOGRAPHY FOLLOWS |============================================
%=====================-------------------------============================================
%******************************************************************************************

%\newpage
%\addcontentsline{toc}{chapter}{Bibliography}

\end{document}